\documentclass[aps,prl,twocolumn,10pt,superscriptaddress,nofootinbib,nobibnotes,longbibliography]{revtex4-1}

\usepackage{amssymb}
\usepackage{graphicx}
\usepackage{amsmath}
\usepackage{hyperref}
\usepackage{subfigure}
\usepackage{multirow}
\usepackage{setspace}
\usepackage{verbatim}
\usepackage{float}
\usepackage{color}
\usepackage{ulem}
\usepackage[utf8]{inputenc}
\usepackage[table,xcdraw]{xcolor}
\usepackage{makecell}

\newcommand{\md}{\mathrm{d}}

\begin{document}


\title{Bubbles kick off primordial black holes to form more binaries}

\author{Zi-Yan Yuwen}
\email{yuwenziyan@itp.ac.cn}
\affiliation{Institute of Theoretical Physics, Chinese Academy of Sciences, Beijing 100190, China}
\affiliation{School of Physical Sciences, University of Chinese Academy of Sciences (UCAS), Beijing 100049, China}

\author{Cristian Joana}
\email{cristian.joana@itp.ac.cn}
\affiliation{Institute of Theoretical Physics, Chinese Academy of Sciences, Beijing 100190, China}

\author{Shao-Jiang Wang}
\email{schwang@itp.ac.cn}
\affiliation{Institute of Theoretical Physics, Chinese Academy of Sciences, Beijing 100190, China}
\affiliation{Asia Pacific Center for Theoretical Physics (APCTP), Pohang 37673, Korea}

\author{Rong-Gen Cai}
\email{cairg@itp.ac.cn}
\affiliation{Institute of Fundamental Physics and Quantum Technology, Ningbo University, Ningbo, 315211, China}
\affiliation{Institute of Theoretical Physics, Chinese Academy of Sciences, Beijing 100190, China}
\affiliation{School of Fundamental Physics and Mathematical Sciences, Hangzhou Institute for Advanced Study (HIAS), University of Chinese Academy of Sciences (UCAS), Hangzhou 310024, China}

\begin{abstract}
Primordial black holes (PBHs) may form before cosmological first-order phase transitions, leading to inevitable collisions between PBHs and bubble walls. In this Letter, we have simulated for the first time the co-evolution of an expanding scalar wall passing through a black hole with full numerical relativity. This black hole-bubble wall collision yields multiple far-reaching phenomena, including the PBH mass growth, gravitational wave radiations, and momentum recoil that endows PBHs with additional velocities, approximately doubling the formation rate for PBH binaries and hence strengthening the observational constraints on the PBH abundances.
\end{abstract}
\maketitle

\textit{\textbf{Introduction.}---} 
The cosmological first-order phase transition (FOPT) \cite{Mazumdar:2018dfl,Hindmarsh:2020hop,Caldwell:2022qsj,Athron:2023xlk} is one of the most inspiring phenomena in the early Universe to probe the new physics~\cite{Cai:2017cbj,Bian:2021ini} beyond the standard model of particle physics with associated stochastic gravitational-wave backgrounds (SGWBs) \cite{Caprini:2015zlo,Caprini:2019egz}. It can be also responsible for the production of the baryon asymmetry \cite{Cohen:1990py,Cohen:1993nk,Cohen:2012zza} and primordial magnetic fields \cite{Hogan:1983zz,Di:2020nny,Yang:2021uid}. In particular, recent pulsar-timing-array (PTA) observations~\cite{NANOGrav:2023gor,EPTA:2023sfo,Reardon:2023gzh,Xu:2023wog} have found marginal evidence of SGWBs and renewed the interest in the strongly-coupled systems~\cite{Schwaller:2015tja,Bigazzi:2020avc,Ares:2020lbt,Bigazzi:2020phm,Zhu:2021vkj,Ares:2021ntv,Ares:2021nap,Morgante:2022zvc,Bea:2021zsu,Janik:2022wsx,Bigazzi:2021ucw,Bea:2022mfb,Li:2023xto,Wang:2023lam} with phase transitions at the quantum chromodynamics (QCD)-like scales~\cite{He:2022amv,He:2023ado,Gouttenoire:2023bqy,Salvio:2023blb} (see also~\cite{NANOGrav:2023hvm,Addazi:2023jvg,Athron:2023mer,Fujikura:2023lkn,Han:2023olf,Franciolini:2023wjm,Bian:2023dnv,Jiang:2023qbm,Ghosh:2023aum,Xiao:2023dbb,Li:2023bxy,DiBari:2023upq,Cruz:2023lnq,Wu:2023hsa,Du:2023qvj,Ahmadvand:2023lpp,Wang:2023bbc}).

Besides the ubiquitous appearances of FOPTs in various aspects of theoretical perspectives, the formation of primordial black holes (PBHs) is also expected as a general theoretical interest of the early Universe. Apart from earlier proposals of PBH formations from bubble wall collisions~\cite{Hawking:1982ga,Crawford:1982yz,Moss:1994pi,Moss:1994iq} and recently proposed model-dependent particle trapping mechanisms~\cite{Baker:2021nyl,Baker:2021sno,Kawana:2021tde,Lu:2022paj,Kawana:2022lba,Huang:2022him}, PBHs can also be generally produced from delayed-decayed false-vacuum islands of  FOPTs~\cite{Kodama:1982sf,Liu:2021svg,Kawana:2022olo,Gouttenoire:2023naa,Lewicki:2023ioy,Kanemura:2024pae,Cai:2024nln}  in a model-independent manner~\cite{He:2022amv,He:2023ado,Gouttenoire:2023bqy,Salvio:2023blb,Hashino:2021qoq,Hashino:2022tcs,Salvio:2023ynn,Gouttenoire:2023pxh,Conaci:2024tlc,Baldes:2023rqv} as long as the concomitant overdensities~\cite{Liu:2022lvz,Elor:2023xbz,Lewicki:2024ghw,Cai:2024nln} are large enough to reach the PBH formation threshold. Moreover, PBHs can also be generated from primordial curvature perturbations and other topological defects in the early Universe other than cosmological FOPTs (see recent reviews~\cite{Carr:2020xqk,Carr:2021bzv,Escriva:2022duf,Ozsoy:2023ryl,LISACosmologyWorkingGroup:2023njw} and references therein).

Provided that FOPTs and PBHs are both frequently anticipated in the early Universe, the encounters between bubbles and PBHs are inevitable with intriguing outcomes of wide interest. Previous attention has only been focused on how the presence of the BH singularity would affect the bubble nucleation rate~ \cite{Gregory:2013hja,Burda:2015isa,Burda:2015yfa,Burda:2016mou,Oshita:2016oqn,Mukaida:2017bgd,Gorbunov:2017fhq,Canko:2017ebb,Kohri:2017ybt,Gregory:2018bdt,Oshita:2019jan,Gregory:2020cvy}, the reverse impact has never been explored before in the literature as far as we know. The closest consideration is the recent numerical simulations~\cite{He:2022sjf,Yin:2023kzr} of the GW (tensor wave) propagation in a fixed BH background. In this \textit{Letter}, we turn to simulate for the first time the bubble wall (scalar wave) propagation passing through a dynamically co-evolving BH background via full numerical relativity code~\texttt{GRChombo}~\cite{Clough:2015sqa,Andrade:2021rbd}. For convenience, we choose $c = \hbar = 1$ throughout this letter and $G=1$ in the simulation.

\textit{\textbf{Simulation.}---} 
We simulate the co-evolution of an encounter between a bubble wall and a BH configuration governed by the total action
\begin{align}
    S=\int\md^4 x\sqrt{-g}\left(\frac{R}{16\pi G} - \frac{1}{2}g^{\mu\nu}\partial_\mu \phi \partial_\nu\phi - V(\phi) \right)
\end{align}
with a scalar potential~\cite{Lewicki:2019gmv}
\begin{align}
\begin{aligned}\label{eq: Scalar_Potential}
    V(\phi) = &\left( 1 + \lambda \tilde{\phi}^2 - (2\lambda+4) \tilde{\phi}^3 + (\lambda+3)  \tilde{\phi}^4 \right)(V_F - V_T) + V_T
\end{aligned}
\end{align}
admitting a false vacuum $V_F$ at $\tilde{\phi} = 0$ and a true vacuum $V_T$ at $\tilde{\phi}\equiv\phi/\phi_0=1$ separated by a potential barrier adjusted by parameter $\lambda$, and equations of motions read
\begin{align}
    R_{\mu\nu} - \frac{1}{2} R g_{\mu\nu} &= 8\pi G T_{\mu\nu}^\phi, \\ T_{\mu\nu}^\phi = \nabla_\mu \phi \nabla_\nu \phi &- \frac{1}{2} g_{\mu\nu} \left( \nabla_{\sigma}\nabla^{\sigma}\phi + 2V \right),\\
    \nabla_\mu \nabla^\mu \phi - \frac{\md V}{\md \phi} &= 0.
\end{align}
The scalar field initially stays at the false vacuum, then suddenly nucleates a true vacuum bubble due to quantum tunnelings. Numerical relativity cannot depict this quantum tunneling process in real space~\footnote{See, however, Refs.~\cite{Braden:2018tky,Blanco-Pillado:2019xny,Wang:2019hjx} for nucleating bubbles in real space from occasionally developed zero-point quantum fluctuations.}, and hence we have to manually input a bubble profile similar to a static bounce solution as the initial condition, with an ansatz of a form~\cite{Cutting:2018tjt,Cutting:2020nla}
\begin{align}\label{eq: bubble initial data}
    \phi_\mathrm{ini}(r) = \frac{\phi_0}{2}\left(1 - \tanh\left( \frac{r-r_0}{D_0} \right) \right),  \quad \dot{\phi}_\mathrm{ini}(r) = 0,
\end{align}
where $r_0$ and $D_0$ are the initial radius and width of the bubble, respectively, and $r_0$ should be slightly larger than the critical radius of the instanton solution for the nucleated bubble to expand. The pre-existing PBH is assumed to be placed in the false vacuum. The initial values for the metric components are given by Hamiltonian and momentum constraints by setting a Schwarzschild BH in the moving-puncture gauge at the center of the grid with a fixed initial value of $\phi$ given in~\eqref{eq: bubble initial data}, and then solving the constraints by iterations until the results converge. Details of the numerical simulation methods are provided in the Supplemental Materials~\cite{footnote}.

The simulation is implemented within a box of size $128 M\times 128 M\times 128M$ for some length scale $M$. PBHs formed via gravitational collapse during radiation domination are expected to have negligible initial velocities, as velocity perturbations are quickly decayed during inflation and are highly suppressed relative to density perturbations in this era~\cite{Carr:1975qj,Sasaki:2016jop,DeLuca:2020bjf}. Therefore, a PBH of initial mass $m_\mathrm{i} \simeq 0.5M$ is set at rest at the center of the grid, and as vacuum bubble described by the ansatz~\eqref{eq: bubble initial data} is nucleated at $40M$ away along $x$-direction. The model parameters in the scalar potential are fixed at $V_F=10^{-5}M^{-4}$, $V_T=0$, $\phi_0 = 0.01 M$, and $\lambda=10$. The ansatz parameters for the initial bubble profile are fixed at $r_0= 15M$ and $D_0 = 3M$. As our system is not bounded, we have chosen a periodic boundary condition in the $x$-direction. Further accounting for the rotational symmetry, the reflective boundary conditions are applied on the $y$ and $z$ boundaries.

\begin{figure}[t]
    \centering
    \includegraphics[width=0.48\textwidth]{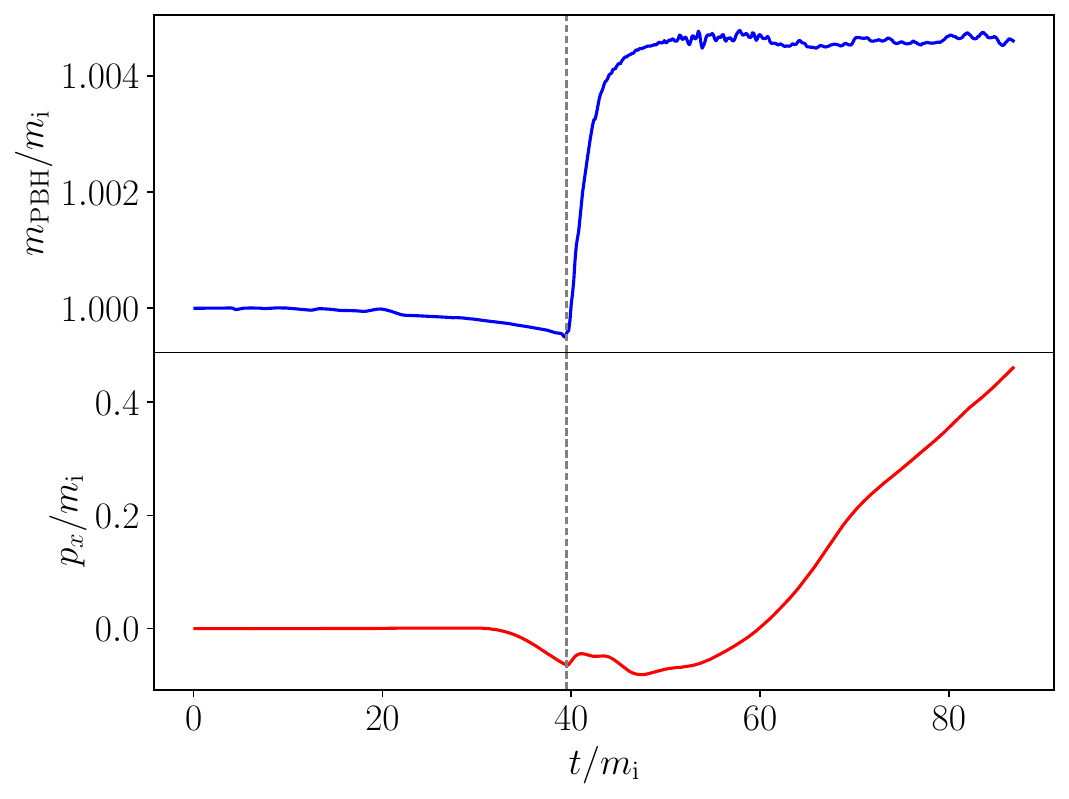}
    \caption{The time evolution for the PBH mass $m_{\mathrm{PBH}}$ and momentum $p_x$ along $x$-direction with the vertical gray dashed line denoting the collision time between the bubble and PBH.}
    \label{fig:mass and momentum}
\end{figure}

\begin{figure*}[t]
    \centering
    \includegraphics[width = 0.95\textwidth]{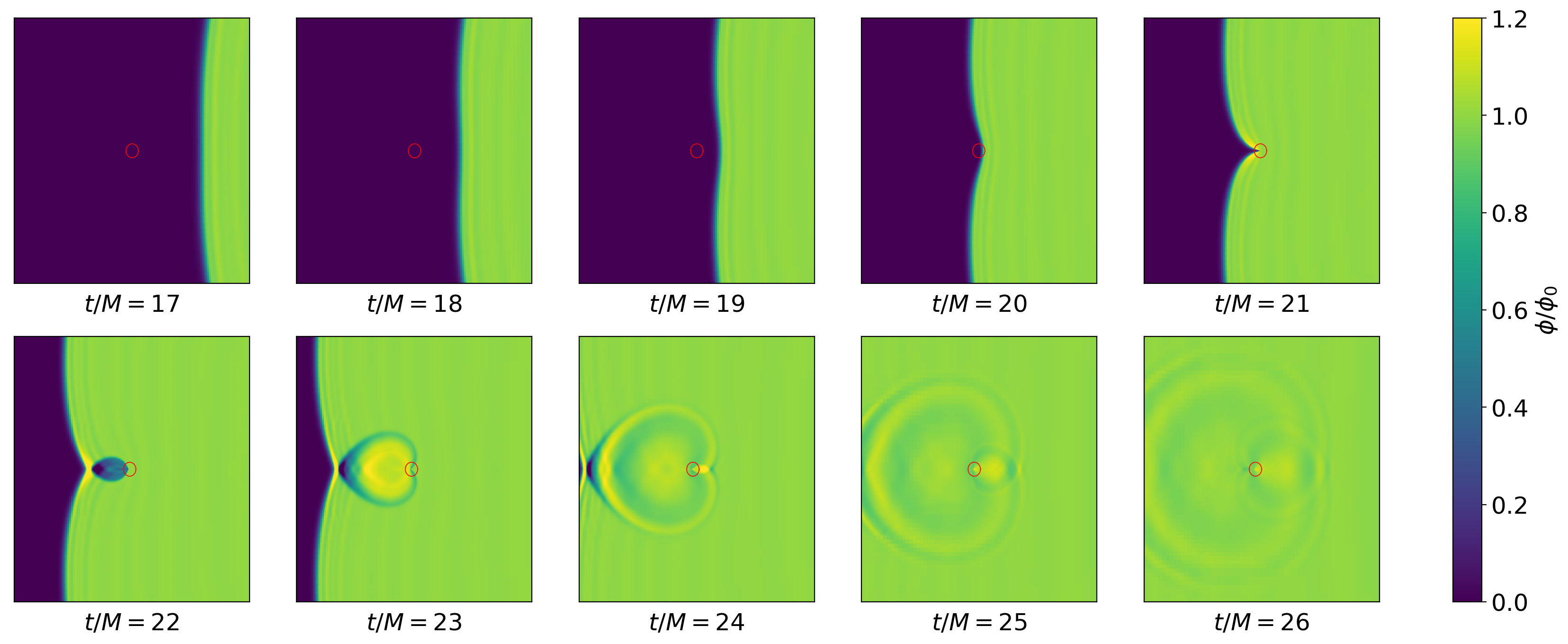}
    \caption{The time slices for an expanding scalar wall passing through a BH with its apparent horizon shown with a red circle.}
    \label{fig:time slice for bubble}
\end{figure*}

\textit{\textbf{Phenomenology.}---} 
The PBH mass and momentum along $x$-direction are shown in Fig.~\ref{fig:mass and momentum} as functions of time, while the time evolution for the scalar field profile passing through a PBH is depicted in Fig.~\ref{fig:time slice for bubble}. It is easy to see that the bubble expands and collides with the PBH roughly at $t=20M$. Multiple far-reaching phenomena can be found during this PBH-wall collision including
\begin{itemize}
    \item \textbf{PBH mass growth:} The PBH extracts energy from the bubble wall passing by, leading to a mass enhancement as shown in the top panel of Fig.~\ref{fig:mass and momentum}, where the mass of this dynamically co-evolving PBH is defined within the apparent horizon. Although the mass only increases by about $0.5\%$ in our simulation, this phenomenon is expected to be more significant at a later stage of bubble expansion as the energy density in the bubble wall grows with its radius. Moreover, multi-step FOPTs can further increase the PBH mass for each encounter with bubbles from different FOPTs.
    \item \textbf{PBH velocity recoil:} The strong collision with the bubble wall would transfer some of its momentum to the PBH as shown in the bottom panel of Fig.~\ref{fig:mass and momentum}, potentially increasing the probability of a PBH running into the other one to form a binary as estimated shortly below. An interesting feature of this momentum transfer is that the bubble wall does not simply push the black hole away but eventually pulls it toward the center of the bubble, indicating that one PBH can only be affected by one vacuum bubble alone.
    \item \textbf{GW radiations:} Non-linear interactions between the PBH and the scalar field disrupt the spherical symmetry of the bubble wall, leading to GW radiations from both the bubble and PBH. Furthermore, the scalar field also oscillates at a length scale comparable to the PBH radius as shown in Fig.~\ref{fig:time slice for bubble} with an expanding heart-shaped profile, and hence this part of GW contribution is produced in the high-frequency band, and the amplitude of this oscillating scalar field is comparable to that of bubble collisions as we have checked in the numerical simulation. Since accurately resolving this high-frequency GW background from numerical simulations is computationally demanding, and should better be done within a radiation fluid background, and hence this task will be left for future work.

\end{itemize}

Here we provide an estimation for the peak frequency of GWs from bubble-PBH collisions. Let us consider a typical case where PBHs were produced during an electro-weak (EW) PT at $T_{\mathrm{EW}}\simeq 100\mathrm{GeV}$ with efficiency factor $\gamma= 0.2$~\cite{Carr:1975qj,Carr:2009jm}, colliding with vacuum bubbles from a QCD phase transition at $T_{\mathrm{QCD}}\simeq 150 \mathrm{MeV}$ with nucleation rate $\beta/H_{\mathrm{PT}}= 10$. Then, we can estimate the BH radius $r_s = 2m_{\mathrm{PBH}} \simeq 2(\gamma/2) H^{-1}_{\mathrm{PBH}} = \gamma H^{-1}_{\mathrm{PBH}}$ and the averaged initial bubble separation $d\simeq \beta^{-1} \simeq H^{-1}_{\mathrm{PT}}/10$, and hence the peak frequency ratio would be 
\begin{align}
    \frac{f_{\mathrm{PBH-b}}}{f_{\mathrm{b-b}}} \simeq \frac{d}{r_s} = 0.5
    \frac{H_{\mathrm{PBH}}}{H_{\mathrm{PT}}} \simeq 0.5 \left(\frac{T_{\mathrm{EW}}}{T_{\mathrm{QCD}}}\right)^2 \simeq 5\times 10^5.
\end{align}
The characteristic peak frequency for GWs from QCD phase transition lies within the nano-Hertz band that can be detected by PTA, indicating the peak frequency for GWs from PBH-bubble collisions is approximately within $\mathcal{O}(10^{-4}) \sim \mathcal{O}(10^{-3})$ Hz, falling into the observation band of future space-borne GW detections such as LISA, Taiji, and TianQin. If we switch the PT model from QCD scale to EW scale with peak frequency of GWs from FOPT as $\mathcal{O}(10^{-4})$, then the peak frequency of GWs from PBH-bubble collisions within $\mathcal{O}(10^{1})\sim\mathcal{O}(10^{2})$ may be detectable by Advanced-LIGO or Einstein Telescope.

\textit{\textbf{PBH binary formation.}---} 
Now let us work out how the velocity recoils induced on PBHs from passing scalar walls would enhance the PBH binary formation rate and correspondingly the PBH abundance constraints. Evaluating the merger rate of PBH binaries is studied in an analytical \cite{Ali-Haimoud:2017rtz,Vaskonen:2019jpv} or numerical way \cite{Delos:2024poq} by considering the conditions for forming bounded orbits. Here we provide a relatively rough estimation. First, let us look at the case of PBH binary formation without initial velocity. A pair of neighboring PBHs should decouple from the cosmic expansion and become gravitationally bounded when their gravitational interaction is strong enough~\cite{Sasaki:2016jop, DeLuca:2020bjf}. As our PBHs encounter bubbles from the early Universe, we will not consider the binary formation at the present time~\cite{Bird:2016dcv}. More specifically, for a pair of PBHs with an equal mass $m_{\mathrm{PBH}}$ separated by a comoving distance $x$, the decoupling occurs if $m_{\mathrm{PBH}}\cdot (ax)^{-3}>\rho$, where $\rho$ is the total energy density of the background evolution. Let $f$ be the fraction of PBHs in the total dust matter and $\bar{x}$ be the average comoving distance of PBHs. As the energy densities of PBHs are redshifted like pressureless matters, PBHs cannot form binaries in the radiation-dominated epoch. Thus, in the matter-dominated epoch, the above condition for PBH binary formation can be reformulated as
\begin{align}\label{eq: decouple condition}
f \cdot \left(\frac{\bar{x}}{x}\right)^3 > 1.
\end{align}
Assuming the distances between neighboring PBHs are homogeneously distributed within the interval $(0, x_{\mathrm{max}})$, where we have set $x_{\mathrm{max}} = 4\bar{x}/3$ to keep $\bar{x}$ as the average separation. Therefore, for PBHs with no initial velocity, the probability of PBH binary formation can be estimated from Eq.~\eqref{eq: decouple condition} as the fraction of the decoupling region, 
\begin{align}
    P_0 = \frac{\frac43\pi(f^{1/3}\bar{x})^3}{\frac43\pi x_\mathrm{max}^3} = \frac{27}{64}f.
\end{align}

Next, let us turn to the case of binary formation with a modest initial PBH velocity induced by the passing scalar wall. The basic picture is that the induced initial PBH velocity would extend the spherical volume of the decoupling region into a cylinder tube enclosed by two hemispheroids on the two ends. See the Supplemental Material~\cite{footnote} for a visualized demonstration. On the one hand, the length of the cylinder tube can be estimated by looking into the comoving distance $\Delta x$ over which a PBH can move during an expanding background from where it originally formed to today. To estimate the PBH movement in an expanding background with $\md s^2 = -\md t^2 + a^2\md \mathbf{x}^2$, we can solve the geodesic equation for the $x$-directional $4$-velocity of a massive particle as
\begin{align}
    u^\mu \equiv \frac{\md x^\mu}{\md \tau} = \left(\sqrt{\frac{(a_{\mathrm{PT}} v)^2}{(a/a_{\mathrm{PT}})^2}+1}, \frac{v}{(a/a_{\mathrm{PT}})^2}, 0, 0\right),
\end{align}
where $a_{\mathrm{PT}}$ is the scale factor around the FOPT, and $v$ is a constant, while the dimensionless term $a_{\mathrm{PT}} v$ can be interpreted as the ``physical velocity'' of the PBH when its Lorentz factor ${1/\sqrt{1-(a_{\mathrm{PT}} v)^2}\sim O(1)}$ is modestly small. Since the matter-dominated epoch dominates the age of the Universe, then the comoving distance over which a massive particle can move on top of the background from the matter-radiation equality $t_\mathrm{eq}$ to today $t_0$ can be estimated as
\begin{equation}
\begin{split}
    \Delta x &= \int_{t_\mathrm{eq}}^{t_0} \md t \frac{\md x}{\md t} 
    \simeq v t_\mathrm{eq} \int_{t_\mathrm{eq}}^{t_0} \frac{\md (t/t_\mathrm{eq} )}{(a/a_{\mathrm{PT}})^2}  \\
    &= \frac{v }{2H_{\mathrm{eq}}} \int_{a_\mathrm{eq}}^{a_0} \frac{3}{2} \left(\frac{a}{a_\mathrm{eq}^3}\right)^{1/2} \left(\frac{a_\mathrm{PT}}{a}\right)^{2} \md a \\
    &\simeq  \frac{3v}{2H_{\mathrm{eq}}} \left(\frac{a_\mathrm{PT}}{a_\mathrm{eq}}\right)^{2},
\end{split}
\end{equation}
where we have dropped in the last line a term contributed by $a_0$ as $a_0\gg a_\mathrm{eq}$. On the other hand, the radius of the original decoupling sphere, $f^{1/3}\bar{x}$, can be obtained from the average comoving separation $\bar{x}$ according to the definition of $f$ at the matter-radiation equality,
\begin{align}
    f &\equiv \frac{m_\mathrm{PBH}}{\frac{4\pi}{3} (a_{\mathrm{eq}}\bar{x})^3}\left( \frac{1}{2} \frac{3 H^2_\mathrm{eq}}{8\pi G} \right)^{-1} = \frac{2 \gamma H^{-1}_\mathrm{PBH} H^{-2}_\mathrm{eq} }{(a_{\mathrm{eq}}\bar{x})^3},
\end{align}
where $\gamma$ is the usual efficiency factor for the total mass within a Hubble volume to eventually collapse into the PBH at formation time, that is, $m_\mathrm{PBH} = \gamma M_H = \frac{1}{2G}\gamma H^{-1}_\mathrm{PBH}$. Therefore, the ratio
\begin{equation}
\begin{split}
\label{eq: Deltax by xbar}
    \frac{\Delta x}{\bar{x}} = \frac{3}{2}(a_\mathrm{PT}v) \left(\frac{f}{2\gamma}\right)^{1/3} 
    \left(\frac{a_\mathrm{eq}}{a_\mathrm{PBH}}\right)^{2/3} 
    \left(\frac{a_\mathrm{PT}}{a_\mathrm{eq}}\right)
    \equiv \Gamma f^{1/3}
\end{split}
\end{equation}
depicts how far a PBH can move ever since its formation, and we have absorbed all the model dependency into a dimensionless parameter $\Gamma$.

\begin{figure}
    \centering
    \includegraphics[width=0.43\textwidth]{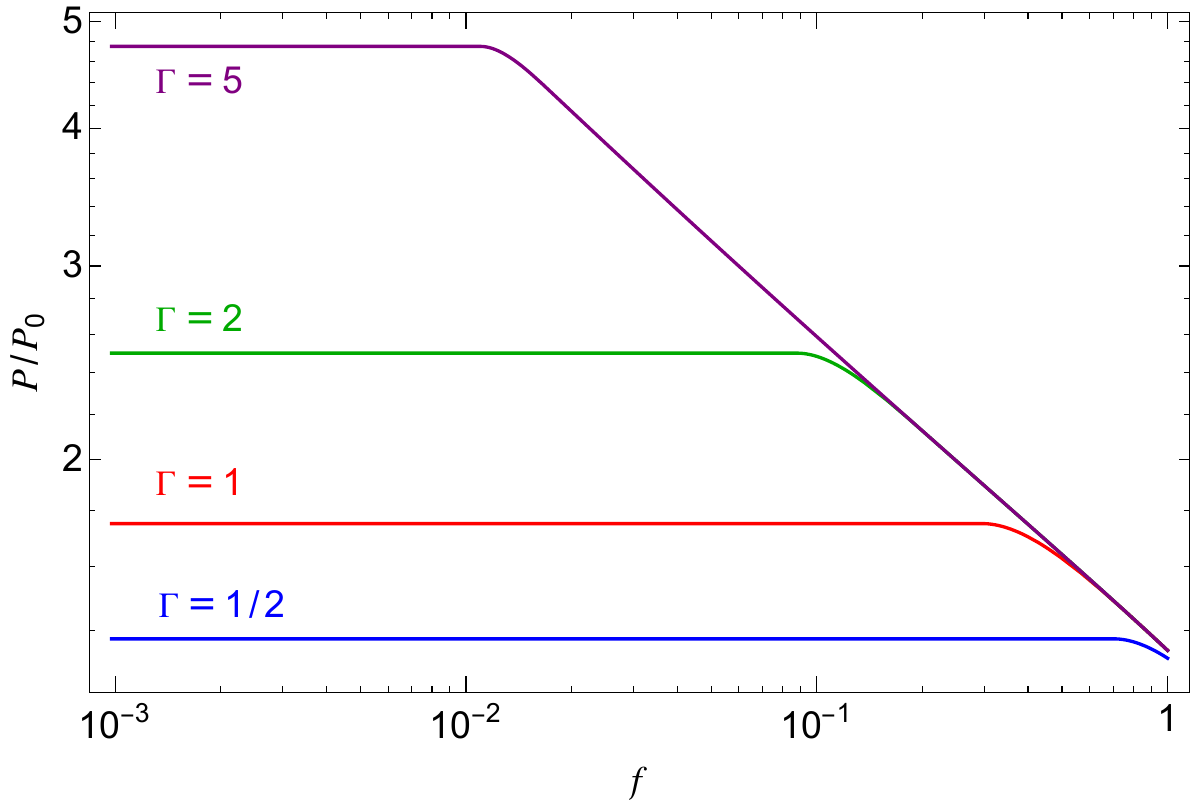}
    \caption{The enhanced probability ratio $P/P_0$ for $\Gamma=1/2, 1, 2, 5$ with respect to PBH fraction $f$ in the total matter.}
    \label{fig: P by P0}
\end{figure}

\begin{figure}[t!]
    \centering
    \includegraphics[width=0.46\textwidth]{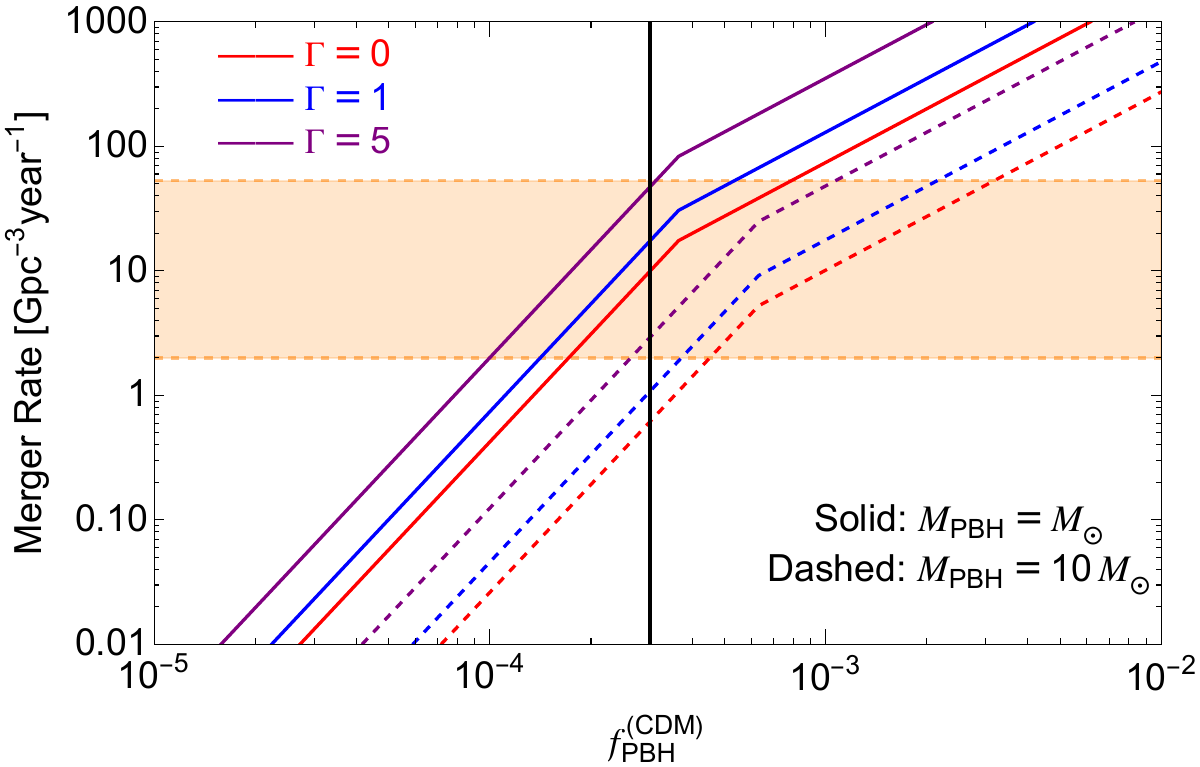}
    \caption{The modified merger rate with respect to the PBH abundance in cold dark matter $f_{\mathrm{PBH}}^{\mathrm{(CDM)}}=f\Omega_{\mathrm{m}}/\Omega_\mathrm{CDM}$ for $\Gamma=0, 1, 5$ as shown in red, blue, and purple, respectively. The horizontal shaded region in orange is the LIGO-Virgo constraint on the event rate $2\sim 53 ~\mathrm{Gpc}^{-3} \mathrm{year}^{-1}$~\cite{LIGOScientific:2016kwr}, while the vertical black line at $f_{\mathrm{PBH}}^{\mathrm{(CDM)}}\simeq 3\times 10^{-4}$ is the upper limit from non-detection of CMB distortion~\cite{Ricotti:2007au}.}
    \label{fig: Merger Rate}
\end{figure}

Finally, the modified probability of PBH binary formation can be measured due to the change in the decoupling region with respect to the static case as detailed in the Supplemental Material~\cite{footnote} for a general $\Gamma$ value. Let us illustrate with a typical example of PBHs formed around the electroweak scale at $a_\mathrm{PBH}/a_\mathrm{eq}\simeq 10^{-12}$ that later encounter with vacuum bubbles from a QCD-like FOPT at $a_\mathrm{PT}/a_\mathrm{eq}\simeq 10^{-8}$, then the typical value of the above dimensionless parameter reads $\Gamma\sim O(1)$ for the usual efficiency factor of PBH formation $\gamma = 0.2$ \cite{Carr:1975qj, Carr:2009jm} and a typical PBH velocity $a_{\mathrm{PT}} v = 0.3$. Taking  $\Gamma=1$ in specific, the ratio of the modified probability $P$ with respect to the static one $P_0$ is given by
\begin{align}
    \label{eq: PbyP0}
    \frac{P}{P_0} = \left\{\begin{array}{ll}
         \cfrac{7}{4}~, & f<\cfrac{8}{27} \\
         \cfrac{1}{2} + \cfrac{64 - \left(16-9f^{2/3}\right)^{3/2}}{54f}~, \quad & \cfrac{8}{27}<f<\cfrac{16\sqrt{2}}{27} \\
         \cfrac{4}{3} + \cfrac{32}{27f} - \cfrac{16}{27f^{4/3}}~, & \cfrac{16\sqrt{2}}{27} < f < 1
    \end{array}
    \right.~.
\end{align}
These three situations are related to the PBHs' relative velocity and abundance, see Supplemental Material~\cite{footnote} for details. For a fixed relative velocity, less PBH abundance leads to a larger enhancement in the PBH binary formation probability.
The ratios $P/P_0$ for general values of $\Gamma$ are shown in Fig.~\ref{fig: P by P0}, which always manifests an enhancement $P/P_0>1$ unless $\Gamma\to 0$ in the limit $v=0$ that recovers the static case $P=P_0$. This enhancement reaches its maximum at a lower limit of PBH fraction, and the maximal enhancement is preferred for an earlier PBH formation time and a relatively later occurrence of the FOPT. This enhancement can be understood physically as below: Consider two neighboring PBHs initially separated by a comoving distance $x$. If the static condition~\eqref{eq: decouple condition} is satisfied, they will simply form a binary at $t_\mathrm{eq}$ as expected. If not, it is still possible for them to form a binary in virtue of the velocities gained from the passing bubble walls. Therefore, the probability of forming binary systems will always be enhanced as well as the modified PBH merger rate obtained from multiplying with this enhanced PBH binary formation rate, leading to stronger constraints on PBH abundances in cold dark matter $f_\mathrm{PBH}^{(\mathrm{CDM})}\equiv f\Omega_\mathrm{m}/\Omega_\mathrm{CDM}$ as shown in Fig.~\ref{fig: Merger Rate} for stellar-mass PBHs as candidates of LIGO-Virgo events.

\textit{\textbf{Conclusions and discussions.}---} 
We have numerically simulated for the first time the collisions between scalar bubble walls and dynamically co-evolving PBHs with full numerical relativity. The collisions with PBHs from vacuum bubble walls passing by bring about several far-reaching phenomena of wide interest, including PBH mass and momentum growth as well as high-frequency GWs from cusps produced at the onset of wall-hole collisions with GW amplitudes comparable to that from pure bubble wall collisions.

In particular, the initial velocities given by the wall-hole collisions can significantly enhance the PBH binary formation rate and hence the subsequent PBH merger rate, strengthening the observational constraints on the PBH abundance. The enhancement of the PBH binary formation rate also makes them easier to decouple from background expansion and generate heavier PBHs that might serve as heavy seeds for further formations of supermassive black holes. It should be reminded that our mechanism mainly work at the period when PBHs were much far away from each other, and the GW radiations and three-body disruption can be safely neglected.

One intriguing generalization of our work is to go beyond the current working assumption that PBHs are uniformly distributed with a number density much smaller than that of vacuum bubbles. Things become more interesting if more PBHs are produced in one Hubble volume but with fewer bubbles nucleated, in this case, the acquired initial PBH velocities towards the bubble center might generate PBH clusterings~\cite{Belotsky:2018wph,Desjacques:2018wuu}. Things become more complicated if PBHs clustered even before the FOPT~\cite{Desjacques:2018wuu,LISACosmologyWorkingGroup:2023njw}, where PBHs are formed from the high-$\sigma$ tails of the random Gaussian density fluctuations~\cite{Sasaki:2016jop}.
Further investigations should be implemented for an extended parameter space with, for example, different magnitudes of the PBH mass. Intuitively,  a PBH of mass large enough with its gravitational radius much larger than the bubble radius might as well be too attractive to swallow the whole bubble, and a PBH of mass small enough may be kicked away instead of being pulled into the true vacuum.

The simulation cannot run for a long time due to two main reasons. Firstly, there is no dissipative effect to settle down the scalar field at the true vacuum but keeps oscillating, leading to numerical instability as the oscillations grow too fast for the mesh refinement rate to keep up. The lack of friction also causes the bubble wall to become thinner, resulting in extra numerical errors. Secondly, the ``moving punctures'' gauge choice becomes ineffective after a BH undergoes a ``phase transition'' from asymptotically de Sitter to asymptotically flat ones. Initially, the shift vector pulls observers away from the BH center to overcome the background inflation. Since the collision time is short, the BH suddenly transits so fast into an asymptotically flat one that the observers cannot slow down in time, causing the coordinate radius of the BH apparent horizon to shrink continuously and increasing numerical errors.

Future investigations are needed to address several issues. These include finding a better gauge choice for BH transition from asymptotically de Sitter to asymptotically flat ones to extend simulation times to see the long-time behavior, defining appropriate extraction scheme for GWs from a non-asymptotically flat unbounded system, and incorporating gravitational interactions between the scalar field and radiation fluids to better represent real-world scenarios during FOPTs. Similar effects might also be anticipated for PBH encounters with sound shells induced by expanding walls on bulk fluids long after bubble collisions. A potentially enhanced primordial magnetic field might also be anticipated around the energy-density spikes~\cite{Di:2020kbw,Yang:2021uid} induced in the vicinity of collision regions between PBHs and bubbles.

\textit{\textbf{Acknowledgments}---}
We thank helpful discussions with Misao Sasaki, Teruaki Suyama, and Jian-hua He. This work is supported by 
the National Key Research and Development Program of China Grants No. 2021YFC2203004, No. 2021YFA0718304, and No. 2020YFC2201502,
the National Natural Science Foundation of China (NSFC) Grants No. 12422502, No. 12105344, No. 12235019, No. 11821505, No. 11991052, No. 12347132, and No. 11947302,
and the Science Research Grants from the China Manned Space Project with No. CMS-CSST-2021-B01 (supported by China Manned Space Program through its Space Application System).
We also acknowledge the use of the HPC Cluster of ITP-CAS.



\onecolumngrid
\newpage
\appendix

\begin{center}
{\LARGE\textbf{Supplemental material: detailed derivations for the enhanced binary formation rate from bubble-PBH collisions}}
\end{center}

\begin{center}
{\Large
Zi-Yan Yuwen,$^{1,2}$ Cristian Joana,$^{1}$ Shao-Jiang Wang,$^{1,3}$ and Rong-Gen Cai$^{4,1,5}$}
\end{center}

\begin{center}
{\large\textit{$^1$Institute of Theoretical Physics, Chinese Academy of Sciences (CAS), Beijing 100190, China}}
\end{center}
\begin{center}
{\large\textit{$^2$University of Chinese Academy of Sciences (UCAS), Beijing 100049, China}}
\end{center}
\begin{center}
{\large\textit{$^3$Asia Pacific Center for Theoretical Physics (APCTP), Pohang 37673, Korea}}
\end{center}
\begin{center}
{\large\textit{$^4$School of Physical Science and Technology, Ningbo University, Ningbo, 315211, China}}
\end{center}
\begin{center}
{\large\textit{$^5$School of Fundamental Physics and Mathematical Sciences, Hangzhou Institute for Advanced Study (HIAS), University of Chinese Academy of Sciences (UCAS), Hangzhou 310024, China}}
\end{center}

\section{Numerical Methods}

In this supplemental material, we will go through the numerical techniques used in the simulation, which is studied detaily in the~\texttt{GRChombo}~code~\cite{Clough:2015sqa,Andrade:2021rbd} (see also the Ph.D. Thesis by K. Clough~\cite{Clough:2017ixw}). In the standard $3+1$ ADM decomposition formalism, the spacetime distance is given by
\begin{align}
    \md s^2 = g_{\mu\nu}\md x^\mu \md x^\nu = -\alpha^2 \md t^2 + \gamma_{ij} 
    \left(\md x^i + \beta^i \md t\right)
    \left(\md x^j + \beta^j \md t\right),
\end{align}
where $\gamma_{ij}$ is the induced metric on the spatial hyper-surface, whose timelike unit normal vector reads
\begin{align}
    n^\mu = \frac{1}{\alpha}\left(\partial_t^\mu - \beta^i \partial_i^\mu \right)
\end{align}
with $\alpha$ and $\beta$ are the lapse function and shift vector respectively. The Greek letters $\mu,~\nu$ denote spacetime indices running from $0$ to $3$, and the Latin letters $i,~j$ denote spatial indices running from $1$ to $3$. The corresponding extrinsic curvature is evaluated from the Lie derivative of spatial metric along the normal direction,
\begin{align}
    K_{ij} = -\frac{1}{2}\mathcal{L}_{n}\gamma_{ij} = -\frac{1}{2\alpha}\left(\partial_t \gamma_{ij} - D_i \beta_j - D_j \beta_i \right),
\end{align}
where $D_i$ is the covariant derivative compatible with the spatial metric $\gamma_{ij}$. The $3+1$ Einstein's Equations then can be subjected to the constraints and time evolution equations. The Hamiltonian and Momentum constraints are given by
\begin{align}
    \mathcal{H} &= R^{(3)} + K^2 - K_{ij}K^{ij} - 16\pi \rho = 0, \\
    \mathcal{M}^i &= D_j \left( \gamma^{ij}K - K^{ij} \right) - 8\pi S^i = 0,
\end{align}
and the evolution equations for metric and extrinsic curvature are given by
\begin{align}
    \partial_t \gamma_{ij} &= -2\alpha K_{ij} + D_i \beta_j + D_j \beta_i, \\
    \partial_t K_{ij} &= \beta^k\partial_k K_{ij} + 2 K_{k\left(i\right.}\partial_{\left.j\right)}\beta^k - D_iD_j \alpha + \alpha \left(
        R^{(3)}_{ij} + KK_{ij} - 2 K_{ik}K^{k}_{j} + 4\pi\gamma_{ij}(S-\rho) - 8\pi S_{ij}
    \right),
\end{align}
where $R^{(3)}_{ij}$ is the spatial intrinsic Ricci tensor, $R^{(3)} = \gamma^{ij}R^{(3)}_{ij}$ and $K = \gamma^{ij}K_{ij}$. The matter stress tensor components enrolled in the equations above are defined as those measured by normal observers,
\begin{align}
    \rho = n_\mu n_\nu T^{\mu\nu}, \quad S_i = -\gamma_{i\mu}n_\nu T^{\mu\nu}, \quad S_{ij} = \gamma_{i\mu} \gamma{j\nu} T^{\mu\nu}, \quad S = \gamma^{ij}S_{ij}.
\end{align}
In our simulation, the matter part only consists of a real scalar field, whose stress tensor is defined as
\begin{align}
    T_{\mu\nu}^\phi = \nabla_\mu \phi \nabla_\nu \phi - \frac{1}{2} g_{\mu\nu} \left( \nabla_{\sigma}\nabla^{\sigma}\phi + 2V \right).
\end{align}

The~\texttt{GRChombo}~code allows us to evolve the system in the BSSN formalism or CCZ4 formalism. We adopted the standard BSSN formalism in our simulation, which introduces the conformal connections $\tilde{\Gamma}^i$ and promotes them to dynamical variables~\cite{Baumgarte:1998te,Shibata:1995we}. In the BSSN system induced metric is decomposed as
\begin{align}
    \gamma_{ij} = \frac{1}{\chi}\tilde{\gamma}_{ij}, \quad \det{\tilde{\gamma}_{ij}}=1, \quad \chi = \left(\det{\gamma_{ij}}\right)^{-1/3}.
\end{align}
The conformal connections are defined as $\tilde{\Gamma}^i = \tilde{\gamma}^{jk} \tilde{\Gamma}^i_{~jk}$ with $\tilde{\Gamma}^i_{~jk}$ the Christoffel symbols associated with $\tilde{\gamma}_{ij}$. The extrinsic curvature is decomposed to the trace and trace-free part,
\begin{align}
    K = \gamma^{ij}K_{ij}, \quad \tilde{A}_{ij} = \chi K_{ij} - \frac{1}{3}K\tilde{\gamma}_{ij}, \quad \tilde{\gamma}^{ij} \tilde{A}_{ij} = 0.
\end{align}
Decomposing the ADM variables into conformal versions and replacing certain variables with multiples of constraints to improve stability results in the following evolution equations,
\begin{align}
    \partial_t \chi &= \frac{2}{3}\alpha \chi K - \frac{2}{3}\chi \partial_i \beta^i + \beta^i \partial_i \chi, \\
    \partial_t \tilde{\gamma}_{ij} &= -2\alpha \tilde{A}_{ij} + 2\tilde{\gamma}_{k\left(i\right.}\partial_{\left.j\right)}\beta^k - \frac{2}{3} \tilde{\gamma}_{ij}\partial_k \beta^k + \beta^k\partial_k \gamma_{ij}, \\
    \partial_t K &= -\gamma^{ij}D_iD_j \alpha + \alpha\left(\tilde{A}_{ij} \tilde{A}^{ij} + \frac{1}{3}K^2 \right) + \beta^i \partial_i K + 4\pi \alpha (\rho+S), \\
    \partial_t \tilde{A}_{ij} &= \chi \left[ -D_iD_j \alpha + \alpha\left( R^{(3)}_{ij} - 8\pi \alpha S_{ij} \right) \right]^{\mathrm{TF}} + \alpha\left( K\tilde{A}_{ij} - 2\tilde{A}_{ik}\tilde{A}^k_j \right) + 2 \tilde{A}_{k\left(i\right.}\partial_{\left.j\right)}\beta^k - \frac{2}{3} \tilde{A}_{ij}\partial_k \beta^k + \beta^k\partial_k A_{ij}, \\
    \partial_t \tilde{\Gamma}^i &= -2\tilde{A}^{ij}\partial_j \alpha + 2\alpha \left( \tilde{\Gamma}^i_{~jk} \tilde{A}^{jk} - \frac{2}{3}\tilde{\gamma}^{ij}\partial_j K - \frac{3}{2} \tilde{A}^{ij} \partial_j \log\chi \right) + \nonumber\\
    &\quad\quad\quad + \beta^k\partial_k\tilde{\Gamma}^i + \tilde{\gamma}^{jk}\partial_j\partial_k \beta^i + \frac{1}{3}\tilde{\gamma}^{ij}\partial_j\partial_k \beta^k + \frac{2}{3} \tilde{\Gamma}^i \partial_k\beta^k - \tilde{\Gamma}^k\partial_k \beta^i - 16\pi\alpha \tilde{\gamma}^{ij}S_j,
\end{align}
where $[...]^\mathrm{TF}$ denotes the trace-free part of the expression inside the parenthesis. The Ricci tensor used in the evolution equation for $\tilde{A}_{ij}$ is also split into the conformal and non-conformal parts $R^{(3)}_{ij} = \tilde{R}^{(3)}_{ij} + R^{(3),\chi}_{ij}$, which are evaluated as
\begin{align}
    \tilde{R}^{(3)}_{ij} &= -\frac{1}{2}\tilde{\gamma}^{kl}\partial_k\partial_l \tilde{\gamma}_{ij} + \tilde{\Gamma}^k\tilde{\Gamma}_{(ij)k} + \tilde{\gamma}^{lm} \left( 2\tilde{\Gamma}^k_{l\left(i\right.}\tilde{\Gamma}_{\left.j\right)km} +  \tilde{\Gamma}^k_{im}\tilde{\Gamma}_{klj} \right), \\
    R^{(3),\chi}_{ij} &= \frac{1}{2\chi}\left( \tilde{D}_i \tilde{D}_j \chi + \tilde{\gamma}_{ij} \tilde{D}^k \tilde{D}_k \chi \right) - \frac{1}{4\chi^2}\left( \tilde{D}_i \chi \tilde{D}_j \chi + 3 \tilde{\gamma}_{ij} \tilde{D}^k \chi \tilde{D}_k \chi \right).
\end{align}
There is one more equation to control the evolution of the scalar field, which is directly derived from the conservation of stress tensor,
\begin{align}
    \partial_t \phi &= \alpha \Pi + \beta^i\partial_i \phi, \label{eq:Pi_def}\\
    \partial_t \Pi &= \beta^i\partial_i \Pi + \alpha \partial_i\partial^i \phi + \partial_i\phi \partial^i \alpha + \alpha \left( K\Pi -\gamma^{ij}\Gamma^k_{ij}\partial_k\phi + \frac{\md V}{\md \phi} \right),
\end{align}
where Eq.~\eqref{eq:Pi_def} serves as the definition of conjugate momentum of the field. 

The dynamical variables for the BSSN system consist of $\{ \chi, \tilde{\gamma}_{ij}, K, \tilde{A}_{ij}, \tilde{\Gamma}^i, \phi, \Pi \}$ such that the system is closed. The lapse $\alpha$ and shift $\beta^i$ are free parameters in the beginning and can be determined only after one specifies their initial value and the gauge choice. We consider the simplest possible initial data for the gauge parameters with $\alpha_{\mathrm{ini}}=1$ and $\beta^i_{\mathrm{ini}} = 0$. As for the gauge choice, we adopt the ``moving punctures method'' \cite{Campanelli:2005dd,Baker:2005vv}, which is a combination of ``$1+\log$ slicing'' for lapse and ``gamma-driver'' for shifts,
\begin{align}
    \partial_t \alpha &= -\mu_{\alpha_1} \alpha^{\mu_{\alpha_2}} K + \mu_{\alpha_3} \beta^i\partial_i \alpha, \\
    \partial_t \beta^i &= \eta_{\beta_1} B^i, \quad \partial_t B^i = \mu_{\beta_1} \alpha^{\beta_2} \partial_t\tilde{\Gamma}^i - \eta_{\beta_2} B^i,
\end{align}
The values of the parameters above are chosen to be
\begin{align}
\begin{aligned}
    \mu_{\alpha_1} = 2, \quad \mu_{\alpha_2} =1, \quad \mu_{\alpha_3} = 1, \quad
    \mu_{\beta_1} = 1, \quad \mu_{\beta_2} =0, \quad \eta_{\beta_2} = 1,
\end{aligned}
\end{align}
while the value of $\eta_{\beta_1}$ is set to be $0.75,~ 0.7$ and $0.6$ for different situations and the comparison is shown below.

Having the well-posed evolution equations and gauge choice at hand, now we turn to the specification of the initial data. The system we are interested in is made up of a black hole and a scalar field, the initial value of which can be generated using the CTTK method \cite{Aurrekoetxea:2022mpw}. In standard PBH formation scenarios the spin can be safely neglected \cite{DeLuca:2020qqa,Blinnikov:2016bxu,Garcia-Bellido:2017fdg,Ivanov:1994pa,Garcia-Bellido:1996mdl,Ivanov:1997ia}. Therefore, a non-spinning PBH with no initial velocity is considered in our simulation. Let us consider the vacuum solution of a non-spinning black hole (without a scalar field), where the Momentum constraints are trivially satisfied
by choosing $K_{ij}^{\mathrm{vac}} = K^{\mathrm{vac}} = 0$, where the suffix ``vac'' stands for ``vacuum'' with no matter distribution. The spatial line-element in isotropic coordinates is therefore given by
\begin{align}\label{eq:BH_initial_data}
    \md l^2 = \left(1+\frac{m_{\mathrm{i}}}{2r}\right)^4\left( \md r^2 + r^2 \md\Omega^2 \right)~,
\end{align}
where $m_\mathrm{i}$ is the initial guess of the black hole mass. The 3-dim line-element given above can be transformed to a Schwarzchild black hole by proper coordinate transformations. Next, we add the real scalar field as matter source into the system, that is, to provide a fixed initial profile for the scalar field $\phi_{\mathrm{ini}}(\mathbf{x})$ and corresponding conjugate momentum. For simplicity, we choose a static vacuum bubble in the initial time so that a vanishing conjugate momentum $\Pi_{\mathrm{ini}}=0$ is imposed, i.e. $\partial_t\phi|_{\mathrm{ini}}=0$. Then, we adopt a ``hybrid CTTK'' formalism by specifying the following choice for the trace of extrinsic curvature
\begin{align}
    K^2 = 24\pi \rho ~.
\end{align} 
With this, the constraint equations reduce to a set of elliptical equations that are numerically solved iteratively. The initial values of other metric components are given by Eq.~\eqref{eq:BH_initial_data}. Note here that for an expanding universe, we should choose the negative solution for the extrinsic curvature's trace, i.e. $K=-\sqrt{K^2}$. The main evolution codes are available at \href{https://https://github.com/GRTLCollaboration/GRChombo}{the Github repository} of \texttt{GRChombo} collaboration, and the initial condition solver and specific example of our simulation are available at \href{https://github.com/Einste11N/GRChombo_ScalarField}{this Github repository}.

\begin{figure*}[htbp]
    \centering
    \includegraphics[width=0.5\textwidth]{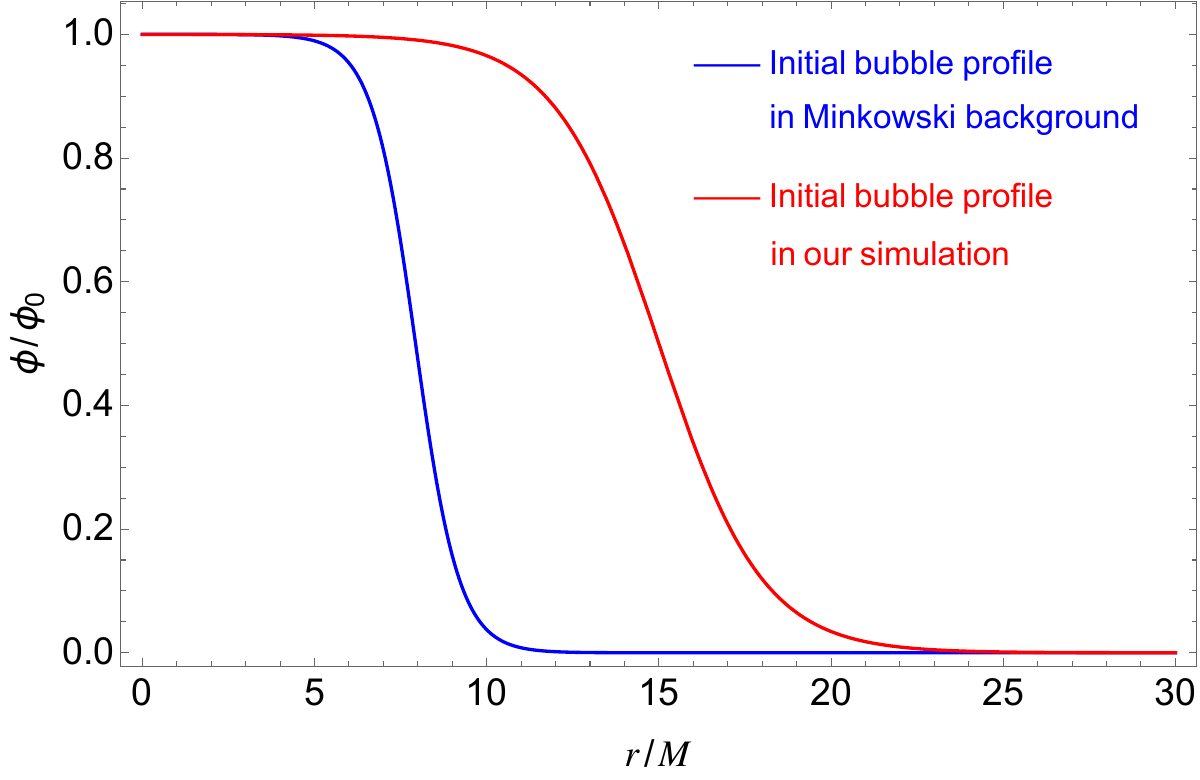}
    \caption{The initial bubble profiles in the Minkowski spacetime (blue) and in our simulation (red), where $r$ denotes the radial distance from the center of the bubble as the origin. The latter has a larger radius for taking into account of the cosmic expansion.}
    \label{fig:initial_profile}
\end{figure*}

The next task is to get reasonable initial data for the scalar field. Assuming initially far away from the black hole, the gravitational effect is weak enough such that it is acceptable to use the initial vacuum bubble profile in a flat spacetime. Such a profile is well-known as the ``static bounce solution'' first introduced by Coleman and Callan in 1970s \cite{Coleman:1977py,Callan:1977pt}. In such formalism, the scalar field is set to be static in the beginning, $\partial_t\phi|_{\mathrm{ini}}=0$. Performing a Wick rotation to the spacetime, one could search for a bounce solution to the equation of motion in Euclidean spacetime, joining the true vacuum and false vacuum, and get the initial profile of the vacuum bubble after rotating back to the Minkowski spacetime. Solving the equations requires knowledge of the scalar potential, and in our case, it is parameterized in a phenomenological way~\cite{Lewicki:2019gmv} as
\begin{align}
\begin{aligned}\label{eq: Scalar_Potential_SM}
    V(\phi) = &\left( 1 + \lambda \tilde{\phi}^2 - (2\lambda+4) \tilde{\phi}^3 + (\lambda+3)  \tilde{\phi}^4 \right)(V_F - V_T) + V_T,
\end{aligned}
\end{align}
admitting a false vacuum $V_F$ at $\tilde{\phi} = 0$ and a true vacuum $V_T$ at $\tilde{\phi}\equiv\phi/\phi_0=1$. The parameters can be determined when considering a concrete particle physics model \cite{Ellis:2020awk}. Usually, such equations with certain boundary conditions are solved with a shooting method, and there are many publicly available software packages aiming at solving this problem, such as \texttt{AnyBubble}\cite{Masoumi:2017trx}, 
\texttt{FindBounce} \cite{Guada:2020xnz}, \texttt{CosmoTransitions} \cite{Wainwright:2011kj}, etc. In our codes, we use our own shooting method that highly agrees with the results given by \texttt{AnyBubble}. Furthermore, taking into account the cosmic expansion, the radius of the initial bubble should be set slightly larger than that determined in the Minkowski background, which is shown in Fig.~\ref{fig:initial_profile}.

\section{Convergence test}

\begin{figure*}[htbp]
    \centering
    \subfigure[]{
    \includegraphics[width=0.47\textwidth]{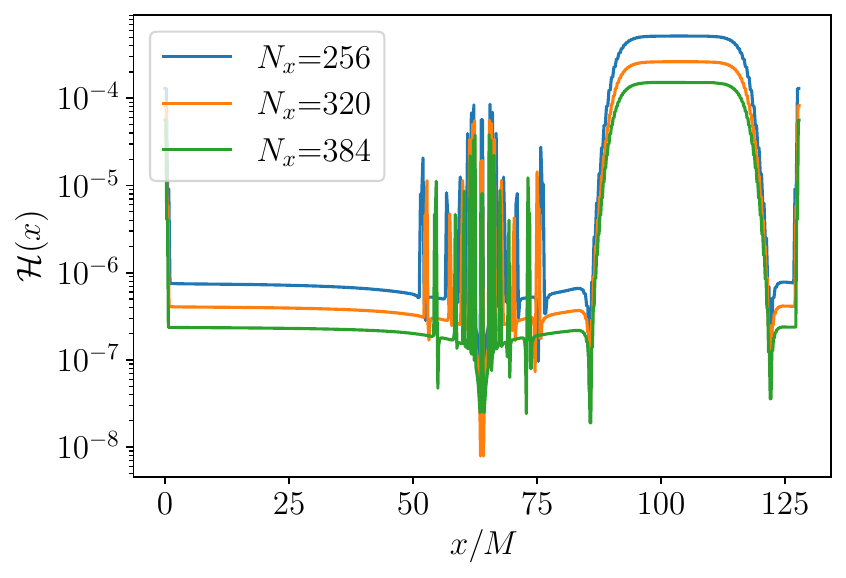}
    }
    \quad
    \subfigure[]{
    \includegraphics[width=0.47\textwidth]{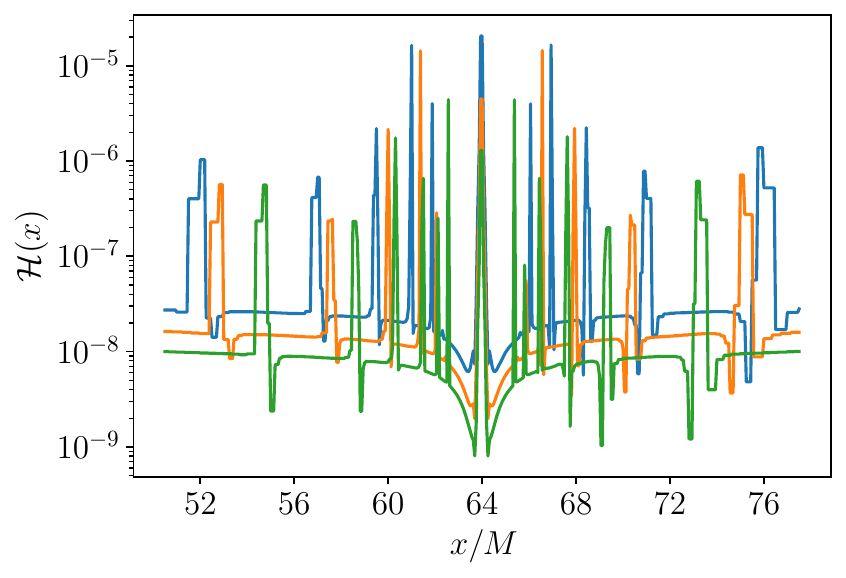}
    }
    \caption{Convergence of $\mathcal{H}$ along $x$-axis for a vacuum scalar bubble generated near a black hole with different grid numbers $N_x = 256$(blue), $320$(orange), and $384$(green). The left panel shows the global convergence along $x$-axis with the black hole centered at $x/M=64$ and bubble centered at $x/M=104$. We zoom in on the nearby region of the black hole in the right panel. The sharp spikes correspond to the location of the refinement boundaries, and the bump from $x\simeq 85$ to $x\simeq 115$ in the local Hamiltonian constraint corresponds to the initial true vacuum region.}
    \label{fig:convergence_ini}
\end{figure*}

\begin{figure*}[htbp]
    \centering
    \includegraphics[width=\textwidth]{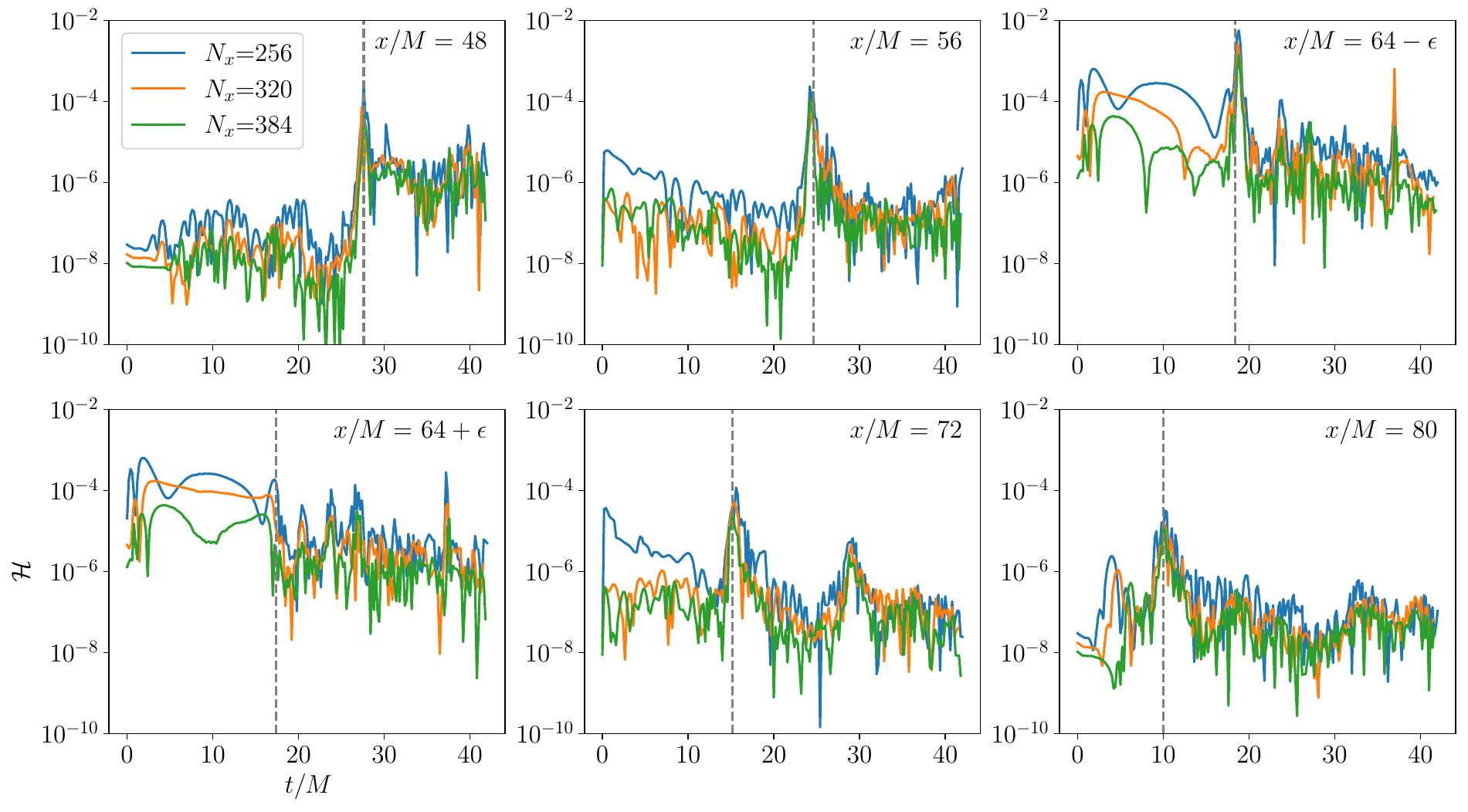}
    \caption{Convergence of $\mathcal{H}$ for some typical points on $x$-axis during the evolution, where $\epsilon = 1/16$. The sharp jumps painted in gray dashed lines in time series are related to (but not equal to) the arrival of the vacuum bubble, which brings updates on the mesh refinement and causes increases in errors.} 
    \label{fig:convergence_time_series}
\end{figure*}

Here we show the convergence test of the simulation with different grid numbers and different gamma-driver for shifts. First of all, for the initial data derived from solving constraint equations, we show here the convergence test of errors of Hamiltonian constraint $\mathcal{H}$ in Fig.~\ref{fig:convergence_ini}, for different grid numbers $N_x = 256$, $320$ and $384$, with $N_x$ defined as the number of boxes on the coarsest adaptive-mesh-refinement level along $x$-direction and $7$ levels in total. The errors during the evolution for three resolutions are shown in Fig.~\ref{fig:convergence_time_series}. The error of the Hamiltonian constraint is normalized by the summation of the absolute value of $\mathcal{H}$ on the coarsest grid level.

\begin{figure*}[htbp]
    \centering
    \subfigure[~$m_\mathrm{PBH}$ as function of $t$]{
    \includegraphics[width=0.47\textwidth]{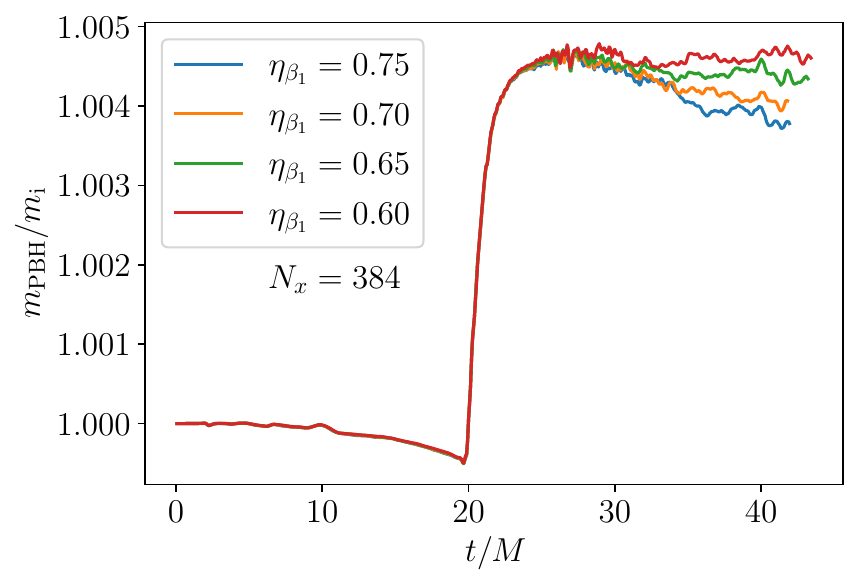}
    }\quad
    \subfigure[~$p_x$ of PBH as function of $t$]{
    \includegraphics[width=0.47\textwidth]{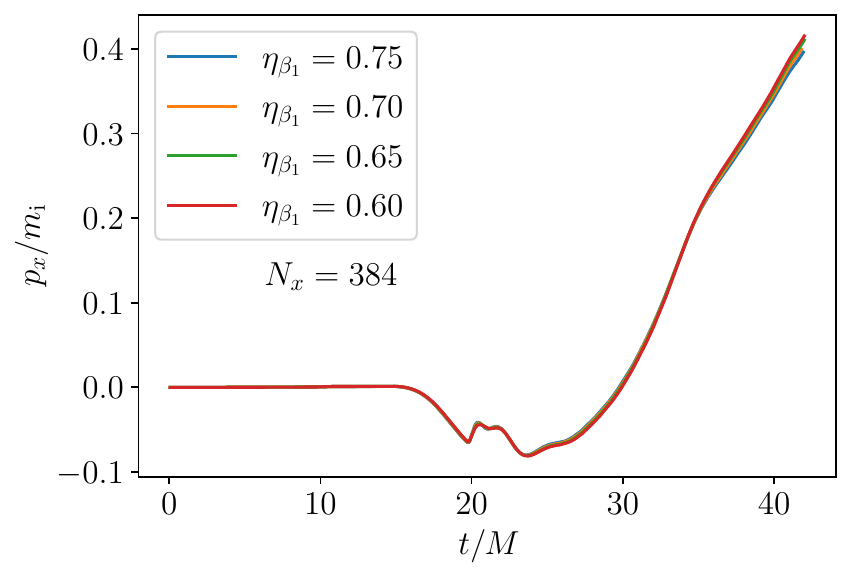}
    }
    \caption{The time evolutions of mass and momentum along $x$-axis of PBH for different gamma-drivers $\eta_{\beta_1}$ with fixed $N_x=384$.}
    \label{fig:Mass_Mom_Evolution}
\end{figure*}

Secondly, for commonly used hyperbolic gamma-driver $\eta_{\beta_1} = 0.75$ \cite{Clough:2017ixw}, although the evolution of momentum seems to converge, we found that the evolutions of black hole mass with different grid numbers agree well only before $t\simeq 30 M \simeq 60 m_\mathrm{i}$, when the coordinate radius begins to shrink rapidly, leading to more and more numerical errors. Thus, for the case with the largest grid number $N_x = 384$, we slowly decrease $\eta_{\beta_1}$ from $0.75$ to $0.6$ in order to maintain more grids inside the black hole apparent horizon after its falling into the true vacuum, maintaining the coordinate radius of the apparent horizon as well. Finally, the mass evolution converges to an almost constant value (with some small fluctuations from numerical errors), providing a reasonable result for the mass growth. The time evolutions of $m_{\mathrm{PBH}}$ and $p_{x}$ with different values of $\eta_{\beta_1}$ are shown in Fig.~\ref{fig:Mass_Mom_Evolution}. The convergence of momentum evolution behaves so well that the PBH velocity recoil effect is accurate and reasonable.

\begin{figure*}[htbp]
    \centering
    \includegraphics[width = \textwidth]{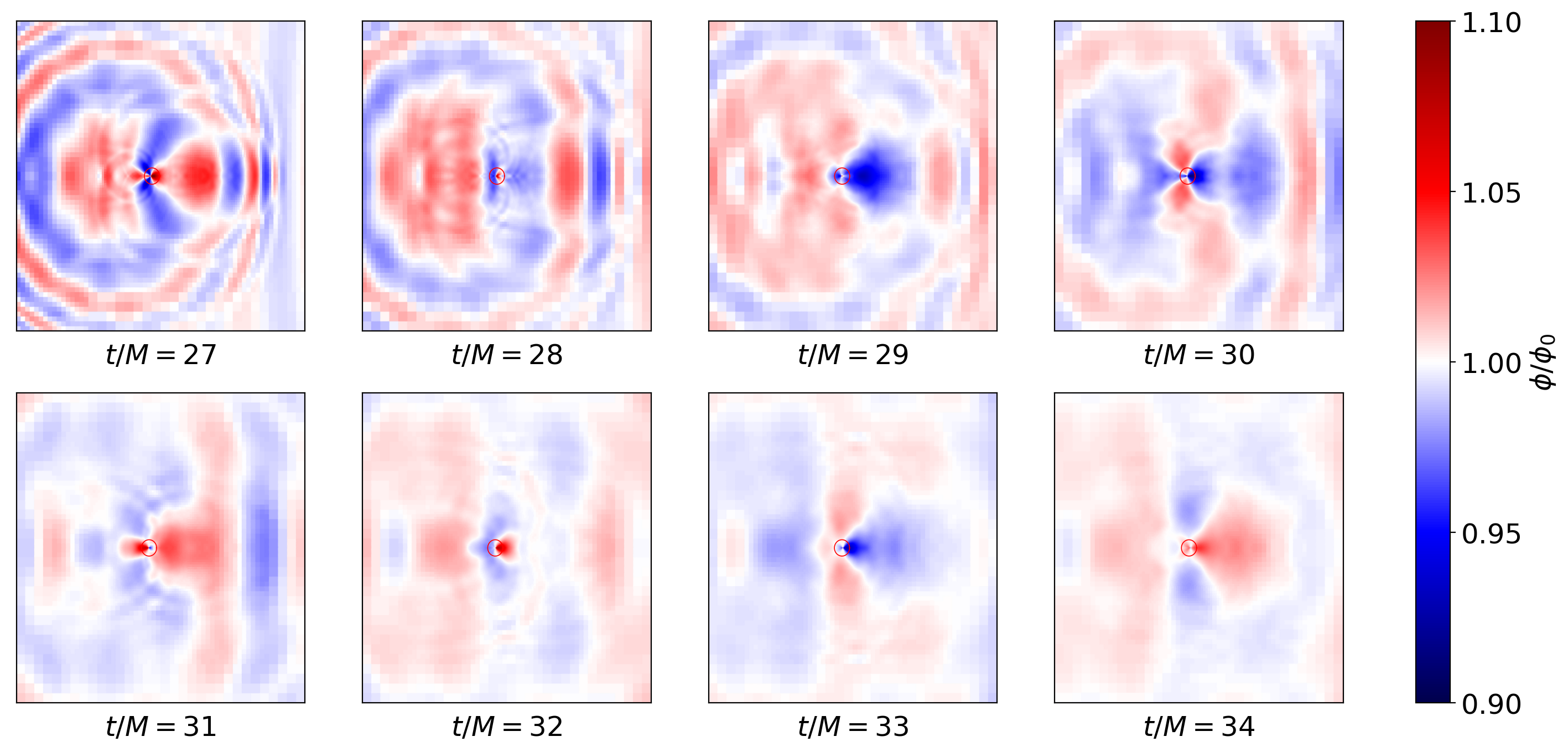}
    \caption{The late time configuration of the scalar field in the vicinity of the black hole.}
    \label{fig:late time slice}
\end{figure*}

\section{Late time evolutions}

The late time configuration of the scalar field in the vicinity of the black hole after PBH-bubble collision is shown in Fig.~\ref{fig:late time slice}, admitting long-term oscillations in the scalar field at true vacuum after the collision. Together with scalar evolutions in the main text, the whole story of momentum recoil is summarized as follows. At the very beginning, there is almost no interaction between the bubble and PBH, leaving the PBH staying at rest. At $t \simeq 20M$ a sudden and strong collision occurs, but the interaction that pushes PBH away to the $-\hat{x}$ direction is quite short, which is represented by a negative $p_x$ value. After bubble wall crossing, the scalar field oscillates to the left of PBH, providing a long-term external force pushing it back to $+\hat{x}$ direction. Although the oscillation amplitude is smaller than bubble-PBH collision, this is an integrated effect of such an interaction that finally leads to a relatively large positive $p_x$ value. A stable phase of $p_x$ at $20M<t<22M$ reflects a competition between these two effects, and an increasing phase at $t>22M$ comes from a cumulative effect from the oscillating scalar field.

\newpage

\section{Forming PBH binaries with relative velocities}

In this supplemental material, we will show how to compute the probability of two PBHs forming a binary when they admit some initial velocity. First, let us revisit the physical picture which consists of two parts. At the very beginning, PBHs were generated by some fluctuations entering the Hubble horizon, but with a very small number density. For example, the number density for PBHs formed during the electro-weak energy scale is about $\mathcal{O}(10^{-9})$ per Hubble volume. Even though it may grow by a factor of $(a/a_\mathrm{PBH})^3$, where $a_\mathrm{PBH}$ is the scale factor of PBH formation time, the number density is still no more than $\mathcal{O}(1)$ per Hubble volume. However, the number density of vacuum bubbles per Hubble volume during a fast FOPT is approximately proportional to $(\beta/H)^3$, which ranges from $\mathcal{O}(10^3)$ to $\mathcal{O}(10^6)$. Once the bubbles collide, the bubble wall will break down to release the latent heat, and the system will fall into the true vacuum. In this case, it is impossible for two PBHs to collide with the same vacuum bubble, which indicates that PBHs were only affected locally during a FOPT. Then the physical picture of a bunch of PBHs staying at rest during the radiation-dominated epoch is changed to PBHs randomly walking around. PBHs were not likely to form binaries because of the Hubble expansion dominated by dense radiation until the dark matter began to take over the universe. In the matter-dominated epoch, the Hubble expansion slowed down, and PBHs with close enough separations and proper relative velocities can form a bound state. 

\begin{figure*}[htbp]
    \centering
    \subfigure[]{
    \includegraphics[width=0.3\textwidth]{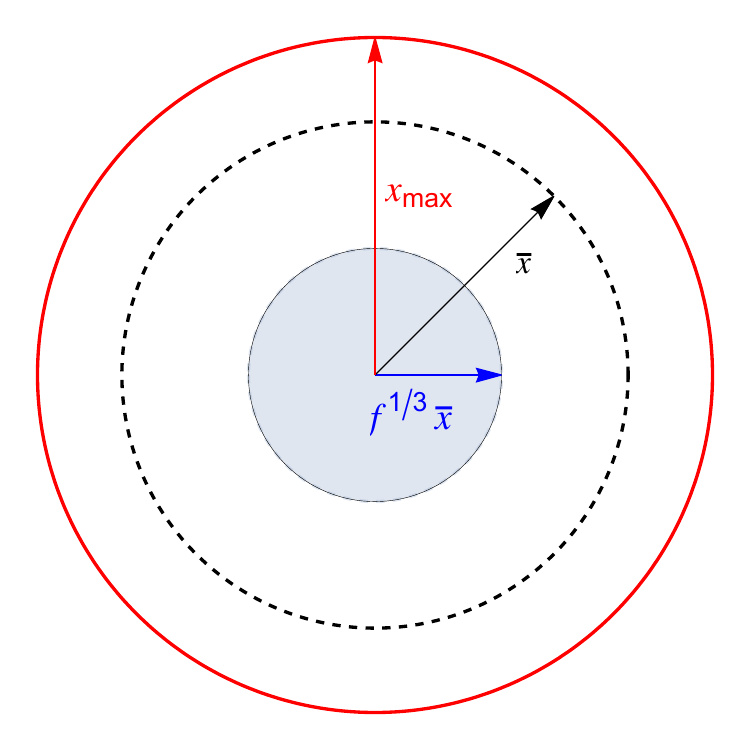}
    \label{fig:marginal sphere}
    }
    \subfigure[]{
    \includegraphics[width=0.3\textwidth]{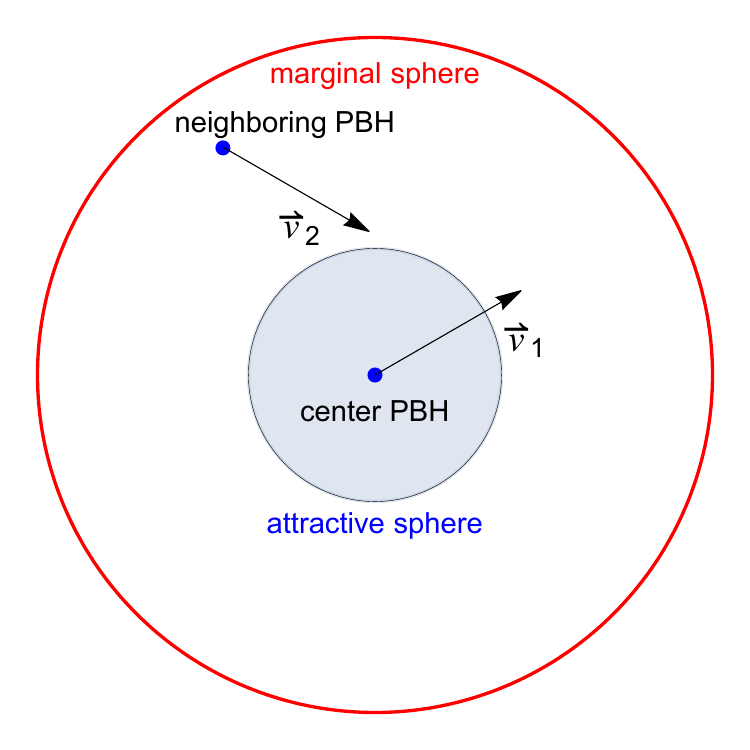}
    \label{fig:two velocities}
    }
    \subfigure[]{
    \includegraphics[width=0.3\textwidth]{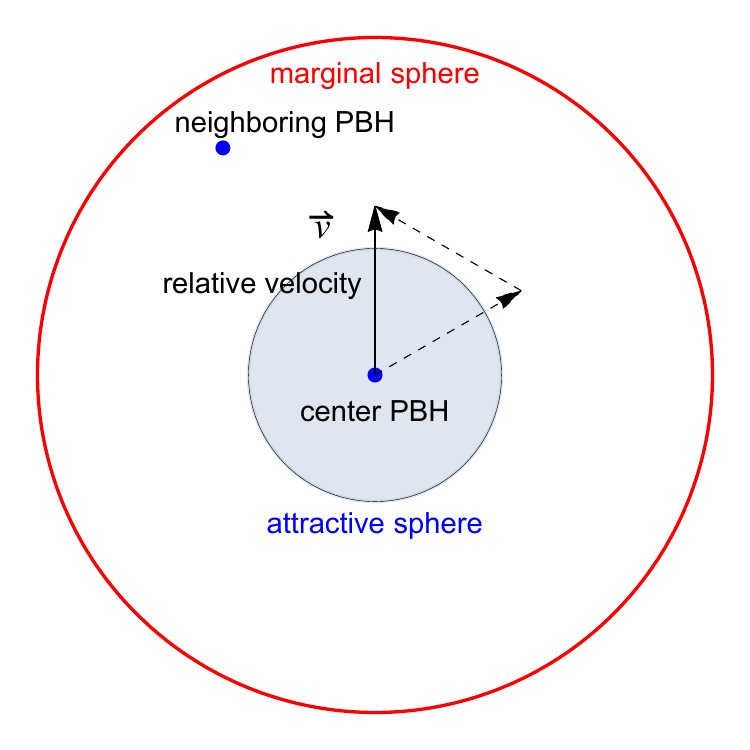}
    \label{fig:the relative velocity}
    }
    \caption{An intuitive picture of the marginal sphere, the attractive sphere of a center PBH, and relative velocity of the two PBHs.}
    \label{fig:intuitive picture}
\end{figure*}

In order to compute the probability of these bound states to form, let us first set up some conventions. A sphere centered around a PBH with a radius $x_\mathrm{max}$ is called a ``marginal sphere'', inside which a neighboring PBH is assumed to exist with a uniform probability. Denoting $\bar{x}$ as the average comoving separation of two neighboring PBHs, there must be a proportional factor $W$ such that $x_\mathrm{max}= W \bar{x}$ as a result of dimensional analysis. Since one PBH is fixed at the center $\Vec{x}_c=0$, and the neighboring PBH appears uniformly probable inside the marginal sphere satisfying $|x|<x_\mathrm{max}$, the value of $\bar{x}$ can be evaluated as
\begin{align}
    \bar{x} = \frac{\displaystyle\int_{|x|<x_\mathrm{max}} |\Vec{x}-\Vec{x}_c| \md^3 x }{\displaystyle\int_{|x|<x_\mathrm{max}} \md^3 x} = \frac{3}{4} x_\mathrm{max},
\end{align}
which simply implies $W=4/3$. Another sphere centered around the PBH with a radius $x=f^{1/3} \bar{x}$ is called an ``attractive sphere'' (or decoupling region), inside which a neighboring PBH will form a binary with the central PBH.  The marginal sphere and attractive sphere are illustrated in Fig.~\ref{fig:marginal sphere} with the red circle and gray shaded region, respectively. The black dashed circle presents an average separation to the central PBH from a neighboring PBH.

Suppose now the two PBHs are gained with a velocity of the same norm $v$ but with two randomly distributed directions as is shown in Fig.~\ref{fig:two velocities}. The relative velocity would be given by the summation of two velocity vectors, where $v$ will be assumed to be small enough so that the special relativity effect is negligible. Thus, an effective picture can be taken that the center PBH moves with a relative velocity and the neighboring PBH stays at rest inside the marginal sphere as is shown in Fig.~\ref{fig:the relative velocity}. Since the norm of the relative velocity is uniformly distributed from $0$ to $2v$, then the center PBH will move with an average relative velocity $v$ in the later discussions.

\begin{figure*}
    \centering
    \subfigure[Black holes with no velocities]{
    \includegraphics[width=0.4\textwidth]{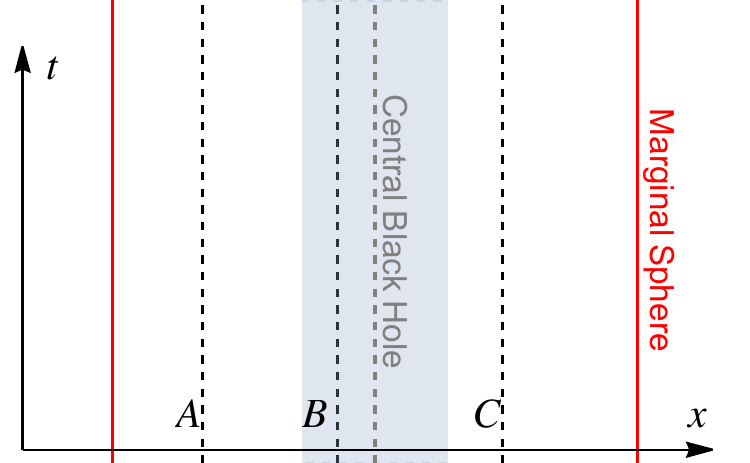}
    \label{fig:Dim11V0}
    }
    \quad
    \subfigure[The central black hole moves with an relative velocity]{
    \includegraphics[width=0.4\textwidth]{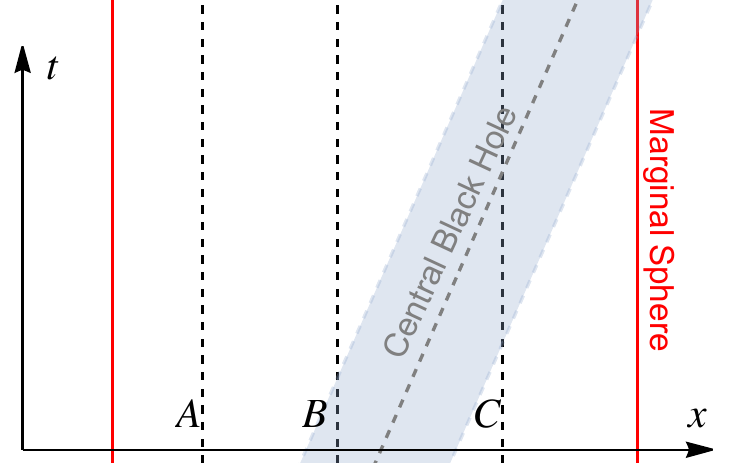}
    \label{fig:Dim11Vrel}
    }
    \caption{An $1+1$ dimensional view of how the relative velocity gained by the central PBH modifies the decoupling region.}
    \label{fig:Dim 1plus1 intuitive picture}
\end{figure*}

It is then easy to see three possible positions for the neighboring PBH inside the marginal sphere with respect to the attractive sphere as shown in Fig.~\ref{fig:Dim 1plus1 intuitive picture} in $1+1$ dimension without loss of generality. In the static case without relative velocity, only the neighboring PBH generated at $B$ within the attractive sphere can form a binary with the central one. However, as the center PBH moves, it will scan over an extended space so that the attractive sphere becomes an ``attractive tube'' with its two ends enclosed with two hemispheroids, respectively. In this case, if the neighboring PBH is generated at $B$, it forms binary at some earlier time since it is inside the attractive sphere. If the neighboring PBH is generated at $C$, the central PBH moves along the gray dashed line and intersects with the world line of the neighboring PBH at $C$, and finally captures it at some later time as is shown in Fig.~\ref{fig:Dim11Vrel}. Although a neighboring PBH generated at $A$ cannot form a binary with the center one in any case, intuitively the binary formation rate is enhanced because of the relative velocity.

\begin{figure*}
    \centering
    \subfigure[$f^{1/3} < \frac{4}{3}(\Gamma + 1)^{-1}$]{
    \includegraphics[width=0.3\textwidth]{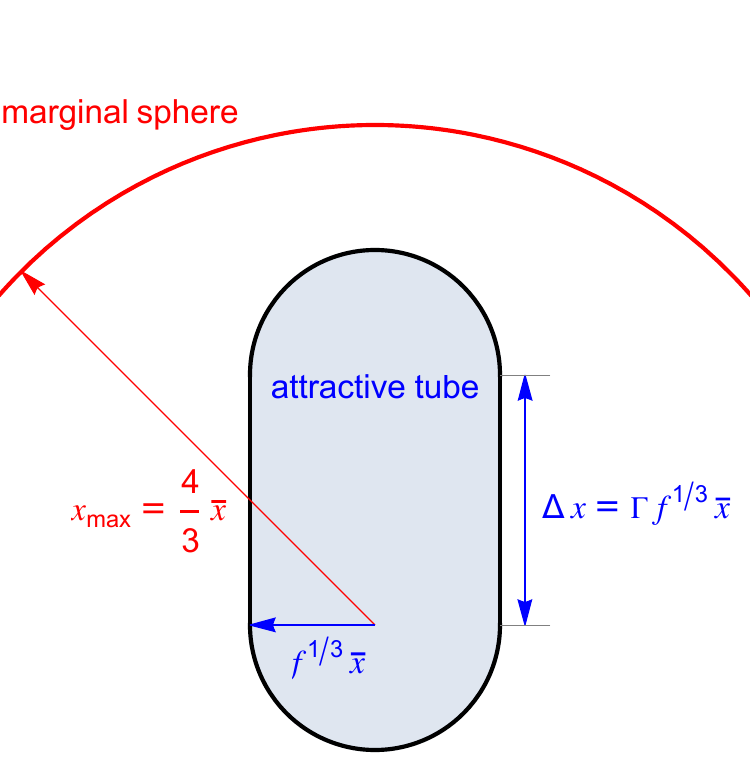}
    \label{fig:attractive tube small}
    }
    \subfigure[$\frac{4}{3}(\Gamma + 1)^{-1} < f^{1/3} < \frac{4}{3}(\Gamma^2 + 1)^{-1/2}$]{
    \includegraphics[width=0.3\textwidth]{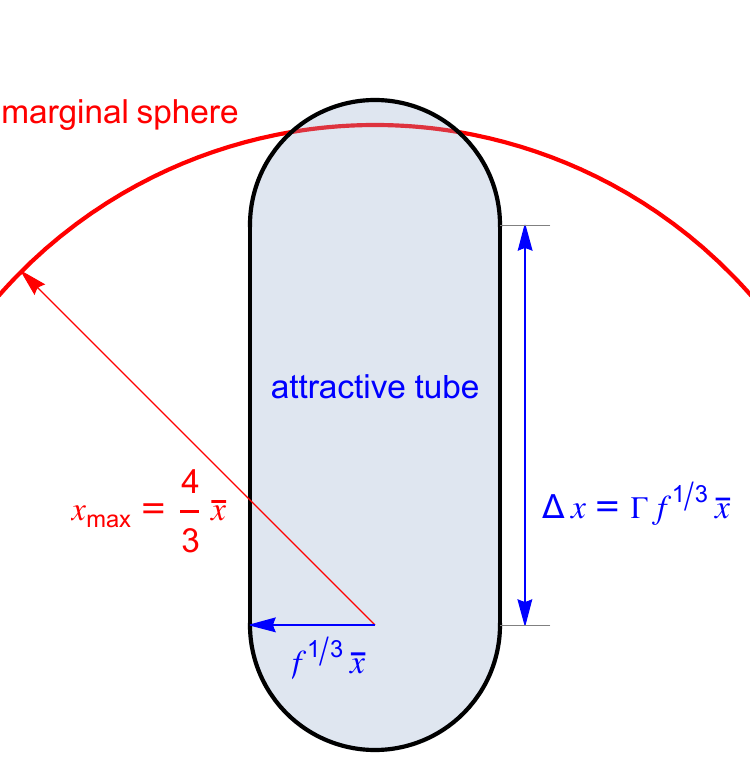}
    \label{fig:attractive tube mid}
    }
    \subfigure[$f^{1/3} > \frac{4}{3}(\Gamma^2 + 1)^{-1/2}$]{
    \includegraphics[width=0.3\textwidth]{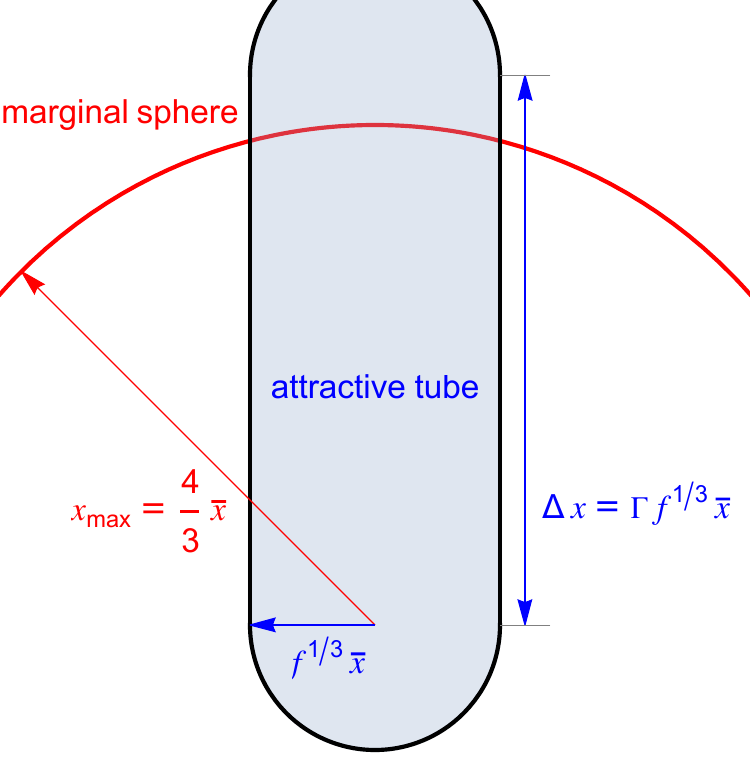}
    \label{fig:attractive tube large}
    }
    \caption{Three cases for how the attractive tube intersects with the marginal sphere.}
    \label{fig:attractive tube}
\end{figure*}

To estimate the probability of the PBH binary formation in the presence of a small initial velocity, we have to compute the volume ratio between the attractive tube and the marginal sphere. The shape of the attractive tube in $2+1$ dimension is shown in Fig.~\ref{fig:attractive tube} with three possible ways to intersect with the marginal sphere. Assuming a PBH with an initial velocity $a_\mathrm{PT} v$ along $x$-direction and a distance shift $\Delta x = \Gamma f^{1/3} \bar{x}$ with $\Gamma \propto a_\mathrm{PT} v$ as estimated in the main context, we can compute the probability for the PBH binary formation as below:

(i) When $f^{1/3} < \frac{4}{3}(\Gamma + 1)^{-1}$, the attractive tube does not touch the boundary of the marginal sphere as shown in Fig.~\ref{fig:attractive tube small}. The probability is then given by the ratio of the volume of the attractive tube to that of the marginal sphere,
\begin{align}
    P = \left(\frac{4\pi}{3}f \bar{x}^3 + \pi \Gamma f \bar{x}^3 \right) \left/ \left(\frac{4\pi}{3} x_\mathrm{max}^3\right)  \right. = \frac{27(4+3\Gamma)}{256}f.
\end{align}

(ii) When $f^{1/3} > \frac{4}{3}(\Gamma^2 + 1)^{-1/2}$, the attractive tube extends with its hemispheroid far outside the marginal sphere, as shown in Fig.~\ref{fig:attractive tube large}. The probability now reads
\begin{align}
\begin{aligned}
    P =& \left(\frac{2\pi}{3}f \bar{x}^3 + \frac{2}{3}\pi f^{2/3} \sqrt{\frac{16}{9} - f^{2/3}} \bar{x}^3 
    + \frac{2\pi}{3}(1-\cos\theta) x_\mathrm{max}^3 \right) \left/ \left(\frac{4\pi}{3} x_\mathrm{max}^3\right)  \right. \\
    = & \frac{27}{128}\left( f+ f^{2/3} \sqrt{\frac{16}{9} - f^{2/3}} \right) + \frac{1-\cos\theta}{2} ,
\end{aligned}
\end{align}
where the angle $\theta$ is given by
\begin{align}
    \theta = \arcsin \left(\frac{3}{4}f^{1/3} \right).
\end{align}

(iii) When $\frac{4}{3}(\Gamma + 1)^{-1} < f^{1/3} < \frac{4}{3}(\Gamma^2 + 1)^{-1/2}$, the attractive tube intersects with the boundary of the marginal sphere at the hemispheroid end of the tube as shown in Fig.~\ref{fig:attractive tube mid}. Now the ratio can be estimated with some effort as
\begin{align}
\begin{aligned}
    P =& \frac{27}{256}\left(2 + 3\Gamma - 2\cos\theta_1 \right) f + \frac{\cos\theta_2 - 1}{2} +  \frac{\sin^2\theta_2 \cos\theta_1}{4}\left(\frac{3f^{1/3}\cos\theta_1}{4\cos\theta_2} - 1 \right),
\end{aligned}
\end{align}
where
\begin{align}
    \theta_1 = \pi - \arccos\left(\frac{(\Gamma^2+1)f^{2/3} - \frac{16}{9}}{2\Gamma f^{2/3}}\right), \\
    \theta_2 = \arccos\left(\frac{\frac{16}{9} + (\Gamma^2-1)f^{2/3}}{\frac{8}{3}\Gamma f^{1/3}}\right).
\end{align}

In the parameter space that is of much interest, the PBH fraction $f$ is much smaller than $1$, thus in most cases, we are interested in the situation (i), that is, $f^{1/3} < \frac{4}{3}(\Gamma + 1)^{-1}$.

\bibliography{ref}

\begin{thebibliography}{138}%
\makeatletter
\providecommand \@ifxundefined [1]{%
 \@ifx{#1\undefined}
}%
\providecommand \@ifnum [1]{%
 \ifnum #1\expandafter \@firstoftwo
 \else \expandafter \@secondoftwo
 \fi
}%
\providecommand \@ifx [1]{%
 \ifx #1\expandafter \@firstoftwo
 \else \expandafter \@secondoftwo
 \fi
}%
\providecommand \natexlab [1]{#1}%
\providecommand \enquote  [1]{``#1''}%
\providecommand \bibnamefont  [1]{#1}%
\providecommand \bibfnamefont [1]{#1}%
\providecommand \citenamefont [1]{#1}%
\providecommand \href@noop [0]{\@secondoftwo}%
\providecommand \href [0]{\begingroup \@sanitize@url \@href}%
\providecommand \@href[1]{\@@startlink{#1}\@@href}%
\providecommand \@@href[1]{\endgroup#1\@@endlink}%
\providecommand \@sanitize@url [0]{\catcode `\\12\catcode `\$12\catcode
  `\&12\catcode `\#12\catcode `\^12\catcode `\_12\catcode `\%12\relax}%
\providecommand \@@startlink[1]{}%
\providecommand \@@endlink[0]{}%
\providecommand \url  [0]{\begingroup\@sanitize@url \@url }%
\providecommand \@url [1]{\endgroup\@href {#1}{\urlprefix }}%
\providecommand \urlprefix  [0]{URL }%
\providecommand \Eprint [0]{\href }%
\providecommand \doibase [0]{http://dx.doi.org/}%
\providecommand \selectlanguage [0]{\@gobble}%
\providecommand \bibinfo  [0]{\@secondoftwo}%
\providecommand \bibfield  [0]{\@secondoftwo}%
\providecommand \translation [1]{[#1]}%
\providecommand \BibitemOpen [0]{}%
\providecommand \bibitemStop [0]{}%
\providecommand \bibitemNoStop [0]{.\EOS\space}%
\providecommand \EOS [0]{\spacefactor3000\relax}%
\providecommand \BibitemShut  [1]{\csname bibitem#1\endcsname}%
\let\auto@bib@innerbib\@empty
\bibitem [{\citenamefont {Mazumdar}\ and\ \citenamefont
  {White}(2019)}]{Mazumdar:2018dfl}%
  \BibitemOpen
  \bibfield  {author} {\bibinfo {author} {\bibfnamefont {Anupam}\ \bibnamefont
  {Mazumdar}}\ and\ \bibinfo {author} {\bibfnamefont {Graham}\ \bibnamefont
  {White}},\ }\bibfield  {title} {\enquote {\bibinfo {title} {{Review of cosmic
  phase transitions: their significance and experimental signatures}},}\ }\href
  {\doibase 10.1088/1361-6633/ab1f55} {\bibfield  {journal} {\bibinfo
  {journal} {Rept. Prog. Phys.}\ }\textbf {\bibinfo {volume} {82}},\ \bibinfo
  {pages} {076901} (\bibinfo {year} {2019})},\ \Eprint
  {http://arxiv.org/abs/1811.01948} {arXiv:1811.01948 [hep-ph]} \BibitemShut
  {NoStop}%
\bibitem [{\citenamefont {Hindmarsh}\ \emph {et~al.}(2021)\citenamefont
  {Hindmarsh}, \citenamefont {L\"uben}, \citenamefont {Lumma},\ and\
  \citenamefont {Pauly}}]{Hindmarsh:2020hop}%
  \BibitemOpen
  \bibfield  {author} {\bibinfo {author} {\bibfnamefont {Mark~B.}\ \bibnamefont
  {Hindmarsh}}, \bibinfo {author} {\bibfnamefont {Marvin}\ \bibnamefont
  {L\"uben}}, \bibinfo {author} {\bibfnamefont {Johannes}\ \bibnamefont
  {Lumma}}, \ and\ \bibinfo {author} {\bibfnamefont {Martin}\ \bibnamefont
  {Pauly}},\ }\bibfield  {title} {\enquote {\bibinfo {title} {{Phase
  transitions in the early universe}},}\ }\href {\doibase
  10.21468/SciPostPhysLectNotes.24} {\bibfield  {journal} {\bibinfo  {journal}
  {SciPost Phys. Lect. Notes}\ }\textbf {\bibinfo {volume} {24}},\ \bibinfo
  {pages} {1} (\bibinfo {year} {2021})},\ \Eprint
  {http://arxiv.org/abs/2008.09136} {arXiv:2008.09136 [astro-ph.CO]}
  \BibitemShut {NoStop}%
\bibitem [{\citenamefont {Caldwell}\ \emph {et~al.}(2022)\citenamefont
  {Caldwell} \emph {et~al.}}]{Caldwell:2022qsj}%
  \BibitemOpen
  \bibfield  {author} {\bibinfo {author} {\bibfnamefont {Robert}\ \bibnamefont
  {Caldwell}} \emph {et~al.},\ }\bibfield  {title} {\enquote {\bibinfo {title}
  {{Detection of early-universe gravitational-wave signatures and fundamental
  physics}},}\ }\href {\doibase 10.1007/s10714-022-03027-x} {\bibfield
  {journal} {\bibinfo  {journal} {Gen. Rel. Grav.}\ }\textbf {\bibinfo {volume}
  {54}},\ \bibinfo {pages} {156} (\bibinfo {year} {2022})},\ \Eprint
  {http://arxiv.org/abs/2203.07972} {arXiv:2203.07972 [gr-qc]} \BibitemShut
  {NoStop}%
\bibitem [{\citenamefont {Athron}\ \emph {et~al.}(2024)\citenamefont {Athron},
  \citenamefont {Bal\'azs}, \citenamefont {Fowlie}, \citenamefont {Morris},\
  and\ \citenamefont {Wu}}]{Athron:2023xlk}%
  \BibitemOpen
  \bibfield  {author} {\bibinfo {author} {\bibfnamefont {Peter}\ \bibnamefont
  {Athron}}, \bibinfo {author} {\bibfnamefont {Csaba}\ \bibnamefont
  {Bal\'azs}}, \bibinfo {author} {\bibfnamefont {Andrew}\ \bibnamefont
  {Fowlie}}, \bibinfo {author} {\bibfnamefont {Lachlan}\ \bibnamefont
  {Morris}}, \ and\ \bibinfo {author} {\bibfnamefont {Lei}\ \bibnamefont
  {Wu}},\ }\bibfield  {title} {\enquote {\bibinfo {title} {{Cosmological phase
  transitions: From perturbative particle physics to gravitational waves}},}\
  }\href {\doibase 10.1016/j.ppnp.2023.104094} {\bibfield  {journal} {\bibinfo
  {journal} {Prog. Part. Nucl. Phys.}\ }\textbf {\bibinfo {volume} {135}},\
  \bibinfo {pages} {104094} (\bibinfo {year} {2024})},\ \Eprint
  {http://arxiv.org/abs/2305.02357} {arXiv:2305.02357 [hep-ph]} \BibitemShut
  {NoStop}%
\bibitem [{\citenamefont {Cai}\ \emph {et~al.}(2017)\citenamefont {Cai},
  \citenamefont {Cao}, \citenamefont {Guo}, \citenamefont {Wang},\ and\
  \citenamefont {Yang}}]{Cai:2017cbj}%
  \BibitemOpen
  \bibfield  {author} {\bibinfo {author} {\bibfnamefont {Rong-Gen}\
  \bibnamefont {Cai}}, \bibinfo {author} {\bibfnamefont {Zhoujian}\
  \bibnamefont {Cao}}, \bibinfo {author} {\bibfnamefont {Zong-Kuan}\
  \bibnamefont {Guo}}, \bibinfo {author} {\bibfnamefont {Shao-Jiang}\
  \bibnamefont {Wang}}, \ and\ \bibinfo {author} {\bibfnamefont {Tao}\
  \bibnamefont {Yang}},\ }\bibfield  {title} {\enquote {\bibinfo {title} {{The
  Gravitational-Wave Physics}},}\ }\href {\doibase 10.1093/nsr/nwx029}
  {\bibfield  {journal} {\bibinfo  {journal} {Natl. Sci. Rev.}\ }\textbf
  {\bibinfo {volume} {4}},\ \bibinfo {pages} {687--706} (\bibinfo {year}
  {2017})},\ \Eprint {http://arxiv.org/abs/1703.00187} {arXiv:1703.00187
  [gr-qc]} \BibitemShut {NoStop}%
\bibitem [{\citenamefont {Bian}\ \emph {et~al.}(2021)\citenamefont {Bian} \emph
  {et~al.}}]{Bian:2021ini}%
  \BibitemOpen
  \bibfield  {author} {\bibinfo {author} {\bibfnamefont {Ligong}\ \bibnamefont
  {Bian}} \emph {et~al.},\ }\bibfield  {title} {\enquote {\bibinfo {title}
  {{The Gravitational-wave physics II: Progress}},}\ }\href {\doibase
  10.1007/s11433-021-1781-x} {\bibfield  {journal} {\bibinfo  {journal} {Sci.
  China Phys. Mech. Astron.}\ }\textbf {\bibinfo {volume} {64}},\ \bibinfo
  {pages} {120401} (\bibinfo {year} {2021})},\ \Eprint
  {http://arxiv.org/abs/2106.10235} {arXiv:2106.10235 [gr-qc]} \BibitemShut
  {NoStop}%
\bibitem [{\citenamefont {Caprini}\ \emph {et~al.}(2016)\citenamefont {Caprini}
  \emph {et~al.}}]{Caprini:2015zlo}%
  \BibitemOpen
  \bibfield  {author} {\bibinfo {author} {\bibfnamefont {Chiara}\ \bibnamefont
  {Caprini}} \emph {et~al.},\ }\bibfield  {title} {\enquote {\bibinfo {title}
  {{Science with the space-based interferometer eLISA. II: Gravitational waves
  from cosmological phase transitions}},}\ }\href {\doibase
  10.1088/1475-7516/2016/04/001} {\bibfield  {journal} {\bibinfo  {journal}
  {JCAP}\ }\textbf {\bibinfo {volume} {1604}},\ \bibinfo {pages} {001}
  (\bibinfo {year} {2016})},\ \Eprint {http://arxiv.org/abs/1512.06239}
  {arXiv:1512.06239 [astro-ph.CO]} \BibitemShut {NoStop}%
\bibitem [{\citenamefont {Caprini}\ \emph {et~al.}(2020)\citenamefont {Caprini}
  \emph {et~al.}}]{Caprini:2019egz}%
  \BibitemOpen
  \bibfield  {author} {\bibinfo {author} {\bibfnamefont {Chiara}\ \bibnamefont
  {Caprini}} \emph {et~al.},\ }\bibfield  {title} {\enquote {\bibinfo {title}
  {{Detecting gravitational waves from cosmological phase transitions with
  LISA: an update}},}\ }\href {\doibase 10.1088/1475-7516/2020/03/024}
  {\bibfield  {journal} {\bibinfo  {journal} {JCAP}\ }\textbf {\bibinfo
  {volume} {2003}},\ \bibinfo {pages} {024} (\bibinfo {year} {2020})},\ \Eprint
  {http://arxiv.org/abs/1910.13125} {arXiv:1910.13125 [astro-ph.CO]}
  \BibitemShut {NoStop}%
\bibitem [{\citenamefont {Cohen}\ \emph {et~al.}(1990)\citenamefont {Cohen},
  \citenamefont {Kaplan},\ and\ \citenamefont {Nelson}}]{Cohen:1990py}%
  \BibitemOpen
  \bibfield  {author} {\bibinfo {author} {\bibfnamefont {Andrew~G.}\
  \bibnamefont {Cohen}}, \bibinfo {author} {\bibfnamefont {David~B.}\
  \bibnamefont {Kaplan}}, \ and\ \bibinfo {author} {\bibfnamefont {Ann~E.}\
  \bibnamefont {Nelson}},\ }\bibfield  {title} {\enquote {\bibinfo {title}
  {{WEAK SCALE BARYOGENESIS}},}\ }\href {\doibase 10.1016/0370-2693(90)90690-8}
  {\bibfield  {journal} {\bibinfo  {journal} {Phys. Lett.}\ }\textbf {\bibinfo
  {volume} {B245}},\ \bibinfo {pages} {561--564} (\bibinfo {year}
  {1990})}\BibitemShut {NoStop}%
\bibitem [{\citenamefont {Cohen}\ \emph {et~al.}(1993)\citenamefont {Cohen},
  \citenamefont {Kaplan},\ and\ \citenamefont {Nelson}}]{Cohen:1993nk}%
  \BibitemOpen
  \bibfield  {author} {\bibinfo {author} {\bibfnamefont {Andrew~G.}\
  \bibnamefont {Cohen}}, \bibinfo {author} {\bibfnamefont {D.~B.}\ \bibnamefont
  {Kaplan}}, \ and\ \bibinfo {author} {\bibfnamefont {A.~E.}\ \bibnamefont
  {Nelson}},\ }\bibfield  {title} {\enquote {\bibinfo {title} {{Progress in
  electroweak baryogenesis}},}\ }\href {\doibase
  10.1146/annurev.ns.43.120193.000331} {\bibfield  {journal} {\bibinfo
  {journal} {Ann. Rev. Nucl. Part. Sci.}\ }\textbf {\bibinfo {volume} {43}},\
  \bibinfo {pages} {27--70} (\bibinfo {year} {1993})},\ \Eprint
  {http://arxiv.org/abs/hep-ph/9302210} {arXiv:hep-ph/9302210 [hep-ph]}
  \BibitemShut {NoStop}%
\bibitem [{\citenamefont {Cohen}\ \emph {et~al.}(2012)\citenamefont {Cohen},
  \citenamefont {Morrissey},\ and\ \citenamefont {Pierce}}]{Cohen:2012zza}%
  \BibitemOpen
  \bibfield  {author} {\bibinfo {author} {\bibfnamefont {Timothy}\ \bibnamefont
  {Cohen}}, \bibinfo {author} {\bibfnamefont {David~E.}\ \bibnamefont
  {Morrissey}}, \ and\ \bibinfo {author} {\bibfnamefont {Aaron}\ \bibnamefont
  {Pierce}},\ }\bibfield  {title} {\enquote {\bibinfo {title} {{Electroweak
  Baryogenesis and Higgs Signatures}},}\ }\href {\doibase
  10.1103/PhysRevD.86.013009} {\bibfield  {journal} {\bibinfo  {journal} {Phys.
  Rev.}\ }\textbf {\bibinfo {volume} {D86}},\ \bibinfo {pages} {013009}
  (\bibinfo {year} {2012})},\ \Eprint {http://arxiv.org/abs/1203.2924}
  {arXiv:1203.2924 [hep-ph]} \BibitemShut {NoStop}%
\bibitem [{\citenamefont {Hogan}(1983)}]{Hogan:1983zz}%
  \BibitemOpen
  \bibfield  {author} {\bibinfo {author} {\bibfnamefont {Craig~J.}\
  \bibnamefont {Hogan}},\ }\bibfield  {title} {\enquote {\bibinfo {title}
  {{Magnetohydrodynamic Effects of a First-Order Cosmological Phase
  Transition}},}\ }\href {\doibase 10.1103/PhysRevLett.51.1488} {\bibfield
  {journal} {\bibinfo  {journal} {Phys. Rev. Lett.}\ }\textbf {\bibinfo
  {volume} {51}},\ \bibinfo {pages} {1488--1491} (\bibinfo {year}
  {1983})}\BibitemShut {NoStop}%
\bibitem [{\citenamefont {Di}\ \emph {et~al.}(2020)\citenamefont {Di},
  \citenamefont {Wang}, \citenamefont {Zhou}, \citenamefont {Bian},
  \citenamefont {Cai},\ and\ \citenamefont {Liu}}]{Di:2020nny}%
  \BibitemOpen
  \bibfield  {author} {\bibinfo {author} {\bibfnamefont {Yuefeng}\ \bibnamefont
  {Di}}, \bibinfo {author} {\bibfnamefont {Jialong}\ \bibnamefont {Wang}},
  \bibinfo {author} {\bibfnamefont {Ruiyu}\ \bibnamefont {Zhou}}, \bibinfo
  {author} {\bibfnamefont {Ligong}\ \bibnamefont {Bian}}, \bibinfo {author}
  {\bibfnamefont {Rong-Gen}\ \bibnamefont {Cai}}, \ and\ \bibinfo {author}
  {\bibfnamefont {Jing}\ \bibnamefont {Liu}},\ }\bibfield  {title} {\enquote
  {\bibinfo {title} {{Magnetic field and gravitational waves from the
  first-order Phase Transition}},}\ }\href@noop {} {\  (\bibinfo {year}
  {2020})},\ \Eprint {http://arxiv.org/abs/2012.15625} {arXiv:2012.15625
  [astro-ph.CO]} \BibitemShut {NoStop}%
\bibitem [{\citenamefont {Yang}\ and\ \citenamefont
  {Bian}(2022)}]{Yang:2021uid}%
  \BibitemOpen
  \bibfield  {author} {\bibinfo {author} {\bibfnamefont {Jing}\ \bibnamefont
  {Yang}}\ and\ \bibinfo {author} {\bibfnamefont {Ligong}\ \bibnamefont
  {Bian}},\ }\bibfield  {title} {\enquote {\bibinfo {title} {{Magnetic field
  generation from bubble collisions during first-order phase transition}},}\
  }\href {\doibase 10.1103/PhysRevD.106.023510} {\bibfield  {journal} {\bibinfo
   {journal} {Phys. Rev. D}\ }\textbf {\bibinfo {volume} {106}},\ \bibinfo
  {pages} {023510} (\bibinfo {year} {2022})},\ \Eprint
  {http://arxiv.org/abs/2102.01398} {arXiv:2102.01398 [astro-ph.CO]}
  \BibitemShut {NoStop}%
\bibitem [{\citenamefont {Agazie}\ \emph {et~al.}(2023)\citenamefont {Agazie}
  \emph {et~al.}}]{NANOGrav:2023gor}%
  \BibitemOpen
  \bibfield  {author} {\bibinfo {author} {\bibfnamefont {Gabriella}\
  \bibnamefont {Agazie}} \emph {et~al.} (\bibinfo {collaboration} {NANOGrav}),\
  }\bibfield  {title} {\enquote {\bibinfo {title} {{The NANOGrav 15 yr Data
  Set: Evidence for a Gravitational-wave Background}},}\ }\href {\doibase
  10.3847/2041-8213/acdac6} {\bibfield  {journal} {\bibinfo  {journal}
  {Astrophys. J. Lett.}\ }\textbf {\bibinfo {volume} {951}},\ \bibinfo {pages}
  {L8} (\bibinfo {year} {2023})},\ \Eprint {http://arxiv.org/abs/2306.16213}
  {arXiv:2306.16213 [astro-ph.HE]} \BibitemShut {NoStop}%
\bibitem [{\citenamefont {Antoniadis}\ \emph {et~al.}(2023)\citenamefont
  {Antoniadis} \emph {et~al.}}]{EPTA:2023sfo}%
  \BibitemOpen
  \bibfield  {author} {\bibinfo {author} {\bibfnamefont {J.}~\bibnamefont
  {Antoniadis}} \emph {et~al.} (\bibinfo {collaboration} {EPTA}),\ }\bibfield
  {title} {\enquote {\bibinfo {title} {{The second data release from the
  European Pulsar Timing Array - I. The dataset and timing analysis}},}\ }\href
  {\doibase 10.1051/0004-6361/202346841} {\bibfield  {journal} {\bibinfo
  {journal} {Astron. Astrophys.}\ }\textbf {\bibinfo {volume} {678}},\ \bibinfo
  {pages} {A48} (\bibinfo {year} {2023})},\ \Eprint
  {http://arxiv.org/abs/2306.16224} {arXiv:2306.16224 [astro-ph.HE]}
  \BibitemShut {NoStop}%
\bibitem [{\citenamefont {Reardon}\ \emph {et~al.}(2023)\citenamefont {Reardon}
  \emph {et~al.}}]{Reardon:2023gzh}%
  \BibitemOpen
  \bibfield  {author} {\bibinfo {author} {\bibfnamefont {Daniel~J.}\
  \bibnamefont {Reardon}} \emph {et~al.},\ }\bibfield  {title} {\enquote
  {\bibinfo {title} {{Search for an Isotropic Gravitational-wave Background
  with the Parkes Pulsar Timing Array}},}\ }\href {\doibase
  10.3847/2041-8213/acdd02} {\bibfield  {journal} {\bibinfo  {journal}
  {Astrophys. J. Lett.}\ }\textbf {\bibinfo {volume} {951}},\ \bibinfo {pages}
  {L6} (\bibinfo {year} {2023})},\ \Eprint {http://arxiv.org/abs/2306.16215}
  {arXiv:2306.16215 [astro-ph.HE]} \BibitemShut {NoStop}%
\bibitem [{\citenamefont {Xu}\ \emph {et~al.}(2023)\citenamefont {Xu} \emph
  {et~al.}}]{Xu:2023wog}%
  \BibitemOpen
  \bibfield  {author} {\bibinfo {author} {\bibfnamefont {Heng}\ \bibnamefont
  {Xu}} \emph {et~al.},\ }\bibfield  {title} {\enquote {\bibinfo {title}
  {{Searching for the Nano-Hertz Stochastic Gravitational Wave Background with
  the Chinese Pulsar Timing Array Data Release I}},}\ }\href {\doibase
  10.1088/1674-4527/acdfa5} {\bibfield  {journal} {\bibinfo  {journal} {Res.
  Astron. Astrophys.}\ }\textbf {\bibinfo {volume} {23}},\ \bibinfo {pages}
  {075024} (\bibinfo {year} {2023})},\ \Eprint
  {http://arxiv.org/abs/2306.16216} {arXiv:2306.16216 [astro-ph.HE]}
  \BibitemShut {NoStop}%
\bibitem [{\citenamefont {Schwaller}(2015)}]{Schwaller:2015tja}%
  \BibitemOpen
  \bibfield  {author} {\bibinfo {author} {\bibfnamefont {Pedro}\ \bibnamefont
  {Schwaller}},\ }\bibfield  {title} {\enquote {\bibinfo {title}
  {{Gravitational Waves from a Dark Phase Transition}},}\ }\href {\doibase
  10.1103/PhysRevLett.115.181101} {\bibfield  {journal} {\bibinfo  {journal}
  {Phys. Rev. Lett.}\ }\textbf {\bibinfo {volume} {115}},\ \bibinfo {pages}
  {181101} (\bibinfo {year} {2015})},\ \Eprint
  {http://arxiv.org/abs/1504.07263} {arXiv:1504.07263 [hep-ph]} \BibitemShut
  {NoStop}%
\bibitem [{\citenamefont {Bigazzi}\ \emph
  {et~al.}(2021{\natexlab{a}})\citenamefont {Bigazzi}, \citenamefont {Caddeo},
  \citenamefont {Cotrone},\ and\ \citenamefont {Paredes}}]{Bigazzi:2020avc}%
  \BibitemOpen
  \bibfield  {author} {\bibinfo {author} {\bibfnamefont {Francesco}\
  \bibnamefont {Bigazzi}}, \bibinfo {author} {\bibfnamefont {Alessio}\
  \bibnamefont {Caddeo}}, \bibinfo {author} {\bibfnamefont {Aldo~L.}\
  \bibnamefont {Cotrone}}, \ and\ \bibinfo {author} {\bibfnamefont {Angel}\
  \bibnamefont {Paredes}},\ }\bibfield  {title} {\enquote {\bibinfo {title}
  {{Dark Holograms and Gravitational Waves}},}\ }\href {\doibase
  10.1007/JHEP04(2021)094} {\bibfield  {journal} {\bibinfo  {journal} {JHEP}\
  }\textbf {\bibinfo {volume} {04}},\ \bibinfo {pages} {094} (\bibinfo {year}
  {2021}{\natexlab{a}})},\ \Eprint {http://arxiv.org/abs/2011.08757}
  {arXiv:2011.08757 [hep-ph]} \BibitemShut {NoStop}%
\bibitem [{\citenamefont {Ares}\ \emph {et~al.}(2020)\citenamefont {Ares},
  \citenamefont {Hindmarsh}, \citenamefont {Hoyos},\ and\ \citenamefont
  {Jokela}}]{Ares:2020lbt}%
  \BibitemOpen
  \bibfield  {author} {\bibinfo {author} {\bibfnamefont {F\"eanor~Reuben}\
  \bibnamefont {Ares}}, \bibinfo {author} {\bibfnamefont {Mark}\ \bibnamefont
  {Hindmarsh}}, \bibinfo {author} {\bibfnamefont {Carlos}\ \bibnamefont
  {Hoyos}}, \ and\ \bibinfo {author} {\bibfnamefont {Niko}\ \bibnamefont
  {Jokela}},\ }\bibfield  {title} {\enquote {\bibinfo {title} {{Gravitational
  waves from a holographic phase transition}},}\ }\href {\doibase
  10.1007/JHEP04(2021)100} {\bibfield  {journal} {\bibinfo  {journal} {JHEP}\
  }\textbf {\bibinfo {volume} {21}},\ \bibinfo {pages} {100} (\bibinfo {year}
  {2020})},\ \Eprint {http://arxiv.org/abs/2011.12878} {arXiv:2011.12878
  [hep-th]} \BibitemShut {NoStop}%
\bibitem [{\citenamefont {Bigazzi}\ \emph {et~al.}(2020)\citenamefont
  {Bigazzi}, \citenamefont {Caddeo}, \citenamefont {Cotrone},\ and\
  \citenamefont {Paredes}}]{Bigazzi:2020phm}%
  \BibitemOpen
  \bibfield  {author} {\bibinfo {author} {\bibfnamefont {Francesco}\
  \bibnamefont {Bigazzi}}, \bibinfo {author} {\bibfnamefont {Alessio}\
  \bibnamefont {Caddeo}}, \bibinfo {author} {\bibfnamefont {Aldo~L.}\
  \bibnamefont {Cotrone}}, \ and\ \bibinfo {author} {\bibfnamefont {Angel}\
  \bibnamefont {Paredes}},\ }\bibfield  {title} {\enquote {\bibinfo {title}
  {{Fate of false vacua in holographic first-order phase transitions}},}\
  }\href {\doibase 10.1007/JHEP12(2020)200} {\bibfield  {journal} {\bibinfo
  {journal} {JHEP}\ }\textbf {\bibinfo {volume} {12}},\ \bibinfo {pages} {200}
  (\bibinfo {year} {2020})},\ \Eprint {http://arxiv.org/abs/2008.02579}
  {arXiv:2008.02579 [hep-th]} \BibitemShut {NoStop}%
\bibitem [{\citenamefont {Zhu}\ \emph {et~al.}(2022)\citenamefont {Zhu},
  \citenamefont {Chen},\ and\ \citenamefont {Hou}}]{Zhu:2021vkj}%
  \BibitemOpen
  \bibfield  {author} {\bibinfo {author} {\bibfnamefont {Zhou-Run}\
  \bibnamefont {Zhu}}, \bibinfo {author} {\bibfnamefont {Jun}\ \bibnamefont
  {Chen}}, \ and\ \bibinfo {author} {\bibfnamefont {Defu}\ \bibnamefont
  {Hou}},\ }\bibfield  {title} {\enquote {\bibinfo {title} {{Gravitational
  waves from holographic QCD phase transition with gluon condensate}},}\ }\href
  {\doibase 10.1140/epja/s10050-022-00754-2} {\bibfield  {journal} {\bibinfo
  {journal} {Eur. Phys. J. A}\ }\textbf {\bibinfo {volume} {58}},\ \bibinfo
  {pages} {104} (\bibinfo {year} {2022})},\ \Eprint
  {http://arxiv.org/abs/2109.09933} {arXiv:2109.09933 [hep-ph]} \BibitemShut
  {NoStop}%
\bibitem [{\citenamefont {Ares}\ \emph
  {et~al.}(2022{\natexlab{a}})\citenamefont {Ares}, \citenamefont {Henriksson},
  \citenamefont {Hindmarsh}, \citenamefont {Hoyos},\ and\ \citenamefont
  {Jokela}}]{Ares:2021ntv}%
  \BibitemOpen
  \bibfield  {author} {\bibinfo {author} {\bibfnamefont {F\"eanor~Reuben}\
  \bibnamefont {Ares}}, \bibinfo {author} {\bibfnamefont {Oscar}\ \bibnamefont
  {Henriksson}}, \bibinfo {author} {\bibfnamefont {Mark}\ \bibnamefont
  {Hindmarsh}}, \bibinfo {author} {\bibfnamefont {Carlos}\ \bibnamefont
  {Hoyos}}, \ and\ \bibinfo {author} {\bibfnamefont {Niko}\ \bibnamefont
  {Jokela}},\ }\bibfield  {title} {\enquote {\bibinfo {title} {{Effective
  actions and bubble nucleation from holography}},}\ }\href {\doibase
  10.1103/PhysRevD.105.066020} {\bibfield  {journal} {\bibinfo  {journal}
  {Phys. Rev. D}\ }\textbf {\bibinfo {volume} {105}},\ \bibinfo {pages}
  {066020} (\bibinfo {year} {2022}{\natexlab{a}})},\ \Eprint
  {http://arxiv.org/abs/2109.13784} {arXiv:2109.13784 [hep-th]} \BibitemShut
  {NoStop}%
\bibitem [{\citenamefont {Ares}\ \emph
  {et~al.}(2022{\natexlab{b}})\citenamefont {Ares}, \citenamefont {Henriksson},
  \citenamefont {Hindmarsh}, \citenamefont {Hoyos},\ and\ \citenamefont
  {Jokela}}]{Ares:2021nap}%
  \BibitemOpen
  \bibfield  {author} {\bibinfo {author} {\bibfnamefont {F\"eanor~Reuben}\
  \bibnamefont {Ares}}, \bibinfo {author} {\bibfnamefont {Oscar}\ \bibnamefont
  {Henriksson}}, \bibinfo {author} {\bibfnamefont {Mark}\ \bibnamefont
  {Hindmarsh}}, \bibinfo {author} {\bibfnamefont {Carlos}\ \bibnamefont
  {Hoyos}}, \ and\ \bibinfo {author} {\bibfnamefont {Niko}\ \bibnamefont
  {Jokela}},\ }\bibfield  {title} {\enquote {\bibinfo {title} {{Gravitational
  Waves at Strong Coupling from an Effective Action}},}\ }\href {\doibase
  10.1103/PhysRevLett.128.131101} {\bibfield  {journal} {\bibinfo  {journal}
  {Phys. Rev. Lett.}\ }\textbf {\bibinfo {volume} {128}},\ \bibinfo {pages}
  {131101} (\bibinfo {year} {2022}{\natexlab{b}})},\ \Eprint
  {http://arxiv.org/abs/2110.14442} {arXiv:2110.14442 [hep-th]} \BibitemShut
  {NoStop}%
\bibitem [{\citenamefont {Morgante}\ \emph {et~al.}(2023)\citenamefont
  {Morgante}, \citenamefont {Ramberg},\ and\ \citenamefont
  {Schwaller}}]{Morgante:2022zvc}%
  \BibitemOpen
  \bibfield  {author} {\bibinfo {author} {\bibfnamefont {Enrico}\ \bibnamefont
  {Morgante}}, \bibinfo {author} {\bibfnamefont {Nicklas}\ \bibnamefont
  {Ramberg}}, \ and\ \bibinfo {author} {\bibfnamefont {Pedro}\ \bibnamefont
  {Schwaller}},\ }\bibfield  {title} {\enquote {\bibinfo {title}
  {{Gravitational waves from dark SU(3) Yang-Mills theory}},}\ }\href {\doibase
  10.1103/PhysRevD.107.036010} {\bibfield  {journal} {\bibinfo  {journal}
  {Phys. Rev. D}\ }\textbf {\bibinfo {volume} {107}},\ \bibinfo {pages}
  {036010} (\bibinfo {year} {2023})},\ \Eprint
  {http://arxiv.org/abs/2210.11821} {arXiv:2210.11821 [hep-ph]} \BibitemShut
  {NoStop}%
\bibitem [{\citenamefont {Bea}\ \emph {et~al.}(2021)\citenamefont {Bea},
  \citenamefont {Casalderrey-Solana}, \citenamefont {Giannakopoulos},
  \citenamefont {Mateos}, \citenamefont {Sanchez-Garitaonandia},\ and\
  \citenamefont {Zilh\~ao}}]{Bea:2021zsu}%
  \BibitemOpen
  \bibfield  {author} {\bibinfo {author} {\bibfnamefont {Yago}\ \bibnamefont
  {Bea}}, \bibinfo {author} {\bibfnamefont {Jorge}\ \bibnamefont
  {Casalderrey-Solana}}, \bibinfo {author} {\bibfnamefont {Thanasis}\
  \bibnamefont {Giannakopoulos}}, \bibinfo {author} {\bibfnamefont {David}\
  \bibnamefont {Mateos}}, \bibinfo {author} {\bibfnamefont {Mikel}\
  \bibnamefont {Sanchez-Garitaonandia}}, \ and\ \bibinfo {author}
  {\bibfnamefont {Miguel}\ \bibnamefont {Zilh\~ao}},\ }\bibfield  {title}
  {\enquote {\bibinfo {title} {{Bubble wall velocity from holography}},}\
  }\href {\doibase 10.1103/PhysRevD.104.L121903} {\bibfield  {journal}
  {\bibinfo  {journal} {Phys. Rev. D}\ }\textbf {\bibinfo {volume} {104}},\
  \bibinfo {pages} {L121903} (\bibinfo {year} {2021})},\ \Eprint
  {http://arxiv.org/abs/2104.05708} {arXiv:2104.05708 [hep-th]} \BibitemShut
  {NoStop}%
\bibitem [{\citenamefont {Janik}\ \emph {et~al.}(2022)\citenamefont {Janik},
  \citenamefont {Jarvinen}, \citenamefont {Soltanpanahi},\ and\ \citenamefont
  {Sonnenschein}}]{Janik:2022wsx}%
  \BibitemOpen
  \bibfield  {author} {\bibinfo {author} {\bibfnamefont {Romuald~A.}\
  \bibnamefont {Janik}}, \bibinfo {author} {\bibfnamefont {Matti}\ \bibnamefont
  {Jarvinen}}, \bibinfo {author} {\bibfnamefont {Hesam}\ \bibnamefont
  {Soltanpanahi}}, \ and\ \bibinfo {author} {\bibfnamefont {Jacob}\
  \bibnamefont {Sonnenschein}},\ }\bibfield  {title} {\enquote {\bibinfo
  {title} {{Perfect Fluid Hydrodynamic Picture of Domain Wall Velocities at
  Strong Coupling}},}\ }\href {\doibase 10.1103/PhysRevLett.129.081601}
  {\bibfield  {journal} {\bibinfo  {journal} {Phys. Rev. Lett.}\ }\textbf
  {\bibinfo {volume} {129}},\ \bibinfo {pages} {081601} (\bibinfo {year}
  {2022})},\ \Eprint {http://arxiv.org/abs/2205.06274} {arXiv:2205.06274
  [hep-th]} \BibitemShut {NoStop}%
\bibitem [{\citenamefont {Bigazzi}\ \emph
  {et~al.}(2021{\natexlab{b}})\citenamefont {Bigazzi}, \citenamefont {Caddeo},
  \citenamefont {Canneti},\ and\ \citenamefont {Cotrone}}]{Bigazzi:2021ucw}%
  \BibitemOpen
  \bibfield  {author} {\bibinfo {author} {\bibfnamefont {Francesco}\
  \bibnamefont {Bigazzi}}, \bibinfo {author} {\bibfnamefont {Alessio}\
  \bibnamefont {Caddeo}}, \bibinfo {author} {\bibfnamefont {Tommaso}\
  \bibnamefont {Canneti}}, \ and\ \bibinfo {author} {\bibfnamefont {Aldo~L.}\
  \bibnamefont {Cotrone}},\ }\bibfield  {title} {\enquote {\bibinfo {title}
  {{Bubble wall velocity at strong coupling}},}\ }\href {\doibase
  10.1007/JHEP08(2021)090} {\bibfield  {journal} {\bibinfo  {journal} {JHEP}\
  }\textbf {\bibinfo {volume} {08}},\ \bibinfo {pages} {090} (\bibinfo {year}
  {2021}{\natexlab{b}})},\ \Eprint {http://arxiv.org/abs/2104.12817}
  {arXiv:2104.12817 [hep-ph]} \BibitemShut {NoStop}%
\bibitem [{\citenamefont {Bea}\ \emph {et~al.}(2022)\citenamefont {Bea},
  \citenamefont {Casalderrey-Solana}, \citenamefont {Giannakopoulos},
  \citenamefont {Jansen}, \citenamefont {Mateos}, \citenamefont
  {Sanchez-Garitaonandia},\ and\ \citenamefont {Zilh\~ao}}]{Bea:2022mfb}%
  \BibitemOpen
  \bibfield  {author} {\bibinfo {author} {\bibfnamefont {Yago}\ \bibnamefont
  {Bea}}, \bibinfo {author} {\bibfnamefont {Jorge}\ \bibnamefont
  {Casalderrey-Solana}}, \bibinfo {author} {\bibfnamefont {Thanasis}\
  \bibnamefont {Giannakopoulos}}, \bibinfo {author} {\bibfnamefont {Aron}\
  \bibnamefont {Jansen}}, \bibinfo {author} {\bibfnamefont {David}\
  \bibnamefont {Mateos}}, \bibinfo {author} {\bibfnamefont {Mikel}\
  \bibnamefont {Sanchez-Garitaonandia}}, \ and\ \bibinfo {author}
  {\bibfnamefont {Miguel}\ \bibnamefont {Zilh\~ao}},\ }\bibfield  {title}
  {\enquote {\bibinfo {title} {{Holographic bubbles with Jecco: expanding,
  collapsing and critical}},}\ }\href {\doibase 10.1007/JHEP09(2022)008}
  {\bibfield  {journal} {\bibinfo  {journal} {JHEP}\ }\textbf {\bibinfo
  {volume} {09}},\ \bibinfo {pages} {008} (\bibinfo {year} {2022})},\ \bibinfo
  {note} {[Erratum: JHEP 03, 225 (2023)]},\ \Eprint
  {http://arxiv.org/abs/2202.10503} {arXiv:2202.10503 [hep-th]} \BibitemShut
  {NoStop}%
\bibitem [{\citenamefont {Li}\ \emph {et~al.}(2023)\citenamefont {Li},
  \citenamefont {Wang},\ and\ \citenamefont {Yuwen}}]{Li:2023xto}%
  \BibitemOpen
  \bibfield  {author} {\bibinfo {author} {\bibfnamefont {Li}~\bibnamefont
  {Li}}, \bibinfo {author} {\bibfnamefont {Shao-Jiang}\ \bibnamefont {Wang}}, \
  and\ \bibinfo {author} {\bibfnamefont {Zi-Yan}\ \bibnamefont {Yuwen}},\
  }\bibfield  {title} {\enquote {\bibinfo {title} {{Bubble expansion at strong
  coupling}},}\ }\href {\doibase 10.1103/PhysRevD.108.096033} {\bibfield
  {journal} {\bibinfo  {journal} {Phys. Rev. D}\ }\textbf {\bibinfo {volume}
  {108}},\ \bibinfo {pages} {096033} (\bibinfo {year} {2023})},\ \Eprint
  {http://arxiv.org/abs/2302.10042} {arXiv:2302.10042 [hep-th]} \BibitemShut
  {NoStop}%
\bibitem [{\citenamefont {Wang}\ \emph {et~al.}(2024)\citenamefont {Wang},
  \citenamefont {Yuwen}, \citenamefont {Hao},\ and\ \citenamefont
  {Wang}}]{Wang:2023lam}%
  \BibitemOpen
  \bibfield  {author} {\bibinfo {author} {\bibfnamefont {Jun-Chen}\
  \bibnamefont {Wang}}, \bibinfo {author} {\bibfnamefont {Zi-Yan}\ \bibnamefont
  {Yuwen}}, \bibinfo {author} {\bibfnamefont {Yu-Shi}\ \bibnamefont {Hao}}, \
  and\ \bibinfo {author} {\bibfnamefont {Shao-Jiang}\ \bibnamefont {Wang}},\
  }\bibfield  {title} {\enquote {\bibinfo {title} {{General bubble expansion at
  strong coupling}},}\ }\href {\doibase 10.1103/PhysRevD.109.096012} {\bibfield
   {journal} {\bibinfo  {journal} {Phys. Rev. D}\ }\textbf {\bibinfo {volume}
  {109}},\ \bibinfo {pages} {096012} (\bibinfo {year} {2024})},\ \Eprint
  {http://arxiv.org/abs/2311.07347} {arXiv:2311.07347 [hep-ph]} \BibitemShut
  {NoStop}%
\bibitem [{\citenamefont {He}\ \emph {et~al.}(2022)\citenamefont {He},
  \citenamefont {Li}, \citenamefont {Li},\ and\ \citenamefont
  {Wang}}]{He:2022amv}%
  \BibitemOpen
  \bibfield  {author} {\bibinfo {author} {\bibfnamefont {Song}\ \bibnamefont
  {He}}, \bibinfo {author} {\bibfnamefont {Li}~\bibnamefont {Li}}, \bibinfo
  {author} {\bibfnamefont {Zhibin}\ \bibnamefont {Li}}, \ and\ \bibinfo
  {author} {\bibfnamefont {Shao-Jiang}\ \bibnamefont {Wang}},\ }\bibfield
  {title} {\enquote {\bibinfo {title} {{Gravitational Waves and Primordial
  Black Hole Productions from Gluodynamics}},}\ }\href@noop {} {\  (\bibinfo
  {year} {2022})},\ \Eprint {http://arxiv.org/abs/2210.14094} {arXiv:2210.14094
  [hep-ph]} \BibitemShut {NoStop}%
\bibitem [{\citenamefont {He}\ \emph {et~al.}(2023)\citenamefont {He},
  \citenamefont {Li}, \citenamefont {Wang},\ and\ \citenamefont
  {Wang}}]{He:2023ado}%
  \BibitemOpen
  \bibfield  {author} {\bibinfo {author} {\bibfnamefont {Song}\ \bibnamefont
  {He}}, \bibinfo {author} {\bibfnamefont {Li}~\bibnamefont {Li}}, \bibinfo
  {author} {\bibfnamefont {Sai}\ \bibnamefont {Wang}}, \ and\ \bibinfo {author}
  {\bibfnamefont {Shao-Jiang}\ \bibnamefont {Wang}},\ }\bibfield  {title}
  {\enquote {\bibinfo {title} {{Constraints on holographic QCD phase
  transitions from PTA observations}},}\ }\href@noop {} {\  (\bibinfo {year}
  {2023})},\ \Eprint {http://arxiv.org/abs/2308.07257} {arXiv:2308.07257
  [hep-ph]} \BibitemShut {NoStop}%
\bibitem [{\citenamefont
  {Gouttenoire}(2023{\natexlab{a}})}]{Gouttenoire:2023bqy}%
  \BibitemOpen
  \bibfield  {author} {\bibinfo {author} {\bibfnamefont {Yann}\ \bibnamefont
  {Gouttenoire}},\ }\bibfield  {title} {\enquote {\bibinfo {title}
  {{First-Order Phase Transition Interpretation of Pulsar Timing Array Signal
  Is Consistent with Solar-Mass Black Holes}},}\ }\href {\doibase
  10.1103/PhysRevLett.131.171404} {\bibfield  {journal} {\bibinfo  {journal}
  {Phys. Rev. Lett.}\ }\textbf {\bibinfo {volume} {131}},\ \bibinfo {pages}
  {171404} (\bibinfo {year} {2023}{\natexlab{a}})},\ \Eprint
  {http://arxiv.org/abs/2307.04239} {arXiv:2307.04239 [hep-ph]} \BibitemShut
  {NoStop}%
\bibitem [{\citenamefont {Salvio}(2024)}]{Salvio:2023blb}%
  \BibitemOpen
  \bibfield  {author} {\bibinfo {author} {\bibfnamefont {Alberto}\ \bibnamefont
  {Salvio}},\ }\bibfield  {title} {\enquote {\bibinfo {title} {{Pulsar timing
  arrays and primordial black holes from a supercooled phase transition}},}\
  }\href {\doibase 10.1016/j.physletb.2024.138639} {\bibfield  {journal}
  {\bibinfo  {journal} {Phys. Lett. B}\ }\textbf {\bibinfo {volume} {852}},\
  \bibinfo {pages} {138639} (\bibinfo {year} {2024})},\ \Eprint
  {http://arxiv.org/abs/2312.04628} {arXiv:2312.04628 [hep-ph]} \BibitemShut
  {NoStop}%
\bibitem [{\citenamefont {Afzal}\ \emph {et~al.}(2023)\citenamefont {Afzal}
  \emph {et~al.}}]{NANOGrav:2023hvm}%
  \BibitemOpen
  \bibfield  {author} {\bibinfo {author} {\bibfnamefont {Adeela}\ \bibnamefont
  {Afzal}} \emph {et~al.} (\bibinfo {collaboration} {NANOGrav}),\ }\bibfield
  {title} {\enquote {\bibinfo {title} {{The NANOGrav 15 yr Data Set: Search for
  Signals from New Physics}},}\ }\href {\doibase 10.3847/2041-8213/acdc91}
  {\bibfield  {journal} {\bibinfo  {journal} {Astrophys. J. Lett.}\ }\textbf
  {\bibinfo {volume} {951}},\ \bibinfo {pages} {L11} (\bibinfo {year}
  {2023})},\ \Eprint {http://arxiv.org/abs/2306.16219} {arXiv:2306.16219
  [astro-ph.HE]} \BibitemShut {NoStop}%
\bibitem [{\citenamefont {Addazi}\ \emph {et~al.}(2024)\citenamefont {Addazi},
  \citenamefont {Cai}, \citenamefont {Marciano},\ and\ \citenamefont
  {Visinelli}}]{Addazi:2023jvg}%
  \BibitemOpen
  \bibfield  {author} {\bibinfo {author} {\bibfnamefont {Andrea}\ \bibnamefont
  {Addazi}}, \bibinfo {author} {\bibfnamefont {Yi-Fu}\ \bibnamefont {Cai}},
  \bibinfo {author} {\bibfnamefont {Antonino}\ \bibnamefont {Marciano}}, \ and\
  \bibinfo {author} {\bibfnamefont {Luca}\ \bibnamefont {Visinelli}},\
  }\bibfield  {title} {\enquote {\bibinfo {title} {{Have pulsar timing array
  methods detected a cosmological phase transition?}}}\ }\href {\doibase
  10.1103/PhysRevD.109.015028} {\bibfield  {journal} {\bibinfo  {journal}
  {Phys. Rev. D}\ }\textbf {\bibinfo {volume} {109}},\ \bibinfo {pages}
  {015028} (\bibinfo {year} {2024})},\ \Eprint
  {http://arxiv.org/abs/2306.17205} {arXiv:2306.17205 [astro-ph.CO]}
  \BibitemShut {NoStop}%
\bibitem [{\citenamefont {Athron}\ \emph {et~al.}(2023)\citenamefont {Athron},
  \citenamefont {Fowlie}, \citenamefont {Lu}, \citenamefont {Morris},
  \citenamefont {Wu}, \citenamefont {Wu},\ and\ \citenamefont
  {Xu}}]{Athron:2023mer}%
  \BibitemOpen
  \bibfield  {author} {\bibinfo {author} {\bibfnamefont {Peter}\ \bibnamefont
  {Athron}}, \bibinfo {author} {\bibfnamefont {Andrew}\ \bibnamefont {Fowlie}},
  \bibinfo {author} {\bibfnamefont {Chih-Ting}\ \bibnamefont {Lu}}, \bibinfo
  {author} {\bibfnamefont {Lachlan}\ \bibnamefont {Morris}}, \bibinfo {author}
  {\bibfnamefont {Lei}\ \bibnamefont {Wu}}, \bibinfo {author} {\bibfnamefont
  {Yongcheng}\ \bibnamefont {Wu}}, \ and\ \bibinfo {author} {\bibfnamefont
  {Zhongxiu}\ \bibnamefont {Xu}},\ }\bibfield  {title} {\enquote {\bibinfo
  {title} {{Can supercooled phase transitions explain the gravitational wave
  background observed by pulsar timing arrays?}}}\ }\href@noop {} {\  (\bibinfo
  {year} {2023})},\ \Eprint {http://arxiv.org/abs/2306.17239} {arXiv:2306.17239
  [hep-ph]} \BibitemShut {NoStop}%
\bibitem [{\citenamefont {Fujikura}\ \emph {et~al.}(2023)\citenamefont
  {Fujikura}, \citenamefont {Girmohanta}, \citenamefont {Nakai},\ and\
  \citenamefont {Suzuki}}]{Fujikura:2023lkn}%
  \BibitemOpen
  \bibfield  {author} {\bibinfo {author} {\bibfnamefont {Kohei}\ \bibnamefont
  {Fujikura}}, \bibinfo {author} {\bibfnamefont {Sudhakantha}\ \bibnamefont
  {Girmohanta}}, \bibinfo {author} {\bibfnamefont {Yuichiro}\ \bibnamefont
  {Nakai}}, \ and\ \bibinfo {author} {\bibfnamefont {Motoo}\ \bibnamefont
  {Suzuki}},\ }\bibfield  {title} {\enquote {\bibinfo {title} {{NANOGrav signal
  from a dark conformal phase transition}},}\ }\href {\doibase
  10.1016/j.physletb.2023.138203} {\bibfield  {journal} {\bibinfo  {journal}
  {Phys. Lett. B}\ }\textbf {\bibinfo {volume} {846}},\ \bibinfo {pages}
  {138203} (\bibinfo {year} {2023})},\ \Eprint
  {http://arxiv.org/abs/2306.17086} {arXiv:2306.17086 [hep-ph]} \BibitemShut
  {NoStop}%
\bibitem [{\citenamefont {Han}\ \emph {et~al.}(2024)\citenamefont {Han},
  \citenamefont {Xie}, \citenamefont {Yang},\ and\ \citenamefont
  {Zhang}}]{Han:2023olf}%
  \BibitemOpen
  \bibfield  {author} {\bibinfo {author} {\bibfnamefont {Chengcheng}\
  \bibnamefont {Han}}, \bibinfo {author} {\bibfnamefont {Ke-Pan}\ \bibnamefont
  {Xie}}, \bibinfo {author} {\bibfnamefont {Jin~Min}\ \bibnamefont {Yang}}, \
  and\ \bibinfo {author} {\bibfnamefont {Mengchao}\ \bibnamefont {Zhang}},\
  }\bibfield  {title} {\enquote {\bibinfo {title} {{Self-interacting dark
  matter implied by nano-Hertz gravitational waves}},}\ }\href {\doibase
  10.1103/PhysRevD.109.115025} {\bibfield  {journal} {\bibinfo  {journal}
  {Phys. Rev. D}\ }\textbf {\bibinfo {volume} {109}},\ \bibinfo {pages}
  {115025} (\bibinfo {year} {2024})},\ \Eprint
  {http://arxiv.org/abs/2306.16966} {arXiv:2306.16966 [hep-ph]} \BibitemShut
  {NoStop}%
\bibitem [{\citenamefont {Franciolini}\ \emph {et~al.}(2024)\citenamefont
  {Franciolini}, \citenamefont {Racco},\ and\ \citenamefont
  {Rompineve}}]{Franciolini:2023wjm}%
  \BibitemOpen
  \bibfield  {author} {\bibinfo {author} {\bibfnamefont {Gabriele}\
  \bibnamefont {Franciolini}}, \bibinfo {author} {\bibfnamefont {Davide}\
  \bibnamefont {Racco}}, \ and\ \bibinfo {author} {\bibfnamefont {Fabrizio}\
  \bibnamefont {Rompineve}},\ }\bibfield  {title} {\enquote {\bibinfo {title}
  {{Footprints of the QCD Crossover on Cosmological Gravitational Waves at
  Pulsar Timing Arrays}},}\ }\href {\doibase 10.1103/PhysRevLett.132.081001}
  {\bibfield  {journal} {\bibinfo  {journal} {Phys. Rev. Lett.}\ }\textbf
  {\bibinfo {volume} {132}},\ \bibinfo {pages} {081001} (\bibinfo {year}
  {2024})},\ \Eprint {http://arxiv.org/abs/2306.17136} {arXiv:2306.17136
  [astro-ph.CO]} \BibitemShut {NoStop}%
\bibitem [{\citenamefont {Bian}\ \emph {et~al.}(2024)\citenamefont {Bian},
  \citenamefont {Ge}, \citenamefont {Shu}, \citenamefont {Wang}, \citenamefont
  {Yang},\ and\ \citenamefont {Zong}}]{Bian:2023dnv}%
  \BibitemOpen
  \bibfield  {author} {\bibinfo {author} {\bibfnamefont {Ligong}\ \bibnamefont
  {Bian}}, \bibinfo {author} {\bibfnamefont {Shuailiang}\ \bibnamefont {Ge}},
  \bibinfo {author} {\bibfnamefont {Jing}\ \bibnamefont {Shu}}, \bibinfo
  {author} {\bibfnamefont {Bo}~\bibnamefont {Wang}}, \bibinfo {author}
  {\bibfnamefont {Xing-Yu}\ \bibnamefont {Yang}}, \ and\ \bibinfo {author}
  {\bibfnamefont {Junchao}\ \bibnamefont {Zong}},\ }\bibfield  {title}
  {\enquote {\bibinfo {title} {{Gravitational wave sources for pulsar timing
  arrays}},}\ }\href {\doibase 10.1103/PhysRevD.109.L101301} {\bibfield
  {journal} {\bibinfo  {journal} {Phys. Rev. D}\ }\textbf {\bibinfo {volume}
  {109}},\ \bibinfo {pages} {L101301} (\bibinfo {year} {2024})},\ \Eprint
  {http://arxiv.org/abs/2307.02376} {arXiv:2307.02376 [astro-ph.HE]}
  \BibitemShut {NoStop}%
\bibitem [{\citenamefont {Jiang}\ \emph {et~al.}(2024)\citenamefont {Jiang},
  \citenamefont {Yang}, \citenamefont {Ma},\ and\ \citenamefont
  {Huang}}]{Jiang:2023qbm}%
  \BibitemOpen
  \bibfield  {author} {\bibinfo {author} {\bibfnamefont {Siyu}\ \bibnamefont
  {Jiang}}, \bibinfo {author} {\bibfnamefont {Aidi}\ \bibnamefont {Yang}},
  \bibinfo {author} {\bibfnamefont {Jiucheng}\ \bibnamefont {Ma}}, \ and\
  \bibinfo {author} {\bibfnamefont {Fa~Peng}\ \bibnamefont {Huang}},\
  }\bibfield  {title} {\enquote {\bibinfo {title} {{Implication of nano-Hertz
  stochastic gravitational wave on dynamical dark matter through a dark
  first-order phase transition}},}\ }\href {\doibase 10.1088/1361-6382/ad24c6}
  {\bibfield  {journal} {\bibinfo  {journal} {Class. Quant. Grav.}\ }\textbf
  {\bibinfo {volume} {41}},\ \bibinfo {pages} {065009} (\bibinfo {year}
  {2024})},\ \Eprint {http://arxiv.org/abs/2306.17827} {arXiv:2306.17827
  [hep-ph]} \BibitemShut {NoStop}%
\bibitem [{\citenamefont {Ghosh}\ \emph {et~al.}(2023)\citenamefont {Ghosh},
  \citenamefont {Ghoshal}, \citenamefont {Guo}, \citenamefont {Hajkarim},
  \citenamefont {King}, \citenamefont {Sinha}, \citenamefont {Wang},\ and\
  \citenamefont {White}}]{Ghosh:2023aum}%
  \BibitemOpen
  \bibfield  {author} {\bibinfo {author} {\bibfnamefont {Tathagata}\
  \bibnamefont {Ghosh}}, \bibinfo {author} {\bibfnamefont {Anish}\ \bibnamefont
  {Ghoshal}}, \bibinfo {author} {\bibfnamefont {Huai-Ke}\ \bibnamefont {Guo}},
  \bibinfo {author} {\bibfnamefont {Fazlollah}\ \bibnamefont {Hajkarim}},
  \bibinfo {author} {\bibfnamefont {Stephen~F.}\ \bibnamefont {King}}, \bibinfo
  {author} {\bibfnamefont {Kuver}\ \bibnamefont {Sinha}}, \bibinfo {author}
  {\bibfnamefont {Xin}\ \bibnamefont {Wang}}, \ and\ \bibinfo {author}
  {\bibfnamefont {Graham}\ \bibnamefont {White}},\ }\bibfield  {title}
  {\enquote {\bibinfo {title} {{Did we hear the sound of the Universe boiling?
  Analysis using the full fluid velocity profiles and NANOGrav 15-year
  data}},}\ }\href@noop {} {\  (\bibinfo {year} {2023})},\ \Eprint
  {http://arxiv.org/abs/2307.02259} {arXiv:2307.02259 [astro-ph.HE]}
  \BibitemShut {NoStop}%
\bibitem [{\citenamefont {Xiao}\ \emph {et~al.}(2023)\citenamefont {Xiao},
  \citenamefont {Yang},\ and\ \citenamefont {Zhang}}]{Xiao:2023dbb}%
  \BibitemOpen
  \bibfield  {author} {\bibinfo {author} {\bibfnamefont {Yang}\ \bibnamefont
  {Xiao}}, \bibinfo {author} {\bibfnamefont {Jin~Min}\ \bibnamefont {Yang}}, \
  and\ \bibinfo {author} {\bibfnamefont {Yang}\ \bibnamefont {Zhang}},\
  }\bibfield  {title} {\enquote {\bibinfo {title} {{Implications of nano-Hertz
  gravitational waves on electroweak phase transition in the singlet dark
  matter model}},}\ }\href {\doibase 10.1016/j.scib.2023.11.025} {\bibfield
  {journal} {\bibinfo  {journal} {Sci. Bull.}\ }\textbf {\bibinfo {volume}
  {68}},\ \bibinfo {pages} {3158--3164} (\bibinfo {year} {2023})},\ \Eprint
  {http://arxiv.org/abs/2307.01072} {arXiv:2307.01072 [hep-ph]} \BibitemShut
  {NoStop}%
\bibitem [{\citenamefont {Li}\ and\ \citenamefont {Xie}(2023)}]{Li:2023bxy}%
  \BibitemOpen
  \bibfield  {author} {\bibinfo {author} {\bibfnamefont {Shao-Ping}\
  \bibnamefont {Li}}\ and\ \bibinfo {author} {\bibfnamefont {Ke-Pan}\
  \bibnamefont {Xie}},\ }\bibfield  {title} {\enquote {\bibinfo {title}
  {{Collider test of nano-Hertz gravitational waves from pulsar timing
  arrays}},}\ }\href {\doibase 10.1103/PhysRevD.108.055018} {\bibfield
  {journal} {\bibinfo  {journal} {Phys. Rev. D}\ }\textbf {\bibinfo {volume}
  {108}},\ \bibinfo {pages} {055018} (\bibinfo {year} {2023})},\ \Eprint
  {http://arxiv.org/abs/2307.01086} {arXiv:2307.01086 [hep-ph]} \BibitemShut
  {NoStop}%
\bibitem [{\citenamefont {Di~Bari}\ and\ \citenamefont
  {Rahat}(2023)}]{DiBari:2023upq}%
  \BibitemOpen
  \bibfield  {author} {\bibinfo {author} {\bibfnamefont {Pasquale}\
  \bibnamefont {Di~Bari}}\ and\ \bibinfo {author} {\bibfnamefont
  {Moinul~Hossain}\ \bibnamefont {Rahat}},\ }\bibfield  {title} {\enquote
  {\bibinfo {title} {{The split majoron model confronts the NANOGrav
  signal}},}\ }\href@noop {} {\  (\bibinfo {year} {2023})},\ \Eprint
  {http://arxiv.org/abs/2307.03184} {arXiv:2307.03184 [hep-ph]} \BibitemShut
  {NoStop}%
\bibitem [{\citenamefont {Cruz}\ \emph {et~al.}(2023)\citenamefont {Cruz},
  \citenamefont {Niedermann},\ and\ \citenamefont {Sloth}}]{Cruz:2023lnq}%
  \BibitemOpen
  \bibfield  {author} {\bibinfo {author} {\bibfnamefont {Juan~S.}\ \bibnamefont
  {Cruz}}, \bibinfo {author} {\bibfnamefont {Florian}\ \bibnamefont
  {Niedermann}}, \ and\ \bibinfo {author} {\bibfnamefont {Martin~S.}\
  \bibnamefont {Sloth}},\ }\bibfield  {title} {\enquote {\bibinfo {title}
  {{NANOGrav meets Hot New Early Dark Energy and the origin of neutrino
  mass}},}\ }\href {\doibase 10.1016/j.physletb.2023.138202} {\bibfield
  {journal} {\bibinfo  {journal} {Phys. Lett. B}\ }\textbf {\bibinfo {volume}
  {846}},\ \bibinfo {pages} {138202} (\bibinfo {year} {2023})},\ \Eprint
  {http://arxiv.org/abs/2307.03091} {arXiv:2307.03091 [astro-ph.CO]}
  \BibitemShut {NoStop}%
\bibitem [{\citenamefont {Wu}\ \emph {et~al.}(2024)\citenamefont {Wu},
  \citenamefont {Chen},\ and\ \citenamefont {Huang}}]{Wu:2023hsa}%
  \BibitemOpen
  \bibfield  {author} {\bibinfo {author} {\bibfnamefont {Yu-Mei}\ \bibnamefont
  {Wu}}, \bibinfo {author} {\bibfnamefont {Zu-Cheng}\ \bibnamefont {Chen}}, \
  and\ \bibinfo {author} {\bibfnamefont {Qing-Guo}\ \bibnamefont {Huang}},\
  }\bibfield  {title} {\enquote {\bibinfo {title} {{Cosmological interpretation
  for the stochastic signal in pulsar timing arrays}},}\ }\href {\doibase
  10.1007/s11433-023-2298-7} {\bibfield  {journal} {\bibinfo  {journal} {Sci.
  China Phys. Mech. Astron.}\ }\textbf {\bibinfo {volume} {67}},\ \bibinfo
  {pages} {240412} (\bibinfo {year} {2024})},\ \Eprint
  {http://arxiv.org/abs/2307.03141} {arXiv:2307.03141 [astro-ph.CO]}
  \BibitemShut {NoStop}%
\bibitem [{\citenamefont {Du}\ \emph {et~al.}(2023)\citenamefont {Du},
  \citenamefont {Huang}, \citenamefont {Wang},\ and\ \citenamefont
  {Zhang}}]{Du:2023qvj}%
  \BibitemOpen
  \bibfield  {author} {\bibinfo {author} {\bibfnamefont {Xiao~Kang}\
  \bibnamefont {Du}}, \bibinfo {author} {\bibfnamefont {Ming~Xia}\ \bibnamefont
  {Huang}}, \bibinfo {author} {\bibfnamefont {Fei}\ \bibnamefont {Wang}}, \
  and\ \bibinfo {author} {\bibfnamefont {Ying~Kai}\ \bibnamefont {Zhang}},\
  }\bibfield  {title} {\enquote {\bibinfo {title} {{Did the nHZ Gravitational
  Waves Signatures Observed By NANOGrav Indicate Multiple Sector SUSY
  Breaking?}}}\ }\href@noop {} {\  (\bibinfo {year} {2023})},\ \Eprint
  {http://arxiv.org/abs/2307.02938} {arXiv:2307.02938 [hep-ph]} \BibitemShut
  {NoStop}%
\bibitem [{\citenamefont {Ahmadvand}\ \emph {et~al.}(2023)\citenamefont
  {Ahmadvand}, \citenamefont {Bian},\ and\ \citenamefont
  {Shakeri}}]{Ahmadvand:2023lpp}%
  \BibitemOpen
  \bibfield  {author} {\bibinfo {author} {\bibfnamefont {Moslem}\ \bibnamefont
  {Ahmadvand}}, \bibinfo {author} {\bibfnamefont {Ligong}\ \bibnamefont
  {Bian}}, \ and\ \bibinfo {author} {\bibfnamefont {Soroush}\ \bibnamefont
  {Shakeri}},\ }\bibfield  {title} {\enquote {\bibinfo {title} {{Heavy QCD
  axion model in light of pulsar timing arrays}},}\ }\href {\doibase
  10.1103/PhysRevD.108.115020} {\bibfield  {journal} {\bibinfo  {journal}
  {Phys. Rev. D}\ }\textbf {\bibinfo {volume} {108}},\ \bibinfo {pages}
  {115020} (\bibinfo {year} {2023})},\ \Eprint
  {http://arxiv.org/abs/2307.12385} {arXiv:2307.12385 [hep-ph]} \BibitemShut
  {NoStop}%
\bibitem [{\citenamefont {Wang}(2023)}]{Wang:2023bbc}%
  \BibitemOpen
  \bibfield  {author} {\bibinfo {author} {\bibfnamefont {Deng}\ \bibnamefont
  {Wang}},\ }\bibfield  {title} {\enquote {\bibinfo {title} {{Constraining
  Cosmological Phase Transitions with Chinese Pulsar Timing Array Data Release
  1}},}\ }\href@noop {} {\  (\bibinfo {year} {2023})},\ \Eprint
  {http://arxiv.org/abs/2307.15970} {arXiv:2307.15970 [astro-ph.CO]}
  \BibitemShut {NoStop}%
\bibitem [{\citenamefont {Hawking}\ \emph {et~al.}(1982)\citenamefont
  {Hawking}, \citenamefont {Moss},\ and\ \citenamefont
  {Stewart}}]{Hawking:1982ga}%
  \BibitemOpen
  \bibfield  {author} {\bibinfo {author} {\bibfnamefont {S.~W.}\ \bibnamefont
  {Hawking}}, \bibinfo {author} {\bibfnamefont {I.~G.}\ \bibnamefont {Moss}}, \
  and\ \bibinfo {author} {\bibfnamefont {J.~M.}\ \bibnamefont {Stewart}},\
  }\bibfield  {title} {\enquote {\bibinfo {title} {{Bubble Collisions in the
  Very Early Universe}},}\ }\href {\doibase 10.1103/PhysRevD.26.2681}
  {\bibfield  {journal} {\bibinfo  {journal} {Phys. Rev.}\ }\textbf {\bibinfo
  {volume} {D26}},\ \bibinfo {pages} {2681} (\bibinfo {year}
  {1982})}\BibitemShut {NoStop}%
\bibitem [{\citenamefont {Crawford}\ and\ \citenamefont
  {Schramm}(1982)}]{Crawford:1982yz}%
  \BibitemOpen
  \bibfield  {author} {\bibinfo {author} {\bibfnamefont {Matt}\ \bibnamefont
  {Crawford}}\ and\ \bibinfo {author} {\bibfnamefont {David~N.}\ \bibnamefont
  {Schramm}},\ }\bibfield  {title} {\enquote {\bibinfo {title} {{Spontaneous
  Generation of Density Perturbations in the Early Universe}},}\ }\href
  {\doibase 10.1038/298538a0} {\bibfield  {journal} {\bibinfo  {journal}
  {Nature}\ }\textbf {\bibinfo {volume} {298}},\ \bibinfo {pages} {538--540}
  (\bibinfo {year} {1982})}\BibitemShut {NoStop}%
\bibitem [{\citenamefont {Moss}(1994{\natexlab{a}})}]{Moss:1994pi}%
  \BibitemOpen
  \bibfield  {author} {\bibinfo {author} {\bibfnamefont {Ian~G.}\ \bibnamefont
  {Moss}},\ }\bibfield  {title} {\enquote {\bibinfo {title} {{Black hole
  formation from colliding bubbles}},}\ }\href@noop {} {\  (\bibinfo {year}
  {1994}{\natexlab{a}})},\ \Eprint {http://arxiv.org/abs/gr-qc/9405045}
  {arXiv:gr-qc/9405045} \BibitemShut {NoStop}%
\bibitem [{\citenamefont {Moss}(1994{\natexlab{b}})}]{Moss:1994iq}%
  \BibitemOpen
  \bibfield  {author} {\bibinfo {author} {\bibfnamefont {I.~G.}\ \bibnamefont
  {Moss}},\ }\bibfield  {title} {\enquote {\bibinfo {title} {{Singularity
  formation from colliding bubbles}},}\ }\href {\doibase
  10.1103/PhysRevD.50.676} {\bibfield  {journal} {\bibinfo  {journal} {Phys.
  Rev.}\ }\textbf {\bibinfo {volume} {D50}},\ \bibinfo {pages} {676--681}
  (\bibinfo {year} {1994}{\natexlab{b}})}\BibitemShut {NoStop}%
\bibitem [{\citenamefont {Baker}\ \emph
  {et~al.}(2021{\natexlab{a}})\citenamefont {Baker}, \citenamefont {Breitbach},
  \citenamefont {Kopp},\ and\ \citenamefont {Mittnacht}}]{Baker:2021nyl}%
  \BibitemOpen
  \bibfield  {author} {\bibinfo {author} {\bibfnamefont {Michael~J.}\
  \bibnamefont {Baker}}, \bibinfo {author} {\bibfnamefont {Moritz}\
  \bibnamefont {Breitbach}}, \bibinfo {author} {\bibfnamefont {Joachim}\
  \bibnamefont {Kopp}}, \ and\ \bibinfo {author} {\bibfnamefont {Lukas}\
  \bibnamefont {Mittnacht}},\ }\bibfield  {title} {\enquote {\bibinfo {title}
  {{Primordial Black Holes from First-Order Cosmological Phase Transitions}},}\
  }\href@noop {} {\  (\bibinfo {year} {2021}{\natexlab{a}})},\ \Eprint
  {http://arxiv.org/abs/2105.07481} {arXiv:2105.07481 [astro-ph.CO]}
  \BibitemShut {NoStop}%
\bibitem [{\citenamefont {Baker}\ \emph
  {et~al.}(2021{\natexlab{b}})\citenamefont {Baker}, \citenamefont {Breitbach},
  \citenamefont {Kopp},\ and\ \citenamefont {Mittnacht}}]{Baker:2021sno}%
  \BibitemOpen
  \bibfield  {author} {\bibinfo {author} {\bibfnamefont {Michael~J.}\
  \bibnamefont {Baker}}, \bibinfo {author} {\bibfnamefont {Moritz}\
  \bibnamefont {Breitbach}}, \bibinfo {author} {\bibfnamefont {Joachim}\
  \bibnamefont {Kopp}}, \ and\ \bibinfo {author} {\bibfnamefont {Lukas}\
  \bibnamefont {Mittnacht}},\ }\bibfield  {title} {\enquote {\bibinfo {title}
  {{Detailed Calculation of Primordial Black Hole Formation During First-Order
  Cosmological Phase Transitions}},}\ }\href@noop {} {\  (\bibinfo {year}
  {2021}{\natexlab{b}})},\ \Eprint {http://arxiv.org/abs/2110.00005}
  {arXiv:2110.00005 [astro-ph.CO]} \BibitemShut {NoStop}%
\bibitem [{\citenamefont {Kawana}\ and\ \citenamefont
  {Xie}(2022)}]{Kawana:2021tde}%
  \BibitemOpen
  \bibfield  {author} {\bibinfo {author} {\bibfnamefont {Kiyoharu}\
  \bibnamefont {Kawana}}\ and\ \bibinfo {author} {\bibfnamefont {Ke-Pan}\
  \bibnamefont {Xie}},\ }\bibfield  {title} {\enquote {\bibinfo {title}
  {{Primordial black holes from a cosmic phase transition: The collapse of
  Fermi-balls}},}\ }\href {\doibase 10.1016/j.physletb.2021.136791} {\bibfield
  {journal} {\bibinfo  {journal} {Phys. Lett. B}\ }\textbf {\bibinfo {volume}
  {824}},\ \bibinfo {pages} {136791} (\bibinfo {year} {2022})},\ \Eprint
  {http://arxiv.org/abs/2106.00111} {arXiv:2106.00111 [astro-ph.CO]}
  \BibitemShut {NoStop}%
\bibitem [{\citenamefont {Lu}\ \emph {et~al.}(2022)\citenamefont {Lu},
  \citenamefont {Kawana},\ and\ \citenamefont {Xie}}]{Lu:2022paj}%
  \BibitemOpen
  \bibfield  {author} {\bibinfo {author} {\bibfnamefont {Philip}\ \bibnamefont
  {Lu}}, \bibinfo {author} {\bibfnamefont {Kiyoharu}\ \bibnamefont {Kawana}}, \
  and\ \bibinfo {author} {\bibfnamefont {Ke-Pan}\ \bibnamefont {Xie}},\
  }\bibfield  {title} {\enquote {\bibinfo {title} {{Old phase remnants in
  first-order phase transitions}},}\ }\href {\doibase
  10.1103/PhysRevD.105.123503} {\bibfield  {journal} {\bibinfo  {journal}
  {Phys. Rev. D}\ }\textbf {\bibinfo {volume} {105}},\ \bibinfo {pages}
  {123503} (\bibinfo {year} {2022})},\ \Eprint
  {http://arxiv.org/abs/2202.03439} {arXiv:2202.03439 [astro-ph.CO]}
  \BibitemShut {NoStop}%
\bibitem [{\citenamefont {Kawana}\ \emph {et~al.}(2022)\citenamefont {Kawana},
  \citenamefont {Lu},\ and\ \citenamefont {Xie}}]{Kawana:2022lba}%
  \BibitemOpen
  \bibfield  {author} {\bibinfo {author} {\bibfnamefont {Kiyoharu}\
  \bibnamefont {Kawana}}, \bibinfo {author} {\bibfnamefont {Philip}\
  \bibnamefont {Lu}}, \ and\ \bibinfo {author} {\bibfnamefont {Ke-Pan}\
  \bibnamefont {Xie}},\ }\bibfield  {title} {\enquote {\bibinfo {title}
  {{First-order phase transition and fate of false vacuum remnants}},}\ }\href
  {\doibase 10.1088/1475-7516/2022/10/030} {\bibfield  {journal} {\bibinfo
  {journal} {JCAP}\ }\textbf {\bibinfo {volume} {10}},\ \bibinfo {pages} {030}
  (\bibinfo {year} {2022})},\ \Eprint {http://arxiv.org/abs/2206.09923}
  {arXiv:2206.09923 [astro-ph.CO]} \BibitemShut {NoStop}%
\bibitem [{\citenamefont {Huang}\ and\ \citenamefont
  {Xie}(2022)}]{Huang:2022him}%
  \BibitemOpen
  \bibfield  {author} {\bibinfo {author} {\bibfnamefont {Peisi}\ \bibnamefont
  {Huang}}\ and\ \bibinfo {author} {\bibfnamefont {Ke-Pan}\ \bibnamefont
  {Xie}},\ }\bibfield  {title} {\enquote {\bibinfo {title} {{Primordial black
  holes from an electroweak phase transition}},}\ }\href {\doibase
  10.1103/PhysRevD.105.115033} {\bibfield  {journal} {\bibinfo  {journal}
  {Phys. Rev. D}\ }\textbf {\bibinfo {volume} {105}},\ \bibinfo {pages}
  {115033} (\bibinfo {year} {2022})},\ \Eprint
  {http://arxiv.org/abs/2201.07243} {arXiv:2201.07243 [hep-ph]} \BibitemShut
  {NoStop}%
\bibitem [{\citenamefont {Kodama}\ \emph {et~al.}(1982)\citenamefont {Kodama},
  \citenamefont {Sasaki},\ and\ \citenamefont {Sato}}]{Kodama:1982sf}%
  \BibitemOpen
  \bibfield  {author} {\bibinfo {author} {\bibfnamefont {Hideo}\ \bibnamefont
  {Kodama}}, \bibinfo {author} {\bibfnamefont {Misao}\ \bibnamefont {Sasaki}},
  \ and\ \bibinfo {author} {\bibfnamefont {Katsuhiko}\ \bibnamefont {Sato}},\
  }\bibfield  {title} {\enquote {\bibinfo {title} {{Abundance of Primordial
  Holes Produced by Cosmological First Order Phase Transition}},}\ }\href
  {\doibase 10.1143/PTP.68.1979} {\bibfield  {journal} {\bibinfo  {journal}
  {Prog. Theor. Phys.}\ }\textbf {\bibinfo {volume} {68}},\ \bibinfo {pages}
  {1979} (\bibinfo {year} {1982})}\BibitemShut {NoStop}%
\bibitem [{\citenamefont {Liu}\ \emph {et~al.}(2022)\citenamefont {Liu},
  \citenamefont {Bian}, \citenamefont {Cai}, \citenamefont {Guo},\ and\
  \citenamefont {Wang}}]{Liu:2021svg}%
  \BibitemOpen
  \bibfield  {author} {\bibinfo {author} {\bibfnamefont {Jing}\ \bibnamefont
  {Liu}}, \bibinfo {author} {\bibfnamefont {Ligong}\ \bibnamefont {Bian}},
  \bibinfo {author} {\bibfnamefont {Rong-Gen}\ \bibnamefont {Cai}}, \bibinfo
  {author} {\bibfnamefont {Zong-Kuan}\ \bibnamefont {Guo}}, \ and\ \bibinfo
  {author} {\bibfnamefont {Shao-Jiang}\ \bibnamefont {Wang}},\ }\bibfield
  {title} {\enquote {\bibinfo {title} {{Primordial black hole production during
  first-order phase transitions}},}\ }\href {\doibase
  10.1103/PhysRevD.105.L021303} {\bibfield  {journal} {\bibinfo  {journal}
  {Phys. Rev. D}\ }\textbf {\bibinfo {volume} {105}},\ \bibinfo {pages}
  {L021303} (\bibinfo {year} {2022})},\ \Eprint
  {http://arxiv.org/abs/2106.05637} {arXiv:2106.05637 [astro-ph.CO]}
  \BibitemShut {NoStop}%
\bibitem [{\citenamefont {Kawana}\ \emph {et~al.}(2023)\citenamefont {Kawana},
  \citenamefont {Kim},\ and\ \citenamefont {Lu}}]{Kawana:2022olo}%
  \BibitemOpen
  \bibfield  {author} {\bibinfo {author} {\bibfnamefont {Kiyoharu}\
  \bibnamefont {Kawana}}, \bibinfo {author} {\bibfnamefont {TaeHun}\
  \bibnamefont {Kim}}, \ and\ \bibinfo {author} {\bibfnamefont {Philip}\
  \bibnamefont {Lu}},\ }\bibfield  {title} {\enquote {\bibinfo {title} {{PBH
  formation from overdensities in delayed vacuum transitions}},}\ }\href
  {\doibase 10.1103/PhysRevD.108.103531} {\bibfield  {journal} {\bibinfo
  {journal} {Phys. Rev. D}\ }\textbf {\bibinfo {volume} {108}},\ \bibinfo
  {pages} {103531} (\bibinfo {year} {2023})},\ \Eprint
  {http://arxiv.org/abs/2212.14037} {arXiv:2212.14037 [astro-ph.CO]}
  \BibitemShut {NoStop}%
\bibitem [{\citenamefont {Gouttenoire}\ and\ \citenamefont
  {Volansky}(2023)}]{Gouttenoire:2023naa}%
  \BibitemOpen
  \bibfield  {author} {\bibinfo {author} {\bibfnamefont {Yann}\ \bibnamefont
  {Gouttenoire}}\ and\ \bibinfo {author} {\bibfnamefont {Tomer}\ \bibnamefont
  {Volansky}},\ }\bibfield  {title} {\enquote {\bibinfo {title} {{Primordial
  Black Holes from Supercooled Phase Transitions}},}\ }\href@noop {} {\
  (\bibinfo {year} {2023})},\ \Eprint {http://arxiv.org/abs/2305.04942}
  {arXiv:2305.04942 [hep-ph]} \BibitemShut {NoStop}%
\bibitem [{\citenamefont {Lewicki}\ \emph {et~al.}(2023)\citenamefont
  {Lewicki}, \citenamefont {Toczek},\ and\ \citenamefont
  {Vaskonen}}]{Lewicki:2023ioy}%
  \BibitemOpen
  \bibfield  {author} {\bibinfo {author} {\bibfnamefont {Marek}\ \bibnamefont
  {Lewicki}}, \bibinfo {author} {\bibfnamefont {Piotr}\ \bibnamefont {Toczek}},
  \ and\ \bibinfo {author} {\bibfnamefont {Ville}\ \bibnamefont {Vaskonen}},\
  }\bibfield  {title} {\enquote {\bibinfo {title} {{Primordial black holes from
  strong first-order phase transitions}},}\ }\href {\doibase
  10.1007/JHEP09(2023)092} {\bibfield  {journal} {\bibinfo  {journal} {JHEP}\
  }\textbf {\bibinfo {volume} {09}},\ \bibinfo {pages} {092} (\bibinfo {year}
  {2023})},\ \Eprint {http://arxiv.org/abs/2305.04924} {arXiv:2305.04924
  [astro-ph.CO]} \BibitemShut {NoStop}%
\bibitem [{\citenamefont {Kanemura}\ \emph {et~al.}(2024)\citenamefont
  {Kanemura}, \citenamefont {Tanaka},\ and\ \citenamefont
  {Xie}}]{Kanemura:2024pae}%
  \BibitemOpen
  \bibfield  {author} {\bibinfo {author} {\bibfnamefont {Shinya}\ \bibnamefont
  {Kanemura}}, \bibinfo {author} {\bibfnamefont {Masanori}\ \bibnamefont
  {Tanaka}}, \ and\ \bibinfo {author} {\bibfnamefont {Ke-Pan}\ \bibnamefont
  {Xie}},\ }\bibfield  {title} {\enquote {\bibinfo {title} {{Primordial black
  holes from slow phase transitions: a model-building perspective}},}\
  }\href@noop {} {\  (\bibinfo {year} {2024})},\ \Eprint
  {http://arxiv.org/abs/2404.00646} {arXiv:2404.00646 [hep-ph]} \BibitemShut
  {NoStop}%
\bibitem [{\citenamefont {Cai}\ \emph {et~al.}(2024)\citenamefont {Cai},
  \citenamefont {Hao},\ and\ \citenamefont {Wang}}]{Cai:2024nln}%
  \BibitemOpen
  \bibfield  {author} {\bibinfo {author} {\bibfnamefont {Rong-Gen}\
  \bibnamefont {Cai}}, \bibinfo {author} {\bibfnamefont {Yu-Shi}\ \bibnamefont
  {Hao}}, \ and\ \bibinfo {author} {\bibfnamefont {Shao-Jiang}\ \bibnamefont
  {Wang}},\ }\bibfield  {title} {\enquote {\bibinfo {title} {{Primordial black
  holes and curvature perturbations from false-vacuum islands}},}\ }\href@noop
  {} {\  (\bibinfo {year} {2024})},\ \Eprint {http://arxiv.org/abs/2404.06506}
  {arXiv:2404.06506 [astro-ph.CO]} \BibitemShut {NoStop}%
\bibitem [{\citenamefont {Hashino}\ \emph {et~al.}(2022)\citenamefont
  {Hashino}, \citenamefont {Kanemura},\ and\ \citenamefont
  {Takahashi}}]{Hashino:2021qoq}%
  \BibitemOpen
  \bibfield  {author} {\bibinfo {author} {\bibfnamefont {Katsuya}\ \bibnamefont
  {Hashino}}, \bibinfo {author} {\bibfnamefont {Shinya}\ \bibnamefont
  {Kanemura}}, \ and\ \bibinfo {author} {\bibfnamefont {Tomo}\ \bibnamefont
  {Takahashi}},\ }\bibfield  {title} {\enquote {\bibinfo {title} {{Primordial
  black holes as a probe of strongly first-order electroweak phase
  transition}},}\ }\href {\doibase 10.1016/j.physletb.2022.137261} {\bibfield
  {journal} {\bibinfo  {journal} {Phys. Lett. B}\ }\textbf {\bibinfo {volume}
  {833}},\ \bibinfo {pages} {137261} (\bibinfo {year} {2022})},\ \Eprint
  {http://arxiv.org/abs/2111.13099} {arXiv:2111.13099 [hep-ph]} \BibitemShut
  {NoStop}%
\bibitem [{\citenamefont {Hashino}\ \emph {et~al.}(2023)\citenamefont
  {Hashino}, \citenamefont {Kanemura}, \citenamefont {Takahashi},\ and\
  \citenamefont {Tanaka}}]{Hashino:2022tcs}%
  \BibitemOpen
  \bibfield  {author} {\bibinfo {author} {\bibfnamefont {Katsuya}\ \bibnamefont
  {Hashino}}, \bibinfo {author} {\bibfnamefont {Shinya}\ \bibnamefont
  {Kanemura}}, \bibinfo {author} {\bibfnamefont {Tomo}\ \bibnamefont
  {Takahashi}}, \ and\ \bibinfo {author} {\bibfnamefont {Masanori}\
  \bibnamefont {Tanaka}},\ }\bibfield  {title} {\enquote {\bibinfo {title}
  {{Probing first-order electroweak phase transition via primordial black holes
  in the effective field theory}},}\ }\href {\doibase
  10.1016/j.physletb.2023.137688} {\bibfield  {journal} {\bibinfo  {journal}
  {Phys. Lett. B}\ }\textbf {\bibinfo {volume} {838}},\ \bibinfo {pages}
  {137688} (\bibinfo {year} {2023})},\ \Eprint
  {http://arxiv.org/abs/2211.16225} {arXiv:2211.16225 [hep-ph]} \BibitemShut
  {NoStop}%
\bibitem [{\citenamefont {Salvio}(2023)}]{Salvio:2023ynn}%
  \BibitemOpen
  \bibfield  {author} {\bibinfo {author} {\bibfnamefont {Alberto}\ \bibnamefont
  {Salvio}},\ }\bibfield  {title} {\enquote {\bibinfo {title} {{Supercooling in
  radiative symmetry breaking: theory extensions, gravitational wave detection
  and primordial black holes}},}\ }\href {\doibase
  10.1088/1475-7516/2023/12/046} {\bibfield  {journal} {\bibinfo  {journal}
  {JCAP}\ }\textbf {\bibinfo {volume} {12}},\ \bibinfo {pages} {046} (\bibinfo
  {year} {2023})},\ \Eprint {http://arxiv.org/abs/2307.04694} {arXiv:2307.04694
  [hep-ph]} \BibitemShut {NoStop}%
\bibitem [{\citenamefont
  {Gouttenoire}(2023{\natexlab{b}})}]{Gouttenoire:2023pxh}%
  \BibitemOpen
  \bibfield  {author} {\bibinfo {author} {\bibfnamefont {Yann}\ \bibnamefont
  {Gouttenoire}},\ }\bibfield  {title} {\enquote {\bibinfo {title} {{Primordial
  Black Holes from Conformal Higgs}},}\ }\href@noop {} {\  (\bibinfo {year}
  {2023}{\natexlab{b}})},\ \Eprint {http://arxiv.org/abs/2311.13640}
  {arXiv:2311.13640 [hep-ph]} \BibitemShut {NoStop}%
\bibitem [{\citenamefont {Conaci}\ \emph {et~al.}(2024)\citenamefont {Conaci},
  \citenamefont {Delle~Rose}, \citenamefont {Dev},\ and\ \citenamefont
  {Ghoshal}}]{Conaci:2024tlc}%
  \BibitemOpen
  \bibfield  {author} {\bibinfo {author} {\bibfnamefont {Angela}\ \bibnamefont
  {Conaci}}, \bibinfo {author} {\bibfnamefont {Luigi}\ \bibnamefont
  {Delle~Rose}}, \bibinfo {author} {\bibfnamefont {P.~S.~Bhupal}\ \bibnamefont
  {Dev}}, \ and\ \bibinfo {author} {\bibfnamefont {Anish}\ \bibnamefont
  {Ghoshal}},\ }\bibfield  {title} {\enquote {\bibinfo {title} {{Slaying
  Axion-Like Particles via Gravitational Waves and Primordial Black Holes from
  Supercooled Phase Transition}},}\ }\href@noop {} {\  (\bibinfo {year}
  {2024})},\ \Eprint {http://arxiv.org/abs/2401.09411} {arXiv:2401.09411
  [astro-ph.CO]} \BibitemShut {NoStop}%
\bibitem [{\citenamefont {Baldes}\ and\ \citenamefont
  {Olea-Romacho}(2024)}]{Baldes:2023rqv}%
  \BibitemOpen
  \bibfield  {author} {\bibinfo {author} {\bibfnamefont {Iason}\ \bibnamefont
  {Baldes}}\ and\ \bibinfo {author} {\bibfnamefont {Mar\'\i{}a~Olalla}\
  \bibnamefont {Olea-Romacho}},\ }\bibfield  {title} {\enquote {\bibinfo
  {title} {{Primordial black holes as dark matter: interferometric tests of
  phase transition origin}},}\ }\href {\doibase 10.1007/JHEP01(2024)133}
  {\bibfield  {journal} {\bibinfo  {journal} {JHEP}\ }\textbf {\bibinfo
  {volume} {01}},\ \bibinfo {pages} {133} (\bibinfo {year} {2024})},\ \Eprint
  {http://arxiv.org/abs/2307.11639} {arXiv:2307.11639 [hep-ph]} \BibitemShut
  {NoStop}%
\bibitem [{\citenamefont {Liu}\ \emph {et~al.}(2023)\citenamefont {Liu},
  \citenamefont {Bian}, \citenamefont {Cai}, \citenamefont {Guo},\ and\
  \citenamefont {Wang}}]{Liu:2022lvz}%
  \BibitemOpen
  \bibfield  {author} {\bibinfo {author} {\bibfnamefont {Jing}\ \bibnamefont
  {Liu}}, \bibinfo {author} {\bibfnamefont {Ligong}\ \bibnamefont {Bian}},
  \bibinfo {author} {\bibfnamefont {Rong-Gen}\ \bibnamefont {Cai}}, \bibinfo
  {author} {\bibfnamefont {Zong-Kuan}\ \bibnamefont {Guo}}, \ and\ \bibinfo
  {author} {\bibfnamefont {Shao-Jiang}\ \bibnamefont {Wang}},\ }\bibfield
  {title} {\enquote {\bibinfo {title} {{Constraining First-Order Phase
  Transitions with Curvature Perturbations}},}\ }\href {\doibase
  10.1103/PhysRevLett.130.051001} {\bibfield  {journal} {\bibinfo  {journal}
  {Phys. Rev. Lett.}\ }\textbf {\bibinfo {volume} {130}},\ \bibinfo {pages}
  {051001} (\bibinfo {year} {2023})},\ \Eprint
  {http://arxiv.org/abs/2208.14086} {arXiv:2208.14086 [astro-ph.CO]}
  \BibitemShut {NoStop}%
\bibitem [{\citenamefont {Elor}\ \emph {et~al.}(2023)\citenamefont {Elor},
  \citenamefont {Jinno}, \citenamefont {Kumar}, \citenamefont {McGehee},\ and\
  \citenamefont {Tsai}}]{Elor:2023xbz}%
  \BibitemOpen
  \bibfield  {author} {\bibinfo {author} {\bibfnamefont {Gilly}\ \bibnamefont
  {Elor}}, \bibinfo {author} {\bibfnamefont {Ryusuke}\ \bibnamefont {Jinno}},
  \bibinfo {author} {\bibfnamefont {Soubhik}\ \bibnamefont {Kumar}}, \bibinfo
  {author} {\bibfnamefont {Robert}\ \bibnamefont {McGehee}}, \ and\ \bibinfo
  {author} {\bibfnamefont {Yuhsin}\ \bibnamefont {Tsai}},\ }\bibfield  {title}
  {\enquote {\bibinfo {title} {{Finite Bubble Statistics Constrain Late
  Cosmological Phase Transitions}},}\ }\href@noop {} {\  (\bibinfo {year}
  {2023})},\ \Eprint {http://arxiv.org/abs/2311.16222} {arXiv:2311.16222
  [hep-ph]} \BibitemShut {NoStop}%
\bibitem [{\citenamefont {Lewicki}\ \emph {et~al.}(2024)\citenamefont
  {Lewicki}, \citenamefont {Toczek},\ and\ \citenamefont
  {Vaskonen}}]{Lewicki:2024ghw}%
  \BibitemOpen
  \bibfield  {author} {\bibinfo {author} {\bibfnamefont {Marek}\ \bibnamefont
  {Lewicki}}, \bibinfo {author} {\bibfnamefont {Piotr}\ \bibnamefont {Toczek}},
  \ and\ \bibinfo {author} {\bibfnamefont {Ville}\ \bibnamefont {Vaskonen}},\
  }\bibfield  {title} {\enquote {\bibinfo {title} {{Black holes and
  gravitational waves from slow phase transitions}},}\ }\href@noop {} {\
  (\bibinfo {year} {2024})},\ \Eprint {http://arxiv.org/abs/2402.04158}
  {arXiv:2402.04158 [astro-ph.CO]} \BibitemShut {NoStop}%
\bibitem [{\citenamefont {Carr}\ and\ \citenamefont
  {Kuhnel}(2020)}]{Carr:2020xqk}%
  \BibitemOpen
  \bibfield  {author} {\bibinfo {author} {\bibfnamefont {Bernard}\ \bibnamefont
  {Carr}}\ and\ \bibinfo {author} {\bibfnamefont {Florian}\ \bibnamefont
  {Kuhnel}},\ }\bibfield  {title} {\enquote {\bibinfo {title} {{Primordial
  Black Holes as Dark Matter: Recent Developments}},}\ }\href {\doibase
  10.1146/annurev-nucl-050520-125911} {\bibfield  {journal} {\bibinfo
  {journal} {Ann. Rev. Nucl. Part. Sci.}\ }\textbf {\bibinfo {volume} {70}},\
  \bibinfo {pages} {355--394} (\bibinfo {year} {2020})},\ \Eprint
  {http://arxiv.org/abs/2006.02838} {arXiv:2006.02838 [astro-ph.CO]}
  \BibitemShut {NoStop}%
\bibitem [{\citenamefont {Carr}\ and\ \citenamefont
  {Kuhnel}(2022)}]{Carr:2021bzv}%
  \BibitemOpen
  \bibfield  {author} {\bibinfo {author} {\bibfnamefont {Bernard}\ \bibnamefont
  {Carr}}\ and\ \bibinfo {author} {\bibfnamefont {Florian}\ \bibnamefont
  {Kuhnel}},\ }\bibfield  {title} {\enquote {\bibinfo {title} {{Primordial
  black holes as dark matter candidates}},}\ }\href {\doibase
  10.21468/SciPostPhysLectNotes.48} {\bibfield  {journal} {\bibinfo  {journal}
  {SciPost Phys. Lect. Notes}\ }\textbf {\bibinfo {volume} {48}},\ \bibinfo
  {pages} {1} (\bibinfo {year} {2022})},\ \Eprint
  {http://arxiv.org/abs/2110.02821} {arXiv:2110.02821 [astro-ph.CO]}
  \BibitemShut {NoStop}%
\bibitem [{\citenamefont {Escriv\`a}\ \emph {et~al.}(2022)\citenamefont
  {Escriv\`a}, \citenamefont {Kuhnel},\ and\ \citenamefont
  {Tada}}]{Escriva:2022duf}%
  \BibitemOpen
  \bibfield  {author} {\bibinfo {author} {\bibfnamefont {Albert}\ \bibnamefont
  {Escriv\`a}}, \bibinfo {author} {\bibfnamefont {Florian}\ \bibnamefont
  {Kuhnel}}, \ and\ \bibinfo {author} {\bibfnamefont {Yuichiro}\ \bibnamefont
  {Tada}},\ }\bibfield  {title} {\enquote {\bibinfo {title} {{Primordial Black
  Holes}},}\ }\href@noop {} {\  (\bibinfo {year} {2022})},\ \Eprint
  {http://arxiv.org/abs/2211.05767} {arXiv:2211.05767 [astro-ph.CO]}
  \BibitemShut {NoStop}%
\bibitem [{\citenamefont {\"Ozsoy}\ and\ \citenamefont
  {Tasinato}(2023)}]{Ozsoy:2023ryl}%
  \BibitemOpen
  \bibfield  {author} {\bibinfo {author} {\bibfnamefont {Ogan}\ \bibnamefont
  {\"Ozsoy}}\ and\ \bibinfo {author} {\bibfnamefont {Gianmassimo}\ \bibnamefont
  {Tasinato}},\ }\bibfield  {title} {\enquote {\bibinfo {title} {{Inflation and
  Primordial Black Holes}},}\ }\href {\doibase 10.3390/universe9050203}
  {\bibfield  {journal} {\bibinfo  {journal} {Universe}\ }\textbf {\bibinfo
  {volume} {9}},\ \bibinfo {pages} {203} (\bibinfo {year} {2023})},\ \Eprint
  {http://arxiv.org/abs/2301.03600} {arXiv:2301.03600 [astro-ph.CO]}
  \BibitemShut {NoStop}%
\bibitem [{\citenamefont {Bagui}\ \emph {et~al.}(2023)\citenamefont {Bagui}
  \emph {et~al.}}]{LISACosmologyWorkingGroup:2023njw}%
  \BibitemOpen
  \bibfield  {author} {\bibinfo {author} {\bibfnamefont {Eleni}\ \bibnamefont
  {Bagui}} \emph {et~al.} (\bibinfo {collaboration} {LISA Cosmology Working
  Group}),\ }\bibfield  {title} {\enquote {\bibinfo {title} {{Primordial black
  holes and their gravitational-wave signatures}},}\ }\href@noop {} {\
  (\bibinfo {year} {2023})},\ \Eprint {http://arxiv.org/abs/2310.19857}
  {arXiv:2310.19857 [astro-ph.CO]} \BibitemShut {NoStop}%
\bibitem [{\citenamefont {Gregory}\ \emph {et~al.}(2014)\citenamefont
  {Gregory}, \citenamefont {Moss},\ and\ \citenamefont
  {Withers}}]{Gregory:2013hja}%
  \BibitemOpen
  \bibfield  {author} {\bibinfo {author} {\bibfnamefont {Ruth}\ \bibnamefont
  {Gregory}}, \bibinfo {author} {\bibfnamefont {Ian~G.}\ \bibnamefont {Moss}},
  \ and\ \bibinfo {author} {\bibfnamefont {Benjamin}\ \bibnamefont {Withers}},\
  }\bibfield  {title} {\enquote {\bibinfo {title} {{Black holes as bubble
  nucleation sites}},}\ }\href {\doibase 10.1007/JHEP03(2014)081} {\bibfield
  {journal} {\bibinfo  {journal} {JHEP}\ }\textbf {\bibinfo {volume} {03}},\
  \bibinfo {pages} {081} (\bibinfo {year} {2014})},\ \Eprint
  {http://arxiv.org/abs/1401.0017} {arXiv:1401.0017 [hep-th]} \BibitemShut
  {NoStop}%
\bibitem [{\citenamefont {Burda}\ \emph
  {et~al.}(2015{\natexlab{a}})\citenamefont {Burda}, \citenamefont {Gregory},\
  and\ \citenamefont {Moss}}]{Burda:2015isa}%
  \BibitemOpen
  \bibfield  {author} {\bibinfo {author} {\bibfnamefont {Philipp}\ \bibnamefont
  {Burda}}, \bibinfo {author} {\bibfnamefont {Ruth}\ \bibnamefont {Gregory}}, \
  and\ \bibinfo {author} {\bibfnamefont {Ian}\ \bibnamefont {Moss}},\
  }\bibfield  {title} {\enquote {\bibinfo {title} {{Gravity and the stability
  of the Higgs vacuum}},}\ }\href {\doibase 10.1103/PhysRevLett.115.071303}
  {\bibfield  {journal} {\bibinfo  {journal} {Phys. Rev. Lett.}\ }\textbf
  {\bibinfo {volume} {115}},\ \bibinfo {pages} {071303} (\bibinfo {year}
  {2015}{\natexlab{a}})},\ \Eprint {http://arxiv.org/abs/1501.04937}
  {arXiv:1501.04937 [hep-th]} \BibitemShut {NoStop}%
\bibitem [{\citenamefont {Burda}\ \emph
  {et~al.}(2015{\natexlab{b}})\citenamefont {Burda}, \citenamefont {Gregory},\
  and\ \citenamefont {Moss}}]{Burda:2015yfa}%
  \BibitemOpen
  \bibfield  {author} {\bibinfo {author} {\bibfnamefont {Philipp}\ \bibnamefont
  {Burda}}, \bibinfo {author} {\bibfnamefont {Ruth}\ \bibnamefont {Gregory}}, \
  and\ \bibinfo {author} {\bibfnamefont {Ian}\ \bibnamefont {Moss}},\
  }\bibfield  {title} {\enquote {\bibinfo {title} {{Vacuum metastability with
  black holes}},}\ }\href {\doibase 10.1007/JHEP08(2015)114} {\bibfield
  {journal} {\bibinfo  {journal} {JHEP}\ }\textbf {\bibinfo {volume} {08}},\
  \bibinfo {pages} {114} (\bibinfo {year} {2015}{\natexlab{b}})},\ \Eprint
  {http://arxiv.org/abs/1503.07331} {arXiv:1503.07331 [hep-th]} \BibitemShut
  {NoStop}%
\bibitem [{\citenamefont {Burda}\ \emph {et~al.}(2016)\citenamefont {Burda},
  \citenamefont {Gregory},\ and\ \citenamefont {Moss}}]{Burda:2016mou}%
  \BibitemOpen
  \bibfield  {author} {\bibinfo {author} {\bibfnamefont {Philipp}\ \bibnamefont
  {Burda}}, \bibinfo {author} {\bibfnamefont {Ruth}\ \bibnamefont {Gregory}}, \
  and\ \bibinfo {author} {\bibfnamefont {Ian}\ \bibnamefont {Moss}},\
  }\bibfield  {title} {\enquote {\bibinfo {title} {{The fate of the Higgs
  vacuum}},}\ }\href {\doibase 10.1007/JHEP06(2016)025} {\bibfield  {journal}
  {\bibinfo  {journal} {JHEP}\ }\textbf {\bibinfo {volume} {06}},\ \bibinfo
  {pages} {025} (\bibinfo {year} {2016})},\ \Eprint
  {http://arxiv.org/abs/1601.02152} {arXiv:1601.02152 [hep-th]} \BibitemShut
  {NoStop}%
\bibitem [{\citenamefont {Oshita}\ and\ \citenamefont
  {Yokoyama}(2016)}]{Oshita:2016oqn}%
  \BibitemOpen
  \bibfield  {author} {\bibinfo {author} {\bibfnamefont {Naritaka}\
  \bibnamefont {Oshita}}\ and\ \bibinfo {author} {\bibfnamefont {Jun'ichi}\
  \bibnamefont {Yokoyama}},\ }\bibfield  {title} {\enquote {\bibinfo {title}
  {{Entropic interpretation of the Hawking\textendash{}Moss bounce}},}\ }\href
  {\doibase 10.1093/ptep/ptw053} {\bibfield  {journal} {\bibinfo  {journal}
  {PTEP}\ }\textbf {\bibinfo {volume} {2016}},\ \bibinfo {pages} {051E02}
  (\bibinfo {year} {2016})},\ \Eprint {http://arxiv.org/abs/1603.06671}
  {arXiv:1603.06671 [hep-th]} \BibitemShut {NoStop}%
\bibitem [{\citenamefont {Mukaida}\ and\ \citenamefont
  {Yamada}(2017)}]{Mukaida:2017bgd}%
  \BibitemOpen
  \bibfield  {author} {\bibinfo {author} {\bibfnamefont {Kyohei}\ \bibnamefont
  {Mukaida}}\ and\ \bibinfo {author} {\bibfnamefont {Masaki}\ \bibnamefont
  {Yamada}},\ }\bibfield  {title} {\enquote {\bibinfo {title} {{False Vacuum
  Decay Catalyzed by Black Holes}},}\ }\href {\doibase
  10.1103/PhysRevD.96.103514} {\bibfield  {journal} {\bibinfo  {journal} {Phys.
  Rev. D}\ }\textbf {\bibinfo {volume} {96}},\ \bibinfo {pages} {103514}
  (\bibinfo {year} {2017})},\ \Eprint {http://arxiv.org/abs/1706.04523}
  {arXiv:1706.04523 [hep-th]} \BibitemShut {NoStop}%
\bibitem [{\citenamefont {Gorbunov}\ \emph {et~al.}(2017)\citenamefont
  {Gorbunov}, \citenamefont {Levkov},\ and\ \citenamefont
  {Panin}}]{Gorbunov:2017fhq}%
  \BibitemOpen
  \bibfield  {author} {\bibinfo {author} {\bibfnamefont {Dmitry}\ \bibnamefont
  {Gorbunov}}, \bibinfo {author} {\bibfnamefont {Dmitry}\ \bibnamefont
  {Levkov}}, \ and\ \bibinfo {author} {\bibfnamefont {Alexander}\ \bibnamefont
  {Panin}},\ }\bibfield  {title} {\enquote {\bibinfo {title} {{Fatal youth of
  the Universe: black hole threat for the electroweak vacuum during
  preheating}},}\ }\href {\doibase 10.1088/1475-7516/2017/10/016} {\bibfield
  {journal} {\bibinfo  {journal} {JCAP}\ }\textbf {\bibinfo {volume} {10}},\
  \bibinfo {pages} {016} (\bibinfo {year} {2017})},\ \Eprint
  {http://arxiv.org/abs/1704.05399} {arXiv:1704.05399 [astro-ph.CO]}
  \BibitemShut {NoStop}%
\bibitem [{\citenamefont {Canko}\ \emph {et~al.}(2018)\citenamefont {Canko},
  \citenamefont {Gialamas}, \citenamefont {Jelic-Cizmek}, \citenamefont
  {Riotto},\ and\ \citenamefont {Tetradis}}]{Canko:2017ebb}%
  \BibitemOpen
  \bibfield  {author} {\bibinfo {author} {\bibfnamefont {D.}~\bibnamefont
  {Canko}}, \bibinfo {author} {\bibfnamefont {I.}~\bibnamefont {Gialamas}},
  \bibinfo {author} {\bibfnamefont {G.}~\bibnamefont {Jelic-Cizmek}}, \bibinfo
  {author} {\bibfnamefont {A.}~\bibnamefont {Riotto}}, \ and\ \bibinfo {author}
  {\bibfnamefont {N.}~\bibnamefont {Tetradis}},\ }\bibfield  {title} {\enquote
  {\bibinfo {title} {{On the Catalysis of the Electroweak Vacuum Decay by Black
  Holes at High Temperature}},}\ }\href {\doibase
  10.1140/epjc/s10052-018-5808-y} {\bibfield  {journal} {\bibinfo  {journal}
  {Eur. Phys. J. C}\ }\textbf {\bibinfo {volume} {78}},\ \bibinfo {pages} {328}
  (\bibinfo {year} {2018})},\ \Eprint {http://arxiv.org/abs/1706.01364}
  {arXiv:1706.01364 [hep-th]} \BibitemShut {NoStop}%
\bibitem [{\citenamefont {Kohri}\ and\ \citenamefont
  {Matsui}(2018)}]{Kohri:2017ybt}%
  \BibitemOpen
  \bibfield  {author} {\bibinfo {author} {\bibfnamefont {Kazunori}\
  \bibnamefont {Kohri}}\ and\ \bibinfo {author} {\bibfnamefont {Hiroki}\
  \bibnamefont {Matsui}},\ }\bibfield  {title} {\enquote {\bibinfo {title}
  {{Electroweak Vacuum Collapse induced by Vacuum Fluctuations of the Higgs
  Field around Evaporating Black Holes}},}\ }\href {\doibase
  10.1103/PhysRevD.98.123509} {\bibfield  {journal} {\bibinfo  {journal} {Phys.
  Rev. D}\ }\textbf {\bibinfo {volume} {98}},\ \bibinfo {pages} {123509}
  (\bibinfo {year} {2018})},\ \Eprint {http://arxiv.org/abs/1708.02138}
  {arXiv:1708.02138 [hep-ph]} \BibitemShut {NoStop}%
\bibitem [{\citenamefont {Gregory}\ \emph {et~al.}(2018)\citenamefont
  {Gregory}, \citenamefont {Marshall}, \citenamefont {Michel},\ and\
  \citenamefont {Moss}}]{Gregory:2018bdt}%
  \BibitemOpen
  \bibfield  {author} {\bibinfo {author} {\bibfnamefont {Ruth}\ \bibnamefont
  {Gregory}}, \bibinfo {author} {\bibfnamefont {Katie~M.}\ \bibnamefont
  {Marshall}}, \bibinfo {author} {\bibfnamefont {Florent}\ \bibnamefont
  {Michel}}, \ and\ \bibinfo {author} {\bibfnamefont {Ian~G.}\ \bibnamefont
  {Moss}},\ }\bibfield  {title} {\enquote {\bibinfo {title} {{Negative modes of
  Coleman\textendash{}De Luccia and black hole bubbles}},}\ }\href {\doibase
  10.1103/PhysRevD.98.085017} {\bibfield  {journal} {\bibinfo  {journal} {Phys.
  Rev. D}\ }\textbf {\bibinfo {volume} {98}},\ \bibinfo {pages} {085017}
  (\bibinfo {year} {2018})},\ \Eprint {http://arxiv.org/abs/1808.02305}
  {arXiv:1808.02305 [hep-th]} \BibitemShut {NoStop}%
\bibitem [{\citenamefont {Oshita}\ \emph {et~al.}(2020)\citenamefont {Oshita},
  \citenamefont {Ueda},\ and\ \citenamefont {Yamaguchi}}]{Oshita:2019jan}%
  \BibitemOpen
  \bibfield  {author} {\bibinfo {author} {\bibfnamefont {Naritaka}\
  \bibnamefont {Oshita}}, \bibinfo {author} {\bibfnamefont {Kazushige}\
  \bibnamefont {Ueda}}, \ and\ \bibinfo {author} {\bibfnamefont {Masahide}\
  \bibnamefont {Yamaguchi}},\ }\bibfield  {title} {\enquote {\bibinfo {title}
  {{Vacuum decays around spinning black holes}},}\ }\href {\doibase
  10.1007/JHEP01(2020)015} {\bibfield  {journal} {\bibinfo  {journal} {JHEP}\
  }\textbf {\bibinfo {volume} {01}},\ \bibinfo {pages} {015} (\bibinfo {year}
  {2020})},\ \bibinfo {note} {[Erratum: JHEP 10, 122 (2020)]},\ \Eprint
  {http://arxiv.org/abs/1909.01378} {arXiv:1909.01378 [hep-th]} \BibitemShut
  {NoStop}%
\bibitem [{\citenamefont {Gregory}\ \emph {et~al.}(2020)\citenamefont
  {Gregory}, \citenamefont {Moss},\ and\ \citenamefont
  {Oshita}}]{Gregory:2020cvy}%
  \BibitemOpen
  \bibfield  {author} {\bibinfo {author} {\bibfnamefont {Ruth}\ \bibnamefont
  {Gregory}}, \bibinfo {author} {\bibfnamefont {Ian~G.}\ \bibnamefont {Moss}},
  \ and\ \bibinfo {author} {\bibfnamefont {Naritaka}\ \bibnamefont {Oshita}},\
  }\bibfield  {title} {\enquote {\bibinfo {title} {{Black Holes, Oscillating
  Instantons, and the Hawking-Moss transition}},}\ }\href {\doibase
  10.1007/JHEP07(2020)024} {\bibfield  {journal} {\bibinfo  {journal} {JHEP}\
  }\textbf {\bibinfo {volume} {07}},\ \bibinfo {pages} {024} (\bibinfo {year}
  {2020})},\ \Eprint {http://arxiv.org/abs/2003.04927} {arXiv:2003.04927
  [hep-th]} \BibitemShut {NoStop}%
\bibitem [{\citenamefont {He}\ and\ \citenamefont {Wu}(2022)}]{He:2022sjf}%
  \BibitemOpen
  \bibfield  {author} {\bibinfo {author} {\bibfnamefont {Jian-hua}\
  \bibnamefont {He}}\ and\ \bibinfo {author} {\bibfnamefont {Zhenyu}\
  \bibnamefont {Wu}},\ }\bibfield  {title} {\enquote {\bibinfo {title}
  {{Simulating gravitational waves passing through the spacetime of a black
  hole}},}\ }\href {\doibase 10.1103/PhysRevD.106.124037} {\bibfield  {journal}
  {\bibinfo  {journal} {Phys. Rev. D}\ }\textbf {\bibinfo {volume} {106}},\
  \bibinfo {pages} {124037} (\bibinfo {year} {2022})},\ \Eprint
  {http://arxiv.org/abs/2208.01621} {arXiv:2208.01621 [gr-qc]} \BibitemShut
  {NoStop}%
\bibitem [{\citenamefont {Yin}\ and\ \citenamefont {He}(2024)}]{Yin:2023kzr}%
  \BibitemOpen
  \bibfield  {author} {\bibinfo {author} {\bibfnamefont {Chengjiang}\
  \bibnamefont {Yin}}\ and\ \bibinfo {author} {\bibfnamefont {Jian-hua}\
  \bibnamefont {He}},\ }\bibfield  {title} {\enquote {\bibinfo {title}
  {{Detectability of Single Spinless Stellar-Mass Black Holes through
  Gravitational Lensing of Gravitational Waves with Advanced LIGO}},}\ }\href
  {\doibase 10.1103/PhysRevLett.132.011401} {\bibfield  {journal} {\bibinfo
  {journal} {Phys. Rev. Lett.}\ }\textbf {\bibinfo {volume} {132}},\ \bibinfo
  {pages} {011401} (\bibinfo {year} {2024})},\ \Eprint
  {http://arxiv.org/abs/2312.12451} {arXiv:2312.12451 [astro-ph.HE]}
  \BibitemShut {NoStop}%
\bibitem [{\citenamefont {Clough}\ \emph {et~al.}(2015)\citenamefont {Clough},
  \citenamefont {Figueras}, \citenamefont {Finkel}, \citenamefont {Kunesch},
  \citenamefont {Lim},\ and\ \citenamefont {Tunyasuvunakool}}]{Clough:2015sqa}%
  \BibitemOpen
  \bibfield  {author} {\bibinfo {author} {\bibfnamefont {Katy}\ \bibnamefont
  {Clough}}, \bibinfo {author} {\bibfnamefont {Pau}\ \bibnamefont {Figueras}},
  \bibinfo {author} {\bibfnamefont {Hal}\ \bibnamefont {Finkel}}, \bibinfo
  {author} {\bibfnamefont {Markus}\ \bibnamefont {Kunesch}}, \bibinfo {author}
  {\bibfnamefont {Eugene~A.}\ \bibnamefont {Lim}}, \ and\ \bibinfo {author}
  {\bibfnamefont {Saran}\ \bibnamefont {Tunyasuvunakool}},\ }\bibfield  {title}
  {\enquote {\bibinfo {title} {{GRChombo : Numerical Relativity with Adaptive
  Mesh Refinement}},}\ }\href {\doibase 10.1088/0264-9381/32/24/245011}
  {\bibfield  {journal} {\bibinfo  {journal} {Class. Quant. Grav.}\ }\textbf
  {\bibinfo {volume} {32}},\ \bibinfo {pages} {245011} (\bibinfo {year}
  {2015})},\ \Eprint {http://arxiv.org/abs/1503.03436} {arXiv:1503.03436
  [gr-qc]} \BibitemShut {NoStop}%
\bibitem [{\citenamefont {Andrade}\ \emph {et~al.}(2021)\citenamefont {Andrade}
  \emph {et~al.}}]{Andrade:2021rbd}%
  \BibitemOpen
  \bibfield  {author} {\bibinfo {author} {\bibfnamefont {Tomas}\ \bibnamefont
  {Andrade}} \emph {et~al.},\ }\bibfield  {title} {\enquote {\bibinfo {title}
  {{GRChombo: An adaptable numerical relativity code for fundamental
  physics}},}\ }\href {\doibase 10.21105/joss.03703} {\bibfield  {journal}
  {\bibinfo  {journal} {J. Open Source Softw.}\ }\textbf {\bibinfo {volume}
  {6}},\ \bibinfo {pages} {3703} (\bibinfo {year} {2021})},\ \Eprint
  {http://arxiv.org/abs/2201.03458} {arXiv:2201.03458 [gr-qc]} \BibitemShut
  {NoStop}%
\bibitem [{\citenamefont {Lewicki}\ and\ \citenamefont
  {Vaskonen}(2020)}]{Lewicki:2019gmv}%
  \BibitemOpen
  \bibfield  {author} {\bibinfo {author} {\bibfnamefont {Marek}\ \bibnamefont
  {Lewicki}}\ and\ \bibinfo {author} {\bibfnamefont {Ville}\ \bibnamefont
  {Vaskonen}},\ }\bibfield  {title} {\enquote {\bibinfo {title} {{On bubble
  collisions in strongly supercooled phase transitions}},}\ }\href {\doibase
  10.1016/j.dark.2020.100672} {\bibfield  {journal} {\bibinfo  {journal} {Phys.
  Dark Univ.}\ }\textbf {\bibinfo {volume} {30}},\ \bibinfo {pages} {100672}
  (\bibinfo {year} {2020})},\ \Eprint {http://arxiv.org/abs/1912.00997}
  {arXiv:1912.00997 [astro-ph.CO]} \BibitemShut {NoStop}%
\bibitem [{\citenamefont {Braden}\ \emph {et~al.}(2019)\citenamefont {Braden},
  \citenamefont {Johnson}, \citenamefont {Peiris}, \citenamefont {Pontzen},\
  and\ \citenamefont {Weinfurtner}}]{Braden:2018tky}%
  \BibitemOpen
  \bibfield  {author} {\bibinfo {author} {\bibfnamefont {Jonathan}\
  \bibnamefont {Braden}}, \bibinfo {author} {\bibfnamefont {Matthew~C.}\
  \bibnamefont {Johnson}}, \bibinfo {author} {\bibfnamefont {Hiranya~V.}\
  \bibnamefont {Peiris}}, \bibinfo {author} {\bibfnamefont {Andrew}\
  \bibnamefont {Pontzen}}, \ and\ \bibinfo {author} {\bibfnamefont {Silke}\
  \bibnamefont {Weinfurtner}},\ }\bibfield  {title} {\enquote {\bibinfo {title}
  {{New Semiclassical Picture of Vacuum Decay}},}\ }\href {\doibase
  10.1103/PhysRevLett.123.031601} {\bibfield  {journal} {\bibinfo  {journal}
  {Phys. Rev. Lett.}\ }\textbf {\bibinfo {volume} {123}},\ \bibinfo {pages}
  {031601} (\bibinfo {year} {2019})},\ \bibinfo {note} {[Erratum:
  Phys.Rev.Lett. 129, 059901 (2022)]},\ \Eprint
  {http://arxiv.org/abs/1806.06069} {arXiv:1806.06069 [hep-th]} \BibitemShut
  {NoStop}%
\bibitem [{\citenamefont {Blanco-Pillado}\ \emph {et~al.}(2019)\citenamefont
  {Blanco-Pillado}, \citenamefont {Deng},\ and\ \citenamefont
  {Vilenkin}}]{Blanco-Pillado:2019xny}%
  \BibitemOpen
  \bibfield  {author} {\bibinfo {author} {\bibfnamefont {Jose~J.}\ \bibnamefont
  {Blanco-Pillado}}, \bibinfo {author} {\bibfnamefont {Heling}\ \bibnamefont
  {Deng}}, \ and\ \bibinfo {author} {\bibfnamefont {Alexander}\ \bibnamefont
  {Vilenkin}},\ }\bibfield  {title} {\enquote {\bibinfo {title} {{Flyover
  vacuum decay}},}\ }\href {\doibase 10.1088/1475-7516/2019/12/001} {\bibfield
  {journal} {\bibinfo  {journal} {JCAP}\ }\textbf {\bibinfo {volume} {12}},\
  \bibinfo {pages} {001} (\bibinfo {year} {2019})},\ \Eprint
  {http://arxiv.org/abs/1906.09657} {arXiv:1906.09657 [hep-th]} \BibitemShut
  {NoStop}%
\bibitem [{\citenamefont {Wang}(2019)}]{Wang:2019hjx}%
  \BibitemOpen
  \bibfield  {author} {\bibinfo {author} {\bibfnamefont {Shao-Jiang}\
  \bibnamefont {Wang}},\ }\bibfield  {title} {\enquote {\bibinfo {title}
  {{Occurrence of semiclassical vacuum decay}},}\ }\href {\doibase
  10.1103/PhysRevD.100.096019} {\bibfield  {journal} {\bibinfo  {journal}
  {Phys. Rev. D}\ }\textbf {\bibinfo {volume} {100}},\ \bibinfo {pages}
  {096019} (\bibinfo {year} {2019})},\ \Eprint
  {http://arxiv.org/abs/1909.11196} {arXiv:1909.11196 [gr-qc]} \BibitemShut
  {NoStop}%
\bibitem [{\citenamefont {Cutting}\ \emph {et~al.}(2018)\citenamefont
  {Cutting}, \citenamefont {Hindmarsh},\ and\ \citenamefont
  {Weir}}]{Cutting:2018tjt}%
  \BibitemOpen
  \bibfield  {author} {\bibinfo {author} {\bibfnamefont {Daniel}\ \bibnamefont
  {Cutting}}, \bibinfo {author} {\bibfnamefont {Mark}\ \bibnamefont
  {Hindmarsh}}, \ and\ \bibinfo {author} {\bibfnamefont {David~J.}\
  \bibnamefont {Weir}},\ }\bibfield  {title} {\enquote {\bibinfo {title}
  {{Gravitational waves from vacuum first-order phase transitions: from the
  envelope to the lattice}},}\ }\href {\doibase 10.1103/PhysRevD.97.123513}
  {\bibfield  {journal} {\bibinfo  {journal} {Phys. Rev. D}\ }\textbf {\bibinfo
  {volume} {97}},\ \bibinfo {pages} {123513} (\bibinfo {year} {2018})},\
  \Eprint {http://arxiv.org/abs/1802.05712} {arXiv:1802.05712 [astro-ph.CO]}
  \BibitemShut {NoStop}%
\bibitem [{\citenamefont {Cutting}\ \emph {et~al.}(2021)\citenamefont
  {Cutting}, \citenamefont {Escartin}, \citenamefont {Hindmarsh},\ and\
  \citenamefont {Weir}}]{Cutting:2020nla}%
  \BibitemOpen
  \bibfield  {author} {\bibinfo {author} {\bibfnamefont {Daniel}\ \bibnamefont
  {Cutting}}, \bibinfo {author} {\bibfnamefont {Elba~Granados}\ \bibnamefont
  {Escartin}}, \bibinfo {author} {\bibfnamefont {Mark}\ \bibnamefont
  {Hindmarsh}}, \ and\ \bibinfo {author} {\bibfnamefont {David~J.}\
  \bibnamefont {Weir}},\ }\bibfield  {title} {\enquote {\bibinfo {title}
  {{Gravitational waves from vacuum first order phase transitions II: from thin
  to thick walls}},}\ }\href {\doibase 10.1103/PhysRevD.103.023531} {\bibfield
  {journal} {\bibinfo  {journal} {Phys. Rev. D}\ }\textbf {\bibinfo {volume}
  {103}},\ \bibinfo {pages} {023531} (\bibinfo {year} {2021})},\ \Eprint
  {http://arxiv.org/abs/2005.13537} {arXiv:2005.13537 [astro-ph.CO]}
  \BibitemShut {NoStop}%
\bibitem [{foo()}]{footnote}%
  \BibitemOpen
  \href@noop {} {}\bibinfo {note} {See the supplemental material for the
  numerical methods, initial data, convergence tests, and analytical
  derivations on the modified possibility of forming a PBH binary.}\BibitemShut
  {Stop}%
\bibitem [{\citenamefont {Carr}(1975)}]{Carr:1975qj}%
  \BibitemOpen
  \bibfield  {author} {\bibinfo {author} {\bibfnamefont {Bernard~J.}\
  \bibnamefont {Carr}},\ }\bibfield  {title} {\enquote {\bibinfo {title} {{The
  Primordial black hole mass spectrum}},}\ }\href {\doibase 10.1086/153853}
  {\bibfield  {journal} {\bibinfo  {journal} {Astrophys. J.}\ }\textbf
  {\bibinfo {volume} {201}},\ \bibinfo {pages} {1--19} (\bibinfo {year}
  {1975})}\BibitemShut {NoStop}%
\bibitem [{\citenamefont {Sasaki}\ \emph {et~al.}(2016)\citenamefont {Sasaki},
  \citenamefont {Suyama}, \citenamefont {Tanaka},\ and\ \citenamefont
  {Yokoyama}}]{Sasaki:2016jop}%
  \BibitemOpen
  \bibfield  {author} {\bibinfo {author} {\bibfnamefont {Misao}\ \bibnamefont
  {Sasaki}}, \bibinfo {author} {\bibfnamefont {Teruaki}\ \bibnamefont
  {Suyama}}, \bibinfo {author} {\bibfnamefont {Takahiro}\ \bibnamefont
  {Tanaka}}, \ and\ \bibinfo {author} {\bibfnamefont {Shuichiro}\ \bibnamefont
  {Yokoyama}},\ }\bibfield  {title} {\enquote {\bibinfo {title} {{Primordial
  Black Hole Scenario for the Gravitational-Wave Event GW150914}},}\ }\href
  {\doibase 10.1103/PhysRevLett.117.061101} {\bibfield  {journal} {\bibinfo
  {journal} {Phys. Rev. Lett.}\ }\textbf {\bibinfo {volume} {117}},\ \bibinfo
  {pages} {061101} (\bibinfo {year} {2016})},\ \bibinfo {note} {[Erratum:
  Phys.Rev.Lett. 121, 059901 (2018)]},\ \Eprint
  {http://arxiv.org/abs/1603.08338} {arXiv:1603.08338 [astro-ph.CO]}
  \BibitemShut {NoStop}%
\bibitem [{\citenamefont {De~Luca}\ \emph
  {et~al.}(2020{\natexlab{a}})\citenamefont {De~Luca}, \citenamefont
  {Franciolini}, \citenamefont {Pani},\ and\ \citenamefont
  {Riotto}}]{DeLuca:2020bjf}%
  \BibitemOpen
  \bibfield  {author} {\bibinfo {author} {\bibfnamefont {V.}~\bibnamefont
  {De~Luca}}, \bibinfo {author} {\bibfnamefont {G.}~\bibnamefont
  {Franciolini}}, \bibinfo {author} {\bibfnamefont {P.}~\bibnamefont {Pani}}, \
  and\ \bibinfo {author} {\bibfnamefont {A.}~\bibnamefont {Riotto}},\
  }\bibfield  {title} {\enquote {\bibinfo {title} {{The evolution of primordial
  black holes and their final observable spins}},}\ }\href {\doibase
  10.1088/1475-7516/2020/04/052} {\bibfield  {journal} {\bibinfo  {journal}
  {JCAP}\ }\textbf {\bibinfo {volume} {04}},\ \bibinfo {pages} {052} (\bibinfo
  {year} {2020}{\natexlab{a}})},\ \Eprint {http://arxiv.org/abs/2003.02778}
  {arXiv:2003.02778 [astro-ph.CO]} \BibitemShut {NoStop}%
\bibitem [{\citenamefont {Carr}\ \emph {et~al.}(2010)\citenamefont {Carr},
  \citenamefont {Kohri}, \citenamefont {Sendouda},\ and\ \citenamefont
  {Yokoyama}}]{Carr:2009jm}%
  \BibitemOpen
  \bibfield  {author} {\bibinfo {author} {\bibfnamefont {B.~J.}\ \bibnamefont
  {Carr}}, \bibinfo {author} {\bibfnamefont {Kazunori}\ \bibnamefont {Kohri}},
  \bibinfo {author} {\bibfnamefont {Yuuiti}\ \bibnamefont {Sendouda}}, \ and\
  \bibinfo {author} {\bibfnamefont {Jun'ichi}\ \bibnamefont {Yokoyama}},\
  }\bibfield  {title} {\enquote {\bibinfo {title} {{New cosmological
  constraints on primordial black holes}},}\ }\href {\doibase
  10.1103/PhysRevD.81.104019} {\bibfield  {journal} {\bibinfo  {journal} {Phys.
  Rev. D}\ }\textbf {\bibinfo {volume} {81}},\ \bibinfo {pages} {104019}
  (\bibinfo {year} {2010})},\ \Eprint {http://arxiv.org/abs/0912.5297}
  {arXiv:0912.5297 [astro-ph.CO]} \BibitemShut {NoStop}%
\bibitem [{\citenamefont {Ali-Ha\"\i{}moud}\ \emph {et~al.}(2017)\citenamefont
  {Ali-Ha\"\i{}moud}, \citenamefont {Kovetz},\ and\ \citenamefont
  {Kamionkowski}}]{Ali-Haimoud:2017rtz}%
  \BibitemOpen
  \bibfield  {author} {\bibinfo {author} {\bibfnamefont {Yacine}\ \bibnamefont
  {Ali-Ha\"\i{}moud}}, \bibinfo {author} {\bibfnamefont {Ely~D.}\ \bibnamefont
  {Kovetz}}, \ and\ \bibinfo {author} {\bibfnamefont {Marc}\ \bibnamefont
  {Kamionkowski}},\ }\bibfield  {title} {\enquote {\bibinfo {title} {{Merger
  rate of primordial black-hole binaries}},}\ }\href {\doibase
  10.1103/PhysRevD.96.123523} {\bibfield  {journal} {\bibinfo  {journal} {Phys.
  Rev. D}\ }\textbf {\bibinfo {volume} {96}},\ \bibinfo {pages} {123523}
  (\bibinfo {year} {2017})},\ \Eprint {http://arxiv.org/abs/1709.06576}
  {arXiv:1709.06576 [astro-ph.CO]} \BibitemShut {NoStop}%
\bibitem [{\citenamefont {Vaskonen}\ and\ \citenamefont
  {Veerm\"ae}(2020)}]{Vaskonen:2019jpv}%
  \BibitemOpen
  \bibfield  {author} {\bibinfo {author} {\bibfnamefont {Ville}\ \bibnamefont
  {Vaskonen}}\ and\ \bibinfo {author} {\bibfnamefont {Hardi}\ \bibnamefont
  {Veerm\"ae}},\ }\bibfield  {title} {\enquote {\bibinfo {title} {{Lower bound
  on the primordial black hole merger rate}},}\ }\href {\doibase
  10.1103/PhysRevD.101.043015} {\bibfield  {journal} {\bibinfo  {journal}
  {Phys. Rev. D}\ }\textbf {\bibinfo {volume} {101}},\ \bibinfo {pages}
  {043015} (\bibinfo {year} {2020})},\ \Eprint
  {http://arxiv.org/abs/1908.09752} {arXiv:1908.09752 [astro-ph.CO]}
  \BibitemShut {NoStop}%
\bibitem [{\citenamefont {Delos}\ \emph {et~al.}(2024)\citenamefont {Delos},
  \citenamefont {Rantala}, \citenamefont {Young},\ and\ \citenamefont
  {Schmidt}}]{Delos:2024poq}%
  \BibitemOpen
  \bibfield  {author} {\bibinfo {author} {\bibfnamefont {M.~Sten}\ \bibnamefont
  {Delos}}, \bibinfo {author} {\bibfnamefont {Antti}\ \bibnamefont {Rantala}},
  \bibinfo {author} {\bibfnamefont {Sam}\ \bibnamefont {Young}}, \ and\
  \bibinfo {author} {\bibfnamefont {Fabian}\ \bibnamefont {Schmidt}},\
  }\bibfield  {title} {\enquote {\bibinfo {title} {{Structure formation with
  primordial black holes: collisional dynamics, binaries, and gravitational
  waves}},}\ }\href {\doibase 10.1088/1475-7516/2024/12/005} {\bibfield
  {journal} {\bibinfo  {journal} {JCAP}\ }\textbf {\bibinfo {volume} {12}},\
  \bibinfo {pages} {005} (\bibinfo {year} {2024})},\ \Eprint
  {http://arxiv.org/abs/2410.01876} {arXiv:2410.01876 [astro-ph.CO]}
  \BibitemShut {NoStop}%
\bibitem [{\citenamefont {Bird}\ \emph {et~al.}(2016)\citenamefont {Bird},
  \citenamefont {Cholis}, \citenamefont {Mu\~noz}, \citenamefont
  {Ali-Ha\"\i{}moud}, \citenamefont {Kamionkowski}, \citenamefont {Kovetz},
  \citenamefont {Raccanelli},\ and\ \citenamefont {Riess}}]{Bird:2016dcv}%
  \BibitemOpen
  \bibfield  {author} {\bibinfo {author} {\bibfnamefont {Simeon}\ \bibnamefont
  {Bird}}, \bibinfo {author} {\bibfnamefont {Ilias}\ \bibnamefont {Cholis}},
  \bibinfo {author} {\bibfnamefont {Julian~B.}\ \bibnamefont {Mu\~noz}},
  \bibinfo {author} {\bibfnamefont {Yacine}\ \bibnamefont {Ali-Ha\"\i{}moud}},
  \bibinfo {author} {\bibfnamefont {Marc}\ \bibnamefont {Kamionkowski}},
  \bibinfo {author} {\bibfnamefont {Ely~D.}\ \bibnamefont {Kovetz}}, \bibinfo
  {author} {\bibfnamefont {Alvise}\ \bibnamefont {Raccanelli}}, \ and\ \bibinfo
  {author} {\bibfnamefont {Adam~G.}\ \bibnamefont {Riess}},\ }\bibfield
  {title} {\enquote {\bibinfo {title} {{Did LIGO detect dark matter?}}}\ }\href
  {\doibase 10.1103/PhysRevLett.116.201301} {\bibfield  {journal} {\bibinfo
  {journal} {Phys. Rev. Lett.}\ }\textbf {\bibinfo {volume} {116}},\ \bibinfo
  {pages} {201301} (\bibinfo {year} {2016})},\ \Eprint
  {http://arxiv.org/abs/1603.00464} {arXiv:1603.00464 [astro-ph.CO]}
  \BibitemShut {NoStop}%
\bibitem [{\citenamefont {Abbott}\ \emph {et~al.}(2016)\citenamefont {Abbott}
  \emph {et~al.}}]{LIGOScientific:2016kwr}%
  \BibitemOpen
  \bibfield  {author} {\bibinfo {author} {\bibfnamefont {B.~P.}\ \bibnamefont
  {Abbott}} \emph {et~al.} (\bibinfo {collaboration} {LIGO Scientific,
  Virgo}),\ }\bibfield  {title} {\enquote {\bibinfo {title} {{The Rate of
  Binary Black Hole Mergers Inferred from Advanced LIGO Observations
  Surrounding GW150914}},}\ }\href {\doibase 10.3847/2041-8205/833/1/L1}
  {\bibfield  {journal} {\bibinfo  {journal} {Astrophys. J. Lett.}\ }\textbf
  {\bibinfo {volume} {833}},\ \bibinfo {pages} {L1} (\bibinfo {year} {2016})},\
  \Eprint {http://arxiv.org/abs/1602.03842} {arXiv:1602.03842 [astro-ph.HE]}
  \BibitemShut {NoStop}%
\bibitem [{\citenamefont {Ricotti}\ \emph {et~al.}(2008)\citenamefont
  {Ricotti}, \citenamefont {Ostriker},\ and\ \citenamefont
  {Mack}}]{Ricotti:2007au}%
  \BibitemOpen
  \bibfield  {author} {\bibinfo {author} {\bibfnamefont {Massimo}\ \bibnamefont
  {Ricotti}}, \bibinfo {author} {\bibfnamefont {Jeremiah~P.}\ \bibnamefont
  {Ostriker}}, \ and\ \bibinfo {author} {\bibfnamefont {Katherine~J.}\
  \bibnamefont {Mack}},\ }\bibfield  {title} {\enquote {\bibinfo {title}
  {{Effect of Primordial Black Holes on the Cosmic Microwave Background and
  Cosmological Parameter Estimates}},}\ }\href {\doibase 10.1086/587831}
  {\bibfield  {journal} {\bibinfo  {journal} {Astrophys. J.}\ }\textbf
  {\bibinfo {volume} {680}},\ \bibinfo {pages} {829} (\bibinfo {year}
  {2008})},\ \Eprint {http://arxiv.org/abs/0709.0524} {arXiv:0709.0524
  [astro-ph]} \BibitemShut {NoStop}%
\bibitem [{\citenamefont {Belotsky}\ \emph {et~al.}(2019)\citenamefont
  {Belotsky}, \citenamefont {Dokuchaev}, \citenamefont {Eroshenko},
  \citenamefont {Esipova}, \citenamefont {Khlopov}, \citenamefont {Khromykh},
  \citenamefont {Kirillov}, \citenamefont {Nikulin}, \citenamefont {Rubin},\
  and\ \citenamefont {Svadkovsky}}]{Belotsky:2018wph}%
  \BibitemOpen
  \bibfield  {author} {\bibinfo {author} {\bibfnamefont {Konstantin~M.}\
  \bibnamefont {Belotsky}}, \bibinfo {author} {\bibfnamefont {Vyacheslav~I.}\
  \bibnamefont {Dokuchaev}}, \bibinfo {author} {\bibfnamefont {Yury~N.}\
  \bibnamefont {Eroshenko}}, \bibinfo {author} {\bibfnamefont {Ekaterina~A.}\
  \bibnamefont {Esipova}}, \bibinfo {author} {\bibfnamefont {Maxim~Yu.}\
  \bibnamefont {Khlopov}}, \bibinfo {author} {\bibfnamefont {Leonid~A.}\
  \bibnamefont {Khromykh}}, \bibinfo {author} {\bibfnamefont {Alexander~A.}\
  \bibnamefont {Kirillov}}, \bibinfo {author} {\bibfnamefont {Valeriy~V.}\
  \bibnamefont {Nikulin}}, \bibinfo {author} {\bibfnamefont {Sergey~G.}\
  \bibnamefont {Rubin}}, \ and\ \bibinfo {author} {\bibfnamefont {Igor~V.}\
  \bibnamefont {Svadkovsky}},\ }\bibfield  {title} {\enquote {\bibinfo {title}
  {{Clusters of primordial black holes}},}\ }\href {\doibase
  10.1140/epjc/s10052-019-6741-4} {\bibfield  {journal} {\bibinfo  {journal}
  {Eur. Phys. J. C}\ }\textbf {\bibinfo {volume} {79}},\ \bibinfo {pages} {246}
  (\bibinfo {year} {2019})},\ \Eprint {http://arxiv.org/abs/1807.06590}
  {arXiv:1807.06590 [astro-ph.CO]} \BibitemShut {NoStop}%
\bibitem [{\citenamefont {Desjacques}\ and\ \citenamefont
  {Riotto}(2018)}]{Desjacques:2018wuu}%
  \BibitemOpen
  \bibfield  {author} {\bibinfo {author} {\bibfnamefont {Vincent}\ \bibnamefont
  {Desjacques}}\ and\ \bibinfo {author} {\bibfnamefont {Antonio}\ \bibnamefont
  {Riotto}},\ }\bibfield  {title} {\enquote {\bibinfo {title} {{Spatial
  clustering of primordial black holes}},}\ }\href {\doibase
  10.1103/PhysRevD.98.123533} {\bibfield  {journal} {\bibinfo  {journal} {Phys.
  Rev. D}\ }\textbf {\bibinfo {volume} {98}},\ \bibinfo {pages} {123533}
  (\bibinfo {year} {2018})},\ \Eprint {http://arxiv.org/abs/1806.10414}
  {arXiv:1806.10414 [astro-ph.CO]} \BibitemShut {NoStop}%
\bibitem [{\citenamefont {Di}\ \emph {et~al.}(2021)\citenamefont {Di},
  \citenamefont {Wang}, \citenamefont {Zhou}, \citenamefont {Bian},
  \citenamefont {Cai},\ and\ \citenamefont {Liu}}]{Di:2020kbw}%
  \BibitemOpen
  \bibfield  {author} {\bibinfo {author} {\bibfnamefont {Yuefeng}\ \bibnamefont
  {Di}}, \bibinfo {author} {\bibfnamefont {Jialong}\ \bibnamefont {Wang}},
  \bibinfo {author} {\bibfnamefont {Ruiyu}\ \bibnamefont {Zhou}}, \bibinfo
  {author} {\bibfnamefont {Ligong}\ \bibnamefont {Bian}}, \bibinfo {author}
  {\bibfnamefont {Rong-Gen}\ \bibnamefont {Cai}}, \ and\ \bibinfo {author}
  {\bibfnamefont {Jing}\ \bibnamefont {Liu}},\ }\bibfield  {title} {\enquote
  {\bibinfo {title} {{Magnetic Field and Gravitational Waves from the
  First-Order Phase Transition}},}\ }\href {\doibase
  10.1103/PhysRevLett.126.251102} {\bibfield  {journal} {\bibinfo  {journal}
  {Phys. Rev. Lett.}\ }\textbf {\bibinfo {volume} {126}},\ \bibinfo {pages}
  {251102} (\bibinfo {year} {2021})},\ \Eprint
  {http://arxiv.org/abs/2012.15625} {arXiv:2012.15625 [astro-ph.CO]}
  \BibitemShut {NoStop}%
\bibitem [{\citenamefont {Clough}(2017)}]{Clough:2017ixw}%
  \BibitemOpen
  \bibfield  {author} {\bibinfo {author} {\bibfnamefont {Katy}\ \bibnamefont
  {Clough}},\ }\emph {\bibinfo {title} {{Scalar Fields in Numerical General
  Relativity: Inhomogeneous inflation and asymmetric bubble collapse}}},\ \href
  {\doibase 10.1007/978-3-319-92672-8} {Ph.D. thesis},\ \bibinfo  {school}
  {King's Coll. London}, \bibinfo {address} {Cham} (\bibinfo {year} {2017}),\
  \Eprint {http://arxiv.org/abs/1704.06811} {arXiv:1704.06811 [gr-qc]}
  \BibitemShut {NoStop}%
\bibitem [{\citenamefont {Baumgarte}\ and\ \citenamefont
  {Shapiro}(1998)}]{Baumgarte:1998te}%
  \BibitemOpen
  \bibfield  {author} {\bibinfo {author} {\bibfnamefont {Thomas~W.}\
  \bibnamefont {Baumgarte}}\ and\ \bibinfo {author} {\bibfnamefont {Stuart~L.}\
  \bibnamefont {Shapiro}},\ }\bibfield  {title} {\enquote {\bibinfo {title}
  {{On the numerical integration of Einstein's field equations}},}\ }\href
  {\doibase 10.1103/PhysRevD.59.024007} {\bibfield  {journal} {\bibinfo
  {journal} {Phys. Rev. D}\ }\textbf {\bibinfo {volume} {59}},\ \bibinfo
  {pages} {024007} (\bibinfo {year} {1998})},\ \Eprint
  {http://arxiv.org/abs/gr-qc/9810065} {arXiv:gr-qc/9810065} \BibitemShut
  {NoStop}%
\bibitem [{\citenamefont {Shibata}\ and\ \citenamefont
  {Nakamura}(1995)}]{Shibata:1995we}%
  \BibitemOpen
  \bibfield  {author} {\bibinfo {author} {\bibfnamefont {Masaru}\ \bibnamefont
  {Shibata}}\ and\ \bibinfo {author} {\bibfnamefont {Takashi}\ \bibnamefont
  {Nakamura}},\ }\bibfield  {title} {\enquote {\bibinfo {title} {{Evolution of
  three-dimensional gravitational waves: Harmonic slicing case}},}\ }\href
  {\doibase 10.1103/PhysRevD.52.5428} {\bibfield  {journal} {\bibinfo
  {journal} {Phys. Rev. D}\ }\textbf {\bibinfo {volume} {52}},\ \bibinfo
  {pages} {5428--5444} (\bibinfo {year} {1995})}\BibitemShut {NoStop}%
\bibitem [{\citenamefont {Campanelli}\ \emph {et~al.}(2006)\citenamefont
  {Campanelli}, \citenamefont {Lousto}, \citenamefont {Marronetti},\ and\
  \citenamefont {Zlochower}}]{Campanelli:2005dd}%
  \BibitemOpen
  \bibfield  {author} {\bibinfo {author} {\bibfnamefont {Manuela}\ \bibnamefont
  {Campanelli}}, \bibinfo {author} {\bibfnamefont {C.~O.}\ \bibnamefont
  {Lousto}}, \bibinfo {author} {\bibfnamefont {P.}~\bibnamefont {Marronetti}},
  \ and\ \bibinfo {author} {\bibfnamefont {Y.}~\bibnamefont {Zlochower}},\
  }\bibfield  {title} {\enquote {\bibinfo {title} {{Accurate evolutions of
  orbiting black-hole binaries without excision}},}\ }\href {\doibase
  10.1103/PhysRevLett.96.111101} {\bibfield  {journal} {\bibinfo  {journal}
  {Phys. Rev. Lett.}\ }\textbf {\bibinfo {volume} {96}},\ \bibinfo {pages}
  {111101} (\bibinfo {year} {2006})},\ \Eprint
  {http://arxiv.org/abs/gr-qc/0511048} {arXiv:gr-qc/0511048} \BibitemShut
  {NoStop}%
\bibitem [{\citenamefont {Baker}\ \emph {et~al.}(2006)\citenamefont {Baker},
  \citenamefont {Centrella}, \citenamefont {Choi}, \citenamefont {Koppitz},\
  and\ \citenamefont {van Meter}}]{Baker:2005vv}%
  \BibitemOpen
  \bibfield  {author} {\bibinfo {author} {\bibfnamefont {John~G.}\ \bibnamefont
  {Baker}}, \bibinfo {author} {\bibfnamefont {Joan}\ \bibnamefont {Centrella}},
  \bibinfo {author} {\bibfnamefont {Dae-Il}\ \bibnamefont {Choi}}, \bibinfo
  {author} {\bibfnamefont {Michael}\ \bibnamefont {Koppitz}}, \ and\ \bibinfo
  {author} {\bibfnamefont {James}\ \bibnamefont {van Meter}},\ }\bibfield
  {title} {\enquote {\bibinfo {title} {{Gravitational wave extraction from an
  inspiraling configuration of merging black holes}},}\ }\href {\doibase
  10.1103/PhysRevLett.96.111102} {\bibfield  {journal} {\bibinfo  {journal}
  {Phys. Rev. Lett.}\ }\textbf {\bibinfo {volume} {96}},\ \bibinfo {pages}
  {111102} (\bibinfo {year} {2006})},\ \Eprint
  {http://arxiv.org/abs/gr-qc/0511103} {arXiv:gr-qc/0511103} \BibitemShut
  {NoStop}%
\bibitem [{\citenamefont {Aurrekoetxea}\ \emph {et~al.}(2023)\citenamefont
  {Aurrekoetxea}, \citenamefont {Clough},\ and\ \citenamefont
  {Lim}}]{Aurrekoetxea:2022mpw}%
  \BibitemOpen
  \bibfield  {author} {\bibinfo {author} {\bibfnamefont {Josu~C.}\ \bibnamefont
  {Aurrekoetxea}}, \bibinfo {author} {\bibfnamefont {Katy}\ \bibnamefont
  {Clough}}, \ and\ \bibinfo {author} {\bibfnamefont {Eugene~A.}\ \bibnamefont
  {Lim}},\ }\bibfield  {title} {\enquote {\bibinfo {title} {{CTTK: a new method
  to solve the initial data constraints in numerical relativity}},}\ }\href
  {\doibase 10.1088/1361-6382/acb883} {\bibfield  {journal} {\bibinfo
  {journal} {Class. Quant. Grav.}\ }\textbf {\bibinfo {volume} {40}},\ \bibinfo
  {pages} {075003} (\bibinfo {year} {2023})},\ \Eprint
  {http://arxiv.org/abs/2207.03125} {arXiv:2207.03125 [gr-qc]} \BibitemShut
  {NoStop}%
\bibitem [{\citenamefont {De~Luca}\ \emph
  {et~al.}(2020{\natexlab{b}})\citenamefont {De~Luca}, \citenamefont
  {Franciolini}, \citenamefont {Pani},\ and\ \citenamefont
  {Riotto}}]{DeLuca:2020qqa}%
  \BibitemOpen
  \bibfield  {author} {\bibinfo {author} {\bibfnamefont {V.}~\bibnamefont
  {De~Luca}}, \bibinfo {author} {\bibfnamefont {G.}~\bibnamefont
  {Franciolini}}, \bibinfo {author} {\bibfnamefont {P.}~\bibnamefont {Pani}}, \
  and\ \bibinfo {author} {\bibfnamefont {A.}~\bibnamefont {Riotto}},\
  }\bibfield  {title} {\enquote {\bibinfo {title} {{Primordial Black Holes
  Confront LIGO/Virgo data: Current situation}},}\ }\href {\doibase
  10.1088/1475-7516/2020/06/044} {\bibfield  {journal} {\bibinfo  {journal}
  {JCAP}\ }\textbf {\bibinfo {volume} {06}},\ \bibinfo {pages} {044} (\bibinfo
  {year} {2020}{\natexlab{b}})},\ \Eprint {http://arxiv.org/abs/2005.05641}
  {arXiv:2005.05641 [astro-ph.CO]} \BibitemShut {NoStop}%
\bibitem [{\citenamefont {Blinnikov}\ \emph {et~al.}(2016)\citenamefont
  {Blinnikov}, \citenamefont {Dolgov}, \citenamefont {Porayko},\ and\
  \citenamefont {Postnov}}]{Blinnikov:2016bxu}%
  \BibitemOpen
  \bibfield  {author} {\bibinfo {author} {\bibfnamefont {S.}~\bibnamefont
  {Blinnikov}}, \bibinfo {author} {\bibfnamefont {A.}~\bibnamefont {Dolgov}},
  \bibinfo {author} {\bibfnamefont {N.~K.}\ \bibnamefont {Porayko}}, \ and\
  \bibinfo {author} {\bibfnamefont {K.}~\bibnamefont {Postnov}},\ }\bibfield
  {title} {\enquote {\bibinfo {title} {{Solving puzzles of GW150914 by
  primordial black holes}},}\ }\href {\doibase 10.1088/1475-7516/2016/11/036}
  {\bibfield  {journal} {\bibinfo  {journal} {JCAP}\ }\textbf {\bibinfo
  {volume} {11}},\ \bibinfo {pages} {036} (\bibinfo {year} {2016})},\ \Eprint
  {http://arxiv.org/abs/1611.00541} {arXiv:1611.00541 [astro-ph.HE]}
  \BibitemShut {NoStop}%
\bibitem [{\citenamefont {Garc\'\i{}a-Bellido}(2017)}]{Garcia-Bellido:2017fdg}%
  \BibitemOpen
  \bibfield  {author} {\bibinfo {author} {\bibfnamefont {Juan}\ \bibnamefont
  {Garc\'\i{}a-Bellido}},\ }\bibfield  {title} {\enquote {\bibinfo {title}
  {{Massive Primordial Black Holes as Dark Matter and their detection with
  Gravitational Waves}},}\ }\href {\doibase 10.1088/1742-6596/840/1/012032}
  {\bibfield  {journal} {\bibinfo  {journal} {J. Phys. Conf. Ser.}\ }\textbf
  {\bibinfo {volume} {840}},\ \bibinfo {pages} {012032} (\bibinfo {year}
  {2017})},\ \Eprint {http://arxiv.org/abs/1702.08275} {arXiv:1702.08275
  [astro-ph.CO]} \BibitemShut {NoStop}%
\bibitem [{\citenamefont {Ivanov}\ \emph {et~al.}(1994)\citenamefont {Ivanov},
  \citenamefont {Naselsky},\ and\ \citenamefont {Novikov}}]{Ivanov:1994pa}%
  \BibitemOpen
  \bibfield  {author} {\bibinfo {author} {\bibfnamefont {P.}~\bibnamefont
  {Ivanov}}, \bibinfo {author} {\bibfnamefont {P.}~\bibnamefont {Naselsky}}, \
  and\ \bibinfo {author} {\bibfnamefont {I.}~\bibnamefont {Novikov}},\
  }\bibfield  {title} {\enquote {\bibinfo {title} {{Inflation and primordial
  black holes as dark matter}},}\ }\href {\doibase 10.1103/PhysRevD.50.7173}
  {\bibfield  {journal} {\bibinfo  {journal} {Phys. Rev. D}\ }\textbf {\bibinfo
  {volume} {50}},\ \bibinfo {pages} {7173--7178} (\bibinfo {year}
  {1994})}\BibitemShut {NoStop}%
\bibitem [{\citenamefont {Garcia-Bellido}\ \emph {et~al.}(1996)\citenamefont
  {Garcia-Bellido}, \citenamefont {Linde},\ and\ \citenamefont
  {Wands}}]{Garcia-Bellido:1996mdl}%
  \BibitemOpen
  \bibfield  {author} {\bibinfo {author} {\bibfnamefont {Juan}\ \bibnamefont
  {Garcia-Bellido}}, \bibinfo {author} {\bibfnamefont {Andrei~D.}\ \bibnamefont
  {Linde}}, \ and\ \bibinfo {author} {\bibfnamefont {David}\ \bibnamefont
  {Wands}},\ }\bibfield  {title} {\enquote {\bibinfo {title} {{Density
  perturbations and black hole formation in hybrid inflation}},}\ }\href
  {\doibase 10.1103/PhysRevD.54.6040} {\bibfield  {journal} {\bibinfo
  {journal} {Phys. Rev. D}\ }\textbf {\bibinfo {volume} {54}},\ \bibinfo
  {pages} {6040--6058} (\bibinfo {year} {1996})},\ \Eprint
  {http://arxiv.org/abs/astro-ph/9605094} {arXiv:astro-ph/9605094} \BibitemShut
  {NoStop}%
\bibitem [{\citenamefont {Ivanov}(1998)}]{Ivanov:1997ia}%
  \BibitemOpen
  \bibfield  {author} {\bibinfo {author} {\bibfnamefont {P.}~\bibnamefont
  {Ivanov}},\ }\bibfield  {title} {\enquote {\bibinfo {title} {{Nonlinear
  metric perturbations and production of primordial black holes}},}\ }\href
  {\doibase 10.1103/PhysRevD.57.7145} {\bibfield  {journal} {\bibinfo
  {journal} {Phys. Rev. D}\ }\textbf {\bibinfo {volume} {57}},\ \bibinfo
  {pages} {7145--7154} (\bibinfo {year} {1998})},\ \Eprint
  {http://arxiv.org/abs/astro-ph/9708224} {arXiv:astro-ph/9708224} \BibitemShut
  {NoStop}%
\bibitem [{\citenamefont {Coleman}(1977)}]{Coleman:1977py}%
  \BibitemOpen
  \bibfield  {author} {\bibinfo {author} {\bibfnamefont {Sidney~R.}\
  \bibnamefont {Coleman}},\ }\bibfield  {title} {\enquote {\bibinfo {title}
  {{The Fate of the False Vacuum. 1. Semiclassical Theory}},}\ }\href {\doibase
  10.1103/PhysRevD.16.1248} {\bibfield  {journal} {\bibinfo  {journal} {Phys.
  Rev. D}\ }\textbf {\bibinfo {volume} {15}},\ \bibinfo {pages} {2929--2936}
  (\bibinfo {year} {1977})},\ \bibinfo {note} {[Erratum: Phys.Rev.D 16, 1248
  (1977)]}\BibitemShut {NoStop}%
\bibitem [{\citenamefont {Callan}\ and\ \citenamefont
  {Coleman}(1977)}]{Callan:1977pt}%
  \BibitemOpen
  \bibfield  {author} {\bibinfo {author} {\bibfnamefont {Curtis~G.}\
  \bibnamefont {Callan}, \bibfnamefont {Jr.}}\ and\ \bibinfo {author}
  {\bibfnamefont {Sidney~R.}\ \bibnamefont {Coleman}},\ }\bibfield  {title}
  {\enquote {\bibinfo {title} {{The Fate of the False Vacuum. 2. First Quantum
  Corrections}},}\ }\href {\doibase 10.1103/PhysRevD.16.1762} {\bibfield
  {journal} {\bibinfo  {journal} {Phys. Rev. D}\ }\textbf {\bibinfo {volume}
  {16}},\ \bibinfo {pages} {1762--1768} (\bibinfo {year} {1977})}\BibitemShut
  {NoStop}%
\bibitem [{\citenamefont {Ellis}\ \emph {et~al.}(2020)\citenamefont {Ellis},
  \citenamefont {Lewicki},\ and\ \citenamefont {No}}]{Ellis:2020awk}%
  \BibitemOpen
  \bibfield  {author} {\bibinfo {author} {\bibfnamefont {John}\ \bibnamefont
  {Ellis}}, \bibinfo {author} {\bibfnamefont {Marek}\ \bibnamefont {Lewicki}},
  \ and\ \bibinfo {author} {\bibfnamefont {Jos\'e~Miguel}\ \bibnamefont {No}},\
  }\bibfield  {title} {\enquote {\bibinfo {title} {{Gravitational waves from
  first-order cosmological phase transitions: lifetime of the sound wave
  source}},}\ }\href {\doibase 10.1088/1475-7516/2020/07/050} {\bibfield
  {journal} {\bibinfo  {journal} {JCAP}\ }\textbf {\bibinfo {volume} {07}},\
  \bibinfo {pages} {050} (\bibinfo {year} {2020})},\ \Eprint
  {http://arxiv.org/abs/2003.07360} {arXiv:2003.07360 [hep-ph]} \BibitemShut
  {NoStop}%
\bibitem [{\citenamefont {Masoumi}\ \emph {et~al.}(2017)\citenamefont
  {Masoumi}, \citenamefont {Olum},\ and\ \citenamefont
  {Wachter}}]{Masoumi:2017trx}%
  \BibitemOpen
  \bibfield  {author} {\bibinfo {author} {\bibfnamefont {Ali}\ \bibnamefont
  {Masoumi}}, \bibinfo {author} {\bibfnamefont {Ken~D.}\ \bibnamefont {Olum}},
  \ and\ \bibinfo {author} {\bibfnamefont {Jeremy~M.}\ \bibnamefont
  {Wachter}},\ }\bibfield  {title} {\enquote {\bibinfo {title} {{Approximating
  tunneling rates in multi-dimensional field spaces}},}\ }\href {\doibase
  10.1088/1475-7516/2017/10/022} {\bibfield  {journal} {\bibinfo  {journal}
  {JCAP}\ }\textbf {\bibinfo {volume} {10}},\ \bibinfo {pages} {022} (\bibinfo
  {year} {2017})},\ \bibinfo {note} {[Erratum: JCAP 05, E01 (2023)]},\ \Eprint
  {http://arxiv.org/abs/1702.00356} {arXiv:1702.00356 [gr-qc]} \BibitemShut
  {NoStop}%
\bibitem [{\citenamefont {Guada}\ \emph {et~al.}(2020)\citenamefont {Guada},
  \citenamefont {Nemev\v{s}ek},\ and\ \citenamefont {Pintar}}]{Guada:2020xnz}%
  \BibitemOpen
  \bibfield  {author} {\bibinfo {author} {\bibfnamefont {Victor}\ \bibnamefont
  {Guada}}, \bibinfo {author} {\bibfnamefont {Miha}\ \bibnamefont
  {Nemev\v{s}ek}}, \ and\ \bibinfo {author} {\bibfnamefont {Matev\v{z}}\
  \bibnamefont {Pintar}},\ }\bibfield  {title} {\enquote {\bibinfo {title}
  {{FindBounce: Package for multi-field bounce actions}},}\ }\href {\doibase
  10.1016/j.cpc.2020.107480} {\bibfield  {journal} {\bibinfo  {journal}
  {Comput. Phys. Commun.}\ }\textbf {\bibinfo {volume} {256}},\ \bibinfo
  {pages} {107480} (\bibinfo {year} {2020})},\ \Eprint
  {http://arxiv.org/abs/2002.00881} {arXiv:2002.00881 [hep-ph]} \BibitemShut
  {NoStop}%
\bibitem [{\citenamefont {Wainwright}(2012)}]{Wainwright:2011kj}%
  \BibitemOpen
  \bibfield  {author} {\bibinfo {author} {\bibfnamefont {Carroll~L.}\
  \bibnamefont {Wainwright}},\ }\bibfield  {title} {\enquote {\bibinfo {title}
  {{CosmoTransitions: Computing Cosmological Phase Transition Temperatures and
  Bubble Profiles with Multiple Fields}},}\ }\href {\doibase
  10.1016/j.cpc.2012.04.004} {\bibfield  {journal} {\bibinfo  {journal}
  {Comput. Phys. Commun.}\ }\textbf {\bibinfo {volume} {183}},\ \bibinfo
  {pages} {2006--2013} (\bibinfo {year} {2012})},\ \Eprint
  {http://arxiv.org/abs/1109.4189} {arXiv:1109.4189 [hep-ph]} \BibitemShut
  {NoStop}%
\end{thebibliography}%


\end{document}